\pdfoutput=1
\documentclass[11pt]{article}
\usepackage[T1]{fontenc}
\usepackage[utf8]{inputenc}
\usepackage{newtxtext,newtxmath}
\usepackage[margin=1in]{geometry}
\usepackage{amsmath}
\usepackage{graphicx}
\usepackage{longtable,booktabs,array,calc}
\usepackage{textcomp}
\usepackage{newunicodechar}
\usepackage[hidelinks]{hyperref}
\usepackage{xurl}
\usepackage{placeins}
\usepackage{needspace}

\newunicodechar{−}{\ensuremath{-}}
\newunicodechar{·}{\textperiodcentered}
\newunicodechar{°}{\textdegree}
\newunicodechar{≥}{\ensuremath{\geq}}
\newunicodechar{≤}{\ensuremath{\leq}}
\newunicodechar{≈}{\ensuremath{\approx}}
\newunicodechar{×}{\texttimes}
\newunicodechar{μ}{\ensuremath{\upmu}}
\newunicodechar{λ}{\ensuremath{\lambda}}
\newunicodechar{±}{\textpm}

\newenvironment{scirefs}{\small\begin{list}{}{\setlength{\leftmargin}{2.6em}\setlength{\labelwidth}{2.3em}\setlength{\labelsep}{0.3em}\setlength{\itemsep}{1pt}\setlength{\parsep}{0pt}}}{\end{list}}
\newcommand{\paperauthors}{Will W. Qiang\textsuperscript{1,*}, Bo Huang\textsuperscript{2}, Xingjian Liu\textsuperscript{3,*}, Harry F. Lee\textsuperscript{1,*}}
\newcommand{\paperaffils}{\footnotesize\textsuperscript{1}Department of Geography and Resource Management, The Chinese University of Hong Kong, Hong Kong, China\\
\footnotesize\textsuperscript{2}Department of Geography, The University of Hong Kong, Hong Kong, China\\
\footnotesize\textsuperscript{3}Department of Urban Planning and Design, The University of Hong Kong, Hong Kong, China\\
\footnotesize\textsuperscript{*}Corresponding authors. Email: qiangwei@link.cuhk.edu.hk (W.W.Q.); xliu6@hku.hk (X.L.); harrylee@cuhk.edu.hk (H.F.L.)}

\hypersetup{pdftitle={Temporary gas turbines powering a US AI data center created a cross-state air pollution hotspot}, pdfauthor={Will W. Qiang, Bo Huang, Xingjian Liu, Harry F. Lee}}
\title{Temporary gas turbines powering a US AI data center created a cross-state air pollution hotspot}
\author{\parbox{0.97\textwidth}{\centering\paperauthors\\[4pt]\paperaffils}}
\date{}

\begin{document}
\maketitle

\begin{abstract}
Data centers increasingly use on-site gas turbines that regulators can treat as temporary. Two satellites show the fleet powering SpaceXAI's Colossus 2 raised local nitrogen dioxide columns by about half as it expanded, the most anomalous increase among 656 urban locations nationwide since 2020. Mississippi authorized the fleet, but two-thirds of the hotspot's 116,000 mostly Black residents live in Tennessee, and the hotspot contains no regulatory nitrogen dioxide monitor. The fleet causes an estimated four premature deaths yearly, valued at \$49 million, over 80\% outside Mississippi and nearly all from units without selective catalytic reduction. On new turbines, this control's benefits exceed costs over the service life, but not over temporary stays or for in-state damages alone. Standards following the equipment could recover these benefits.
\end{abstract}

US data centers used about 4.7\% of the nation's electricity in 2024 and could use 9.5 to 15.3\% by 2030, driven largely by artificial intelligence, and many regional grids cannot accommodate such growth without extensive upgrades (\emph{1}--\emph{4}). Because connecting a large data center to the grid can take years (\emph{2}, \emph{5}, \emph{6}), some developers have turned to on-site generation to start operating sooner. In the first half of 2026, US gas-fired capacity in development to power data centers directly nearly doubled, to 189 GW, and nearly half of the proposed generating technology consists of reciprocating engines or simple-cycle turbines, which emit more per unit of electricity than combined-cycle plants (\emph{7}). On-site generators emit nitrogen oxides (NO\emph{\textsubscript{x}}), which raise nitrogen dioxide (NO\textsubscript{2}) concentrations, contribute to the formation of ozone and fine particulate matter, and harm the health of nearby and downwind residents (\emph{8}--\emph{12}). However, the pollution exposures that data centers and their on-site generators create in host communities have rarely been measured directly (\emph{13}).

The MZX Tech LLC turbine fleet that powers SpaceXAI's Colossus 2 data center is an example of such on-site generation (\emph{14}, \emph{15}). Located in Southaven, Mississippi, near the Tennessee state line, the fleet had grown to an installed capacity of about 1.5 GW by May 2026 and about 1.8 GW by July (\emph{16}, \emph{17}). The May capacity alone was comparable to the nameplate capacity of the largest US nuclear generating unit (\emph{18}). Mississippi regulators allowed the fleet's units to operate as mobile sources exempt from stationary-source permitting and from the emission limits in a power plant permit (\emph{16}).

This arrangement raises two problems for pollution control. First, a turbine counts as temporary according to its period on site, the time it stays at one location. The planned stay must be less than 12 months under Mississippi's mobile-source authorization and no more than 24 months under the federal temporary category (\emph{16}, \emph{19}, \emph{20}). Yet the service life of the equipment can span many such periods. The fleet's lessor depreciates its turbines over 25 years, and one of its contracts to power another data center can run for up to 18 years including an extension option (\emph{21}). Just as vintage-differentiated regulation has distorted capital investment (\emph{22}, \emph{23}), evaluating the benefits of emission controls over a single period on site can make a control appear not cost-beneficial, even when its benefits would exceed its costs over the equipment's service life. Second, because a jurisdiction bears the costs of controlling its emissions but not the damages that cross its borders, pollution is often concentrated near borders (\emph{24}--\emph{29}). Mississippi authorized this fleet, but its pollution may cross into residential neighborhoods in Tennessee.

Existing reporting and monitoring do not track this pollution. The fleet does not publicly report hourly emissions, and the nearest recently active regulatory ground-level NO\textsubscript{2} monitor is about 25 km away (\emph{30}). An estimate of total NO\emph{\textsubscript{x}} emissions alone cannot show whether the resulting pollution is anomalous or whom it reaches. Satellite observations at kilometer-scale resolution can resolve NO\textsubscript{2} plumes from individual point sources (\emph{31}, \emph{32}), and the geostationary Tropospheric Emissions: Monitoring of Pollution (TEMPO) instrument now provides such observations every daylight hour (\emph{33}, \emph{34}).

Here, we compare hourly TEMPO NO\textsubscript{2} columns around the turbines in 2026 with the same days and hours in the two preceding years. Using the multiyear record of the TROPOspheric Monitoring Instrument (TROPOMI) (\emph{35}), a comparison across 656 US urban locations, the fleet's deployment history, and wind fields, we determine whether the change is anomalous and originates from the fleet. We also examine who is exposed, whether monitors detect the increase, where health damages fall, and how the evaluation horizon and the geographic scope of counted damages change the value of controls. Regulatory documents, securities filings, and gas transportation contracts filed with federal regulators show how the temporary classification relates to these horizons and boundaries.

The 2026 increase in NO\textsubscript{2} around the turbines was the most anomalous among 656 US urban locations in any year since 2020. It strengthened as the fleet expanded, extended downwind, and crossed the state line into Tennessee, where two-thirds of the 116,000 residents within the pollution hotspot live. Neither the two nearby power plants nor construction at the site before the turbines arrived can account for it. Using an engineering emission inventory, we estimate that the fleet causes about 4.4 premature deaths per year, more than 80\% outside Mississippi and nearly all from units without selective catalytic reduction (SCR). The temporary authorization of these units does not require SCR and ties each unit to a single period on site, over which SCR avoids less damage than it costs. Over 30 years, SCR on new turbines avoids damages worth 1.49 to 1.62 times its cost when damages in all states are counted. Emission standards that follow the equipment across sites could therefore recover the health benefits forgone under temporary authorization.

\section*{Results}

\subsection*{A nationally anomalous NO\textsubscript{2} increase near the turbines}

Two satellites observed an increase in NO\textsubscript{2} columns around the turbines. Following satellite studies of localized sources (\emph{31}, \emph{36}), we defined the near-field excess as the tropospheric NO\textsubscript{2} column within 5 km of the turbines (the core area) minus the column over a weighted reference area, made up of grid cells 10 to 20 km away that were weighted to resemble the core area in land use, traffic, and predeployment NO\textsubscript{2} columns (materials and methods). We paired each hourly TEMPO scan in 2026 taken between 17 and 20 UTC (noon to 4 p.m. local time) with the scans at the same hour and calendar day in 2024 and 2025 (fig. S1A and table S1). From April to July 2026, the near-field excess increased by 1.10 × 10\textsuperscript{15} molecules cm\textsuperscript{\ensuremath{-}2} (95\% confidence interval, 0.58 to 1.66, materials and methods), equivalent to about 49\% of the 2024--2025 mean column over the core area. Column values below are in units of 10\textsuperscript{15} molecules cm\textsuperscript{\ensuremath{-}2}. Because the retrieval's a priori NO\textsubscript{2} profiles omit the new source, these columns tend to understate the increase (materials and methods). The increase was concentrated north of the turbines and extended across the state line into Tennessee (Fig. 1A). With a fixed reference ring, without reference cells downwind of the nearby Colossus 1 data center, or with the three years reweighted to common meteorological conditions, the estimate ranged from 0.98 to 1.30, and every 95\% confidence interval excluded zero (table S2A and fig. S1B). None of the 15 other cities observed in the same TEMPO scans showed a significant increase from April to July, and with the same fixed reference ring, Southaven's increase was 5.6 times the largest of their changes (table S2C). A predeployment comparison of 2025 with 2024 showed no significant change (−0.09, −0.27 to 0.12, table S2A).

In a specification curve analysis (\emph{37}, \emph{38}), all 300 point estimates across seasons, observation windows, reference areas, and other analytic settings showed an increase, and 25 of the 27 confidence intervals computed from more than one observation date excluded zero (fig. S1C). TROPOMI, which differs from TEMPO in orbit, overpass time, and retrieval algorithm, also showed the near-field excess at the site fluctuating between 0.05 and 0.21 from 2019 to 2025 before rising by 0.79 (0.54 to 1.05) in 2026 relative to the previous two years (Fig. 1B and table S3A).

From April to July, the increase at Southaven was the most anomalous among US urban locations and the largest persistent local increase in the contiguous US. Across the 655 other locations, selected without reference to their 2026 changes (materials and methods), the median TROPOMI change was 0.01, indicating that neither revisions to the satellite product nor nationwide weather produced the increase. With the historical variability at each location taken into account, Southaven had a stabilized anomaly score of 4.63, whereas no other location exceeded 1.82 (Fig. 1C). When each year from 2020 to 2025 was scored in the same way against the other five years, the highest score at any location was 2.68 (table S3B). A search of the contiguous US for persistent multimonth peaks in the TEMPO increase, made without presupposing where emission sources lie (\emph{39}), ranked Southaven first in all 12 rankable specifications that used the full April--July window and first or second in all 52 rankable specifications, trailing only a grid cell over a Colorado wildfire whenever it ranked second (\emph{40}) (Fig. 1C, fig. S1D, table S3C, and materials and methods). In monthly TROPOMI synthetic-control analyses (\emph{41}), Southaven ranked first against 80 comparison locations without documented power-sector changes within 20 km under all four estimators (permutation \emph{P} = 1/81), whereas predeployment placebos were unremarkable (ranks 38 to 51 among the same 81 locations, table S3D).

The increase grew larger in late summer. Because each comparison uses the same calendar days in the previous two years, the seasonal cycle of background NO\textsubscript{2} is removed by design. From August to September 2026, the near-field excess increased by 2.69 (2.20 to 3.22), about 2.4 times the increase from April to July. With 2024 alone as the baseline, before any units arrived, the late-summer increase was 2.68, and excluding scans with Hazard Mapping System smoke over the core area in any paired year (\emph{42}) left increases of 1.00 in spring and 2.10 in late summer (table S2A). In the national TROPOMI comparison, the increase was 2.17 (1.22 to 3.21) over the same period, the largest among the 656 locations and nearly five times the second largest (tables S3A and S3B). In the source-agnostic TEMPO search, Southaven also ranked first among 227,072 eligible grid cells in the contiguous US over these two months (table S3C).

\begin{figure}[p]
\centering
\includegraphics[width=\textwidth,height=0.72\textheight,keepaspectratio]{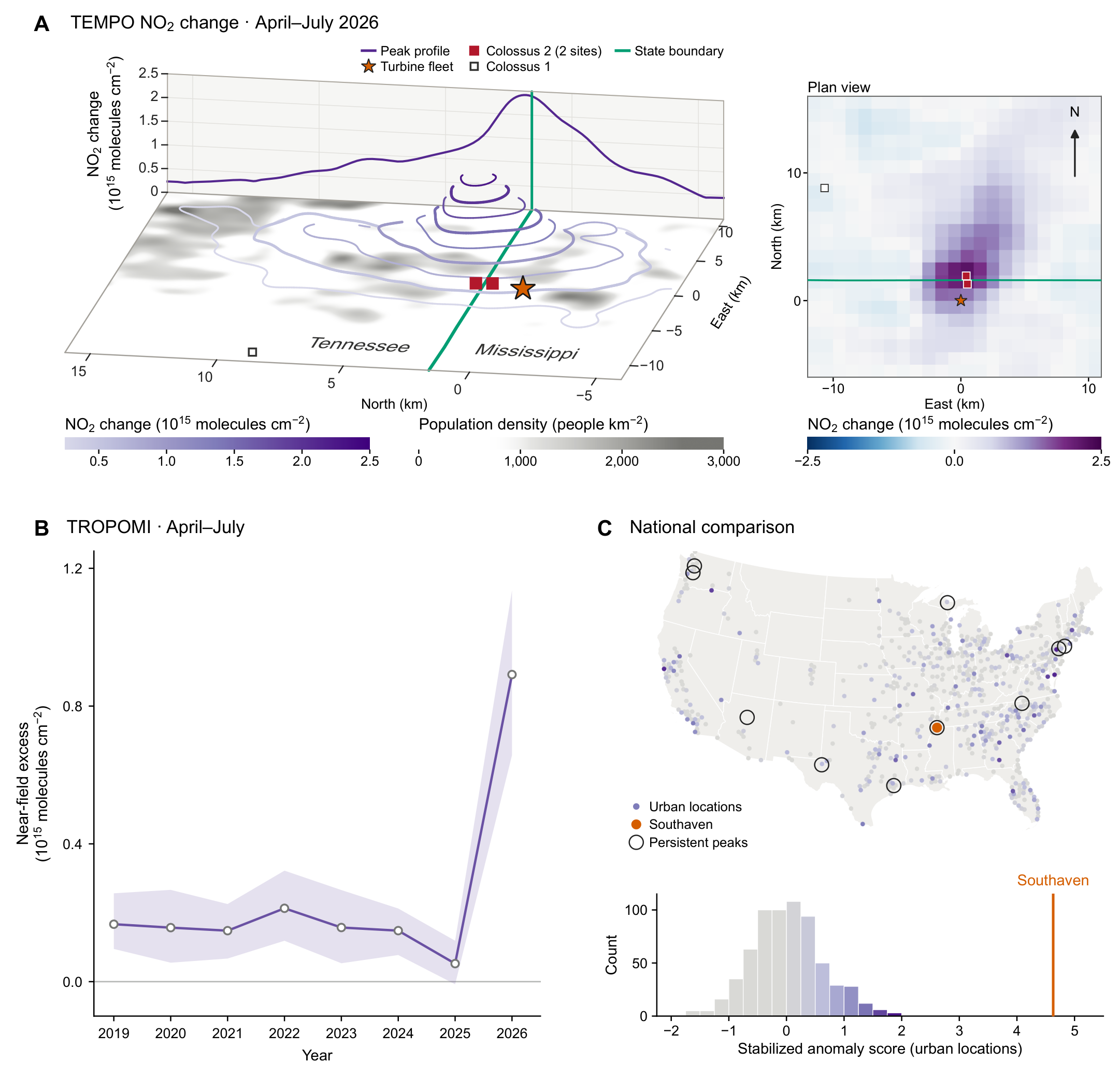}
\par\medskip
\begin{minipage}{\textwidth}\small \textbf{Fig. 1. The NO\textsubscript{2} increase near the turbines stood out nationally.} (\textbf{A}) Change in TEMPO NO\textsubscript{2} columns from April to July 2026 (17 to 20 UTC) relative to the paired 2024--2025 baseline, shown in perspective as contours over 2020 population density with a profile of the largest change along each east--west line (left) and in plan view (right). Green lines mark the Mississippi--Tennessee state line, the star marks the turbine fleet, filled squares mark Colossus 2, and the open square marks Colossus 1. (\textbf{B}) TROPOMI Level-2 near-field excess at the site from April to July, 2019 to 2026, with 95\% confidence intervals. (\textbf{C}) Stabilized anomaly scores for the April to July TROPOMI change at Southaven and 655 other US urban locations, shown on the map and as a histogram. Circles mark the 10 highest persistent multimonth local peaks found by the source-agnostic TEMPO search under the main specification, and orange marks Southaven. NO\textsubscript{2} columns are in units of 10\textsuperscript{15} molecules cm\textsuperscript{\ensuremath{-}2}.\end{minipage}
\end{figure}

\subsection*{An increase that grew with the fleet and extended downwind}

The NO\textsubscript{2} rise began after the first units arrived in August 2025 and strengthened as the fleet grew. While fleet capacity remained below 0.7 GW through March 2026, the increase stayed small (Fig. 2A and supplementary text). In spring 2026, the fleet expanded, reaching 1.0 GW in April and 1.5 GW in May, and the near-field excess relative to the same month in 2024 rose accordingly, from 0.61 in April to between 1.07 and 1.49 from May to July. Because units with SCR emit an order of magnitude less NO\emph{\textsubscript{x}}, we regressed each month's change on the on-site capacity of units without SCR, controlling for season (materials and methods). Across all 14 months since deployment, each additional 100 MW of such capacity corresponded to an increase of 0.16 (heteroskedasticity-robust 95\% interval, 0.05 to 0.26) in the near-field excess (Fig. 2B), and 0.15 with the monthly change in inverse wind speed also held fixed (fig. S2D and supplementary text). Independent of the satellite data, contracted pipeline capacity for the site's delivery point also increased on 7 May 2026, in the month the fleet reached 1.5 GW (\emph{43}).

The NO\textsubscript{2} rise also extended downwind. After each scan was rotated around the turbines to a common wind direction (\emph{32}), the 2026 change showed a band of elevated NO\textsubscript{2} extending downwind (materials and methods), and the column difference between the downwind and upwind boxes increased by 1.46 (1.11 to 1.78) relative to the previous two years (tables S2B and S4). Because year-to-year changes in wind direction could also carry existing urban pollution into the downwind box, we rotated the scans of all three years with the winds of a single year, which holds existing sources in place. The difference then increased by 1.51 (1.13 to 1.86) with the 2026 winds but by only 0.06 and 0.25 with the 2024 and 2025 winds (Fig. 2C and fig. S2A). The added NO\textsubscript{2} thus follows the 2026 winds, consistent with a new emission source at the site.

The added NO\textsubscript{2} was concentrated near the turbines and during periods of weak wind. This pattern is a signature of a local source, whose column enhancement grows as weak winds carry its emissions away more slowly (\emph{36}). Before deployment, the near-field excess was unrelated to inverse wind speed (\emph{r} = −0.008), whereas in 2026 it rose under weak winds (\emph{r} = 0.363, difference 0.371, 95\% confidence interval, 0.138 to 0.596, Fig. 2D). The increase declined with distance, falling halfway to its far-field level at about 3.9 km (2.9 to 5.1 km) from the turbines (Fig. 2E and materials and methods). From April to July, periods with High-Resolution Rapid Refresh (HRRR) wind speeds below 3 m s\textsuperscript{\ensuremath{-}1} accounted for only 37\% of observation opportunities but contributed about three-quarters of the increase (fig. S2B and table S4).

None of the other known sources we examined could account for the increase. The Tennessee Valley Authority (TVA) Southaven combined-cycle plant, 1.5 km from the turbines, and the TVA Allen plant, 14 km to the northwest, report hourly emissions to the U.S. Environmental Protection Agency (EPA) (\emph{44}). Their April to June emissions were about 16\% and 13\% higher, respectively, than in the same months of the previous two years, but their simulated contribution to the near-field excess was close to zero. Under April to June transport conditions, producing the observed increase would take about 75 times their combined emissions (fig. S2E, tables S5A and S5B, and supplementary text). Controlling for roads, traffic, and land use also left the increase at Southaven nearly unchanged (table S5A). Construction had begun at the site before the turbines arrived, yet the change in the near-field excess from May to July 2025 relative to 2024 was −0.14 (−0.33 to 0.13, fig. S2E). At a large data center campus under construction in Richland Parish, Louisiana, the same method detected no increase from April to July 2025 (0.05, −0.08 to 0.18, table S6, fig. S2E, and supplementary text).

\begin{figure}[p]
\centering
\includegraphics[width=\textwidth,height=0.72\textheight,keepaspectratio]{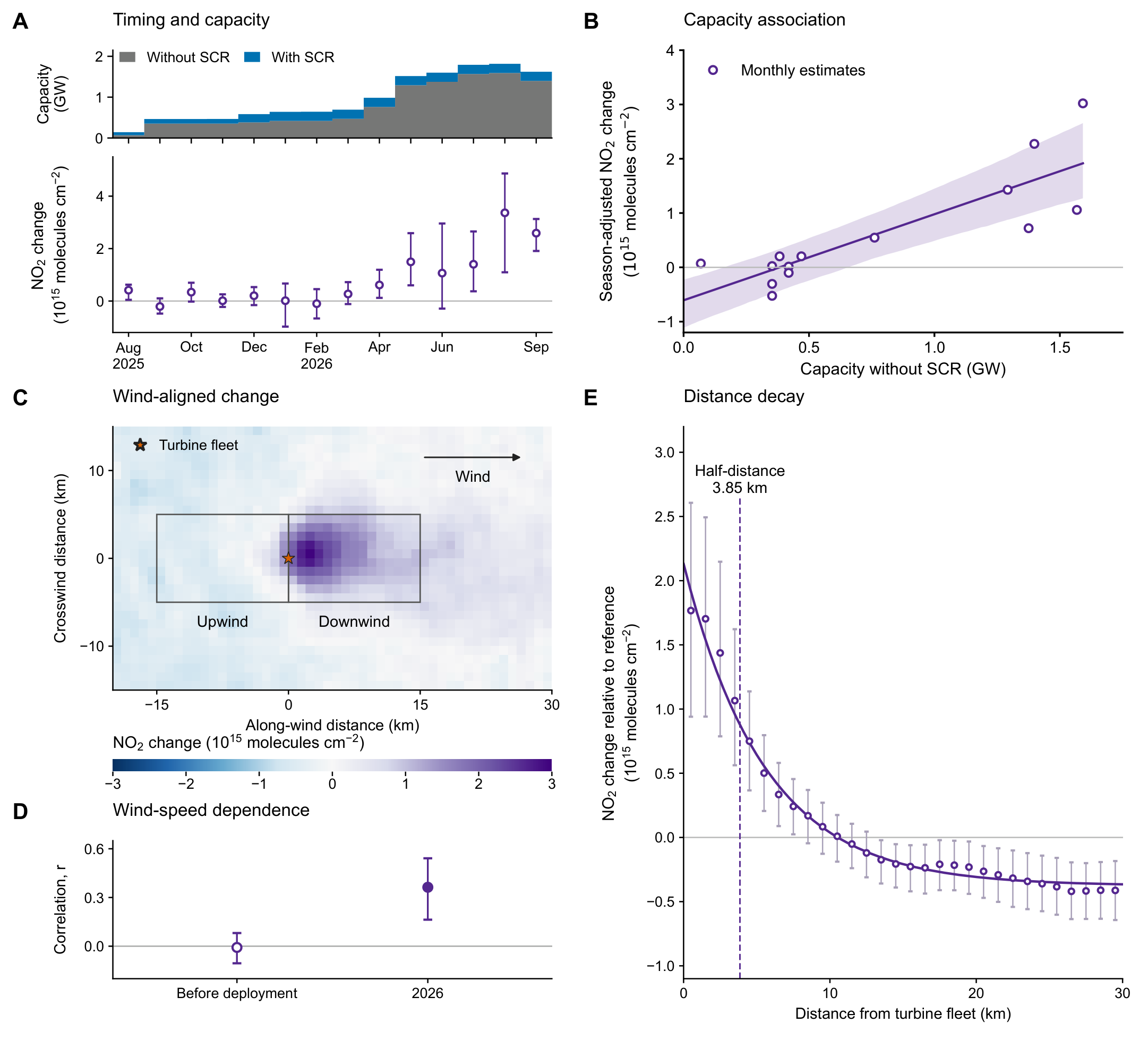}
\par\medskip
\begin{minipage}{\textwidth}\small \textbf{Fig. 2. The NO\textsubscript{2} increase grew with the fleet and extended downwind.} (\textbf{A}) On-site installed capacity by SCR status (top) and monthly change in the near-field excess relative to the same month of 2024 (bottom), August 2025 to September 2026, with 95\% confidence intervals. (\textbf{B}) Season-adjusted monthly change versus the capacity of units without SCR, with the linear fit and a 95\% band from monthly sampling uncertainty. (\textbf{C}) Change from April to July 2026 after rotating scans from all three years to a common wind direction, aligned with the 2026 winds (310 paired scans). The star marks the fleet, and the boxes mark the upwind and downwind comparison areas. (\textbf{D}) April--July correlation between the near-field excess and inverse wind speed before deployment and in 2026, with 95\% confidence intervals. (\textbf{E}) April--July change relative to the weighted reference area by distance from the fleet. Points are annulus means with 95\% confidence intervals. The curve is an exponential fit, and the dashed line marks the distance at which the change falls halfway to its far-field level. Winds are from the High-Resolution Rapid Refresh model in C and the ERA5 reanalysis in D. NO\textsubscript{2} column changes are in 10\textsuperscript{15} molecules cm\textsuperscript{\ensuremath{-}2}.\end{minipage}
\end{figure}

\subsection*{A cross-state hotspot over predominantly Black neighborhoods without NO\textsubscript{2} monitors}

The increase reached densely populated neighborhoods of southern Memphis (Figs. 1A and 3A). About 116,100 people (103,000 to 137,800) live within a hotspot, defined as the area where the April to September increase was at least 0.5 × 10\textsuperscript{15} molecules cm\textsuperscript{\ensuremath{-}2}. Of these residents, 65.5\% (52.4 to 76.3\%) live in Tennessee (Fig. 3A, table S7A, and materials and methods). Hotspot residents include 30,100 children and 18,000 adults aged 65 years and older (table S7B).

Residents within the hotspot are predominantly Black and experience a higher poverty rate than residents of the surrounding counties, consistent with national racial disparities in NO\textsubscript{2} exposure (\emph{45}--\emph{47}). Among them, 76\% are non-Hispanic Black, rising to 93\% on the Tennessee side, and 27\% live below the federal poverty line (table S7A). By comparison, the corresponding shares are 52\% and 18\% in Shelby County, Tennessee, and 33\% and 10\% in DeSoto County, Mississippi (\emph{48}). The disparity reflects the increase crossing the state line into predominantly Black neighborhoods of southern Memphis (table S8).

The regulatory NO\textsubscript{2} monitoring network did not cover the hotspot, even though its measurements determine compliance with the National Ambient Air Quality Standards (\emph{49}, \emph{50}). The two nearest recently active NO\textsubscript{2} monitors are 25 km away in Tennessee and 28 km away in Arkansas, both outside the hotspot (Fig. 3A), and satellite columns above both showed no detectable increase relative to their surroundings (table S2E). In EPA data through May 2026, the Tennessee monitor reported no 2026 data, and the only monitor in Mississippi with data in 2026 was about 530 km away (\emph{30}). Mississippi authorized the fleet and a permanent power plant at the same site under different rules. The permanent plant's Prevention of Significant Deterioration permit of March 2026 requires SCR, continuous NO\emph{\textsubscript{x}} emission monitoring, and ammonia-slip limits (\emph{51}), whereas the fleet operates under a state letter and agreed orders, without construction or operating permits, continuous NO\emph{\textsubscript{x}} emission monitoring, or ambient monitoring (\emph{16}, \emph{17}, \emph{19}).

\subsection*{Health damages borne mostly outside the authorizing state}

Our engineering emission inventory, built from the fleet roster, emission factors for each turbine model, and June 2026 operating fractions applied to each unit's time on site (materials and methods and Data S1), gives an average fleet emission of about 216 kg of NO\emph{\textsubscript{x}} per hour from April to July. This rate is equivalent to about 1,900 metric tons per year, or about five times the combined annual limit for the permanent plant's turbines (table S9). Water-injected TM2500 units of the Gen4, Gen6, and Gen8 models emit 0.36 kg of NO\emph{\textsubscript{x}} per megawatt-hour in the inventory, close to stack-test rates of 0.36 and 0.38 kg for two water-injected TM2500 units at another site (fig. S3A) (\emph{52}). Scheduled pipeline gas deliveries combined with stack-test factors imply 181 to 309 kg of NO\emph{\textsubscript{x}} per hour from mid-July to September, compatible with the inventory (table S9C and supplementary text).

We used two reduced-complexity air quality models, the Intervention Model for Air Pollution (InMAP) and AP4, to convert these emissions into premature deaths from particulate nitrate, a component of fine particulate matter (PM\textsubscript{2.5}) that forms from NO\emph{\textsubscript{x}} in the atmosphere (\emph{53}--\emph{56}). InMAP, the primary model, distributes the emissions among release heights according to hourly plume rise and estimates that the fleet causes about 4.4 premature deaths per year, valued at about \$49 million (2019 dollars) (materials and methods and tables S10A and S10C). We refer to these monetized impacts as damages. With ground-level release, AP4, used for comparison, gives 6.0 deaths and InMAP 5.4 (fig. S3B, table S10A, and supplementary text). With the concentration--response function from a more recent cohort of US adults (\emph{57}) in place of an earlier one (\emph{58}), the InMAP estimate nearly doubles to about 8.5 deaths per year (table S10B). These estimates exclude particulate matter emitted directly by the turbines, health effects associated with NO\textsubscript{2} exposure such as childhood asthma (\emph{8}, \emph{59}), and the contribution of NO\emph{\textsubscript{x}} to ozone.

The NO\textsubscript{2} increase falls mainly on neighborhoods near the turbines on the Tennessee side, whereas most of the damage from particulate nitrate occurs beyond the plume visible from satellites (Fig. 3B). Because particulate nitrate forms over hours of atmospheric transport, InMAP places only half of the deaths within about 150 km of the site (Fig. 3C). Tennessee bears 35.2\% of the deaths, nearly twice the 18.2\% in Mississippi, and other states and the District of Columbia bear 46.6\%. In all, 81.8\% of the deaths occur outside the authorizing state (Fig. 3D and supplementary text). Under AP4 with ground-level release, this share is about 97\% (fig. S3C and supplementary text).

\begin{figure}[p]
\centering
\includegraphics[width=\textwidth,height=0.72\textheight,keepaspectratio]{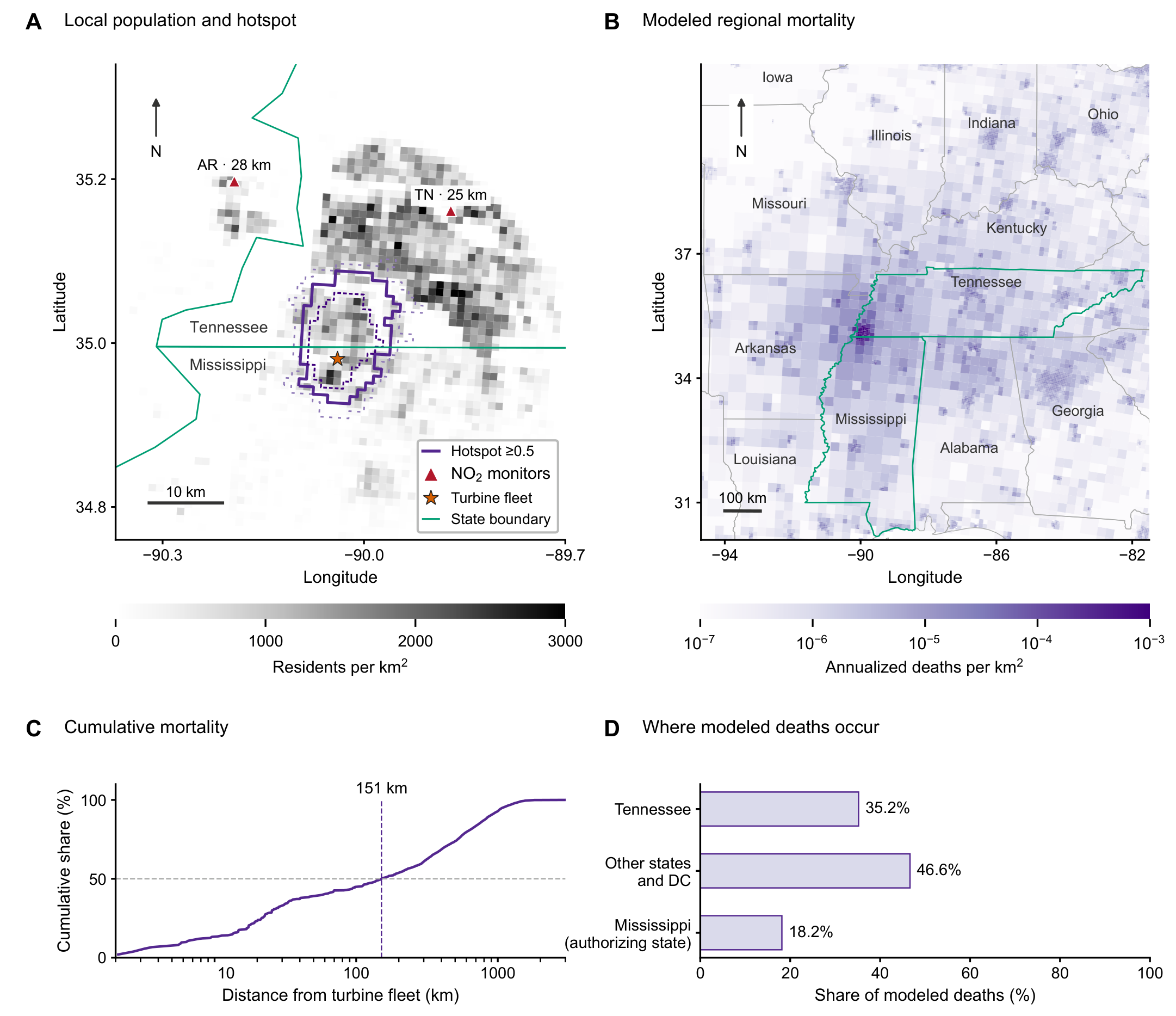}
\par\medskip
\begin{minipage}{\textwidth}\small \textbf{Fig. 3. Residents in the hotspot and the distribution of health damages.} (\textbf{A}) Hotspot contours of the April to September NO\textsubscript{2} increase over 2020 Census population allocated to buildings. The thick line marks 0.5 × 10\textsuperscript{15} molecules cm\textsuperscript{\ensuremath{-}2}, and lighter and darker dashed lines mark 0.25 and 1.0, respectively. Triangles mark the nearest recently active regulatory NO\textsubscript{2} monitors in Arkansas and Tennessee, labeled with their distance from the fleet. The star marks the fleet, and green lines mark state boundaries. (\textbf{B}) Annual premature deaths from particulate nitrate per square kilometer on a logarithmic color scale, estimated by InMAP for the engineering inventory (216 kg of NO\emph{\textsubscript{x}} per hour). (\textbf{C}) Cumulative share of these deaths by distance from the fleet, assigned by receptor grid centroid. The dashed lines mark the 50th percentile and the corresponding distance of 151 km. (\textbf{D}) Distribution of these deaths by state, with Mississippi, the state that authorized the fleet, marked. ``Other states'' includes the District of Columbia. Panels B to D total 4.36 deaths per year.\end{minipage}
\end{figure}

\subsection*{Nearly all damage from units without SCR}

In the engineering inventory for April to July 2026, units without SCR generated about 78\% of the fleet's electricity but emitted about 98\% of its NO\emph{\textsubscript{x}} (Fig. 4A). They emitted an average of about 0.46 kg of NO\emph{\textsubscript{x}} per megawatt-hour, more than twice the marginal emission rate of the regional grid (0.18 kg) (\emph{60}). Units with SCR emitted about 0.04 kg, the rate implied by the 2-ppm limit (parts per million, dry at 15\% oxygen) of the permanent power plant at the same site, and stack tests of seven turbines with SCR at a data center campus in Abilene, Texas, showed about 0.02 kg (fig. S3A) (\emph{51}, \emph{61}, \emph{62}). With SCR, on-site turbines can therefore emit less NO\emph{\textsubscript{x}} per megawatt-hour than marginal grid generation. Per megawatt-hour, the fleet's NO\emph{\textsubscript{x}}-related health damage is about 3.6 times that of grid power. When sulfur dioxide and primary particulate matter are also counted, the fleet's damage in the primary model is about equal to that of grid power, because most grid damage from these pollutants comes from coal-fired units. SCR would bring damages from NO\emph{\textsubscript{x}}, sulfur dioxide, and primary particulate matter together below those of the grid under every model, even with ammonia slip at the limits permitted for the permanent plant (fig. S3D, tables S11A to S11C, and supplementary text).

SCR can be used on mobile turbines. The units with SCR in this fleet are mobile packages covered by the same agreed order as the other units, and seven of the first eight units on site had SCR (fig. S4A and supplementary text). However, the federal new source performance standards allow eligible new natural gas turbines in the temporary category to emit NO\emph{\textsubscript{x}} at concentrations of up to 25 ppm (\emph{20}, \emph{63}), and Mississippi's agreed order does not require SCR on mobile turbines (\emph{16}). Across the state line, the 2025 permit for the Colossus 1 site in Memphis requires 15 SMT-130 turbines, the model fitted with SCR in this fleet, to use SCR and meet the same 2-ppm limit (\emph{64}).

Installing SCR would avoid most of the damage, whereas changing how the fleet operates would have a smaller effect (Fig. 4A and table S12). Retrofitting SCR with a 90\% removal efficiency on all units that lack it (\emph{65}) reduces modeled deaths by 88\%, and bringing all units to the 2-ppm limit reduces them by 91\%. With June delivered electricity held constant, dispatching the lowest-emitting units first reduces deaths by 20 to 34\%, depending on the load fraction (materials and methods and table S12A).

\begin{figure}[p]
\centering
\includegraphics[width=\textwidth,height=0.72\textheight,keepaspectratio]{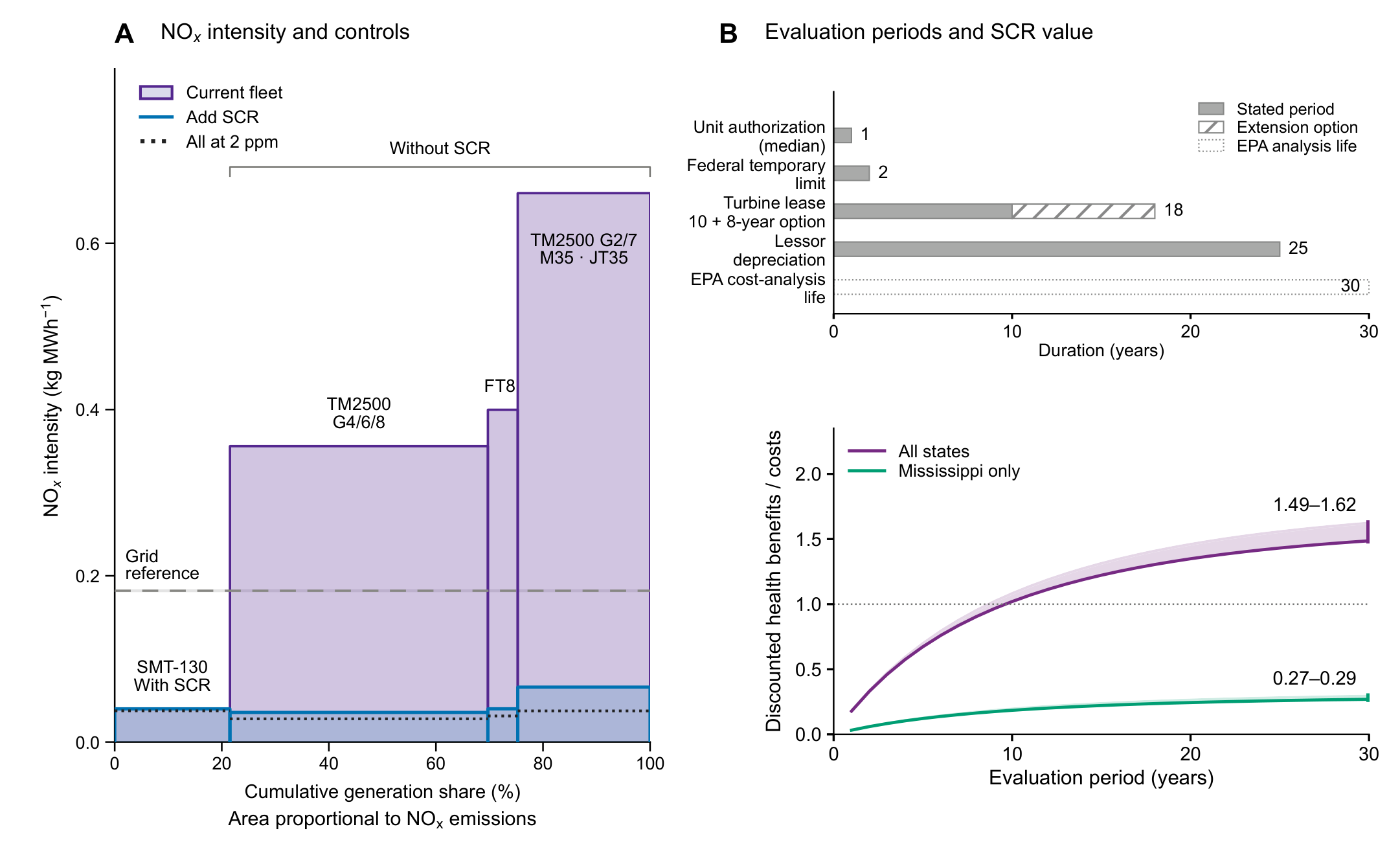}
\par\medskip
\begin{minipage}{\textwidth}\small \textbf{Fig. 4. Nearly all damage comes from units without} \textbf{SCR, and the value of SCR grows with the evaluation horizon.} (\textbf{A}) NO\emph{\textsubscript{x}} emissions per megawatt-hour (bar height) and share of fleet generation (bar width) by unit type in the April to July 2026 engineering inventory. Bar area is NO\emph{\textsubscript{x}} emitted. Blue outlines show emissions with SCR on all units, and dotted lines show all units at the permanent plant's 2-ppm limit. The dashed line marks the regional grid's marginal NO\emph{\textsubscript{x}} emission rate. (\textbf{B}) Regulatory and commercial horizons (top) and the ratio of discounted health benefits to costs of SCR on new turbines over 1 to 30 years (bottom), counting damages in all states or Mississippi only. Regulatory bars show the median one-year authorization and the two-year federal temporary limit. Other bars show the lessor's amended contract with another data center (10 years plus an eight-year extension option), the lessor's 25-year depreciation period, and the 30-year useful life in the EPA's economic impact analysis. Bands span two tax and insurance treatments, and the dotted line marks a ratio of 1. Ratios exclude ammonia-slip damages and use InMAP, 2019 dollars, and a 7\% discount rate.\end{minipage}
\end{figure}

\subsection*{Undervaluation of emission controls during temporary authorization}

Regulatory records treat these turbines as short-term equipment, whereas commercial arrangements span many years (Fig. 4B and fig. S4B). The federal new source performance standards set less stringent limits, based on combustion controls, for temporary turbines, partly because commenters argued that the complexity of SCR would defeat the purpose of bringing in turbines quickly for short-term power (\emph{20}). Mississippi's agreed order authorized most units to stay on site for about one year but allowed 13 of the 14 units with SCR to stay longer, through the end of 2026, in part because they can meet the permanent power plant's emission limits (\emph{16}). Under the amended agreed order of September 2026, the deadlines for most units without SCR fall between March and July 2027, by which time the fleet will have been at the site for nearly two years (fig. S4A) (\emph{17}). The equipment lessor's contract to supply power for this data center has an initial term of seven years, and its two most recent contracts with other data centers have 10-year base terms, with extension options of five and eight years (\emph{21}, \emph{66}). The site operator also holds natural gas transportation rights through 2035 (\emph{43}), and the permanent plant at the same site, about 1,240 MW, is under construction (\emph{16}, \emph{51}) (supplementary text).

These records tie SCR to how long a turbine is expected to stay. The federal standard forgoes SCR for short-term use (\emph{20}), Mississippi's agreed order extends the stay of units that have it (\emph{16}), and the equipment lessor positions its SCR investments for multiyear deployments, noting that multiyear air permits in some jurisdictions require SCR (\emph{67}, \emph{68}). We built a simple accounting framework (fig. S4D, materials and methods, and supplementary text) in which emission controls are evaluated over a turbine's period on site, while the equipment continues to operate and cause damages elsewhere after it leaves. A control fails this evaluation when its cost exceeds the site threshold, the damages it avoids during the period on site. It benefits society when its cost is below the social threshold, the damages it avoids over the full service life of the equipment. Here, SCR on new turbines costs more than the site threshold and less than the social threshold. With InMAP damages and without ammonia slip, it avoids enough damage to cover its cost only after about 9 to 10 years at a 7\% discount rate, longer than the one- or two-year stays allowed under temporary authorization (Fig. 4B). For retrofits, benefits cover costs after about 12 to 14 years, within the 15-year remaining life that the EPA assumed for modified or reconstructed turbines (\emph{69}) but longer than the remaining period on site (table S13F). If the authorizing state counted only in-state damages, even the 30-year ratio for SCR on new turbines would fall to 0.27 to 0.29 (Fig. 4B).

Over the remaining period on site, retrofitting SCR on the 29 evaluated units without SCR would avoid damages worth less than 8\% of the cost (tables S13A and S13F). For SCR installed on new turbines and used for 30 years, the useful life assumed in the EPA's economic impact analysis (\emph{70}), the avoided damages reach 1.49 to 1.62 times the cost, excluding ammonia-slip damages (table S13B). The ratio remains above 1 with the ammonia slip permitted for the permanent plant at the same site (1.20 to 1.47) (\emph{51}), with ammonia slip below about 4.7 to 5.6 ppm dry at 15\% oxygen, and, without ammonia slip, with capital costs up to 82 to 96\% above the estimate (fig. S4C, tables S13B and S13C, and supplementary text). Under the EPA's costing method for its 2026 turbine standards (\emph{69}), SCR on new turbines costs about \$13,100 to \$14,300 (2019 dollars) per metric ton of NO\emph{\textsubscript{x}} removed over 30 years (table S13F). The EPA's economic analysis for these standards did not monetize the health benefits of NO\emph{\textsubscript{x}} reductions (\emph{70}). Without ammonia slip, the 30-year ratio rises to 1.81 to 1.98 when ozone-related benefits are added (table S13E) (\emph{71}), to 2.11 to 2.37 at a 3\% discount rate (table S13B), and to 2.89 to 3.16 with the concentration--response function of Pope et al. (\emph{57}), under which ratios over one or two years remain below 1 (supplementary text).

\section*{Discussion}

The temporarily authorized fleet produced the most anomalous NO\textsubscript{2} increase among 656 US urban locations since 2020. Nearly all of its health damage comes from units without SCR, and most occurs outside the authorizing state. An evaluation limited to a turbine's single period on site or to damage borne within the state rejects SCR on new turbines, which is worth its cost over the full service life of the equipment. Mobile turbines are authorized only for temporary stays but can keep operating elsewhere for years. The equipment lessor holds exclusive rights, subject to certain conditions, to manage and redeploy the assets of its joint venture with the operator (\emph{66}). In South Australia, for example, the nine TM2500 turbines that the state government installed in 2017 as an emergency power plant were leased to private operators and moved by 2022 to two other power stations, where they resumed operation under 25-year leases (\emph{72}--\emph{75}). Temporary units can also stay in place. In California, two TM2500 units certified in 2021 under a state emergency proclamation to operate temporarily at a power plant were brought in 2026 under that plant's regular license, which requires them to operate SCR and, during normal operation, to emit no more than 2.5 ppm of NO\emph{\textsubscript{x}} (\emph{76}, \emph{77}) (supplementary text).

No other known source matches the timing of deployment, the wind direction, the decline with distance, and the weak-wind dependence that appeared only in 2026. Under the observing conditions of the spring sample, the change at this site would be detectable at emissions of about 100 kg per hour, less than half the fleet's rate (materials and methods and supplementary text). The upcoming unit departures will provide further evidence. If the increase comes mainly from units without SCR, the near-field excess should fall clearly below its 2026 level for the same months once those units leave.

Cross-state damage is common in the power sector. About 70\% of the premature deaths from PM\textsubscript{2.5} and ozone caused by power-sector emissions occur outside the emitting state (\emph{78}), and in 35 states, most of the PM\textsubscript{2.5} deaths caused by in-state power plants occur in other states (\emph{79}). A state evaluating a temporarily authorized unit would undervalue emission controls in two ways, by counting benefits over a single period on site and by counting only in-state damage. A common remedy for the second requires the source jurisdiction to curb emissions that significantly contribute to downwind states' failure to meet air quality standards, as the US Clean Air Act's good neighbor provision does (\emph{80}). This case shows, however, that correcting the geographic scope alone is not enough. Counting damages in all states rather than in Mississippi alone raises the benefit--cost ratio for SCR on new turbines about 5.5-fold, yet the ratio remains about 0.18 over one year and 0.33 to 0.34 over two years (table S13F). Extending the horizon to 30 years raises it eight- to ninefold relative to one year, to 1.49 to 1.62 (Fig. 4B).

Fitting SCR to a new turbine before it enters service requires about 20\% less capital than retrofitting it later (\emph{69}), and over 30 years the benefits exceed the cost. Once a turbine enters service without SCR, however, a retrofit fails an evaluation limited to the period on site, because the remaining period is too short for the benefits to cover the cost. If each evaluation counts only one period on site and no separate requirement applies, the turbine may never receive SCR during its service life.

The current arrangement limits how long each unit stays and bars new units after 15 July 2026, but it sets no cap on the fleet's total emissions (\emph{16}, \emph{17}). Deadlines were set mainly by the arrival date of each unit. Under the scheduled deadlines, the authorizations of all units with SCR expire at the end of 2026, and the units still authorized in January 2027 would have a capacity-weighted NO\emph{\textsubscript{x}} emission rate about 16\% higher than that of the units authorized in December (supplementary text). Each unit is reviewed separately, although successive arrivals will have kept the site occupied for nearly two years. In reviewing the permanent plant's permit, the state agency placed the temporary turbines named by commenters outside the review because they did not require a construction permit (\emph{81}).

The pollution largely escapes both regulatory monitoring and the authorizing state's emission accounting. The November 2025 draft ozone maintenance plan for DeSoto County, which the state commission approved for submission to the EPA in January 2026, and its emission inventory workbooks do not include the fleet, even though the plan projects emissions through 2036 (\emph{82}, \emph{83}). Satellite observations can fill the monitoring gap. A retrospective cumulative analysis shows that the change was detectable by May 2026, the second month of the fleet's spring expansion (materials and methods and supplementary text).

Fitting SCR to the units that lack it, a control already used on 14 mobile units in this fleet, would prevent about 88\% of the fleet's estimated deaths (table S12A). An immediate improvement that needs no new equipment, and which the agreed order does not require, is to have the cleanest units run first and leave last (materials and methods and supplementary text) (\emph{84}). Making every extension of a unit's stay conditional on meeting the permanent plant's limits could encourage installing SCR before a turbine enters service. A further step would be to extend review from individual units to the entire site, for example by applying the cumulative counting of same-function replacements used in the 2026 federal turbine standards to state mobile-source authorizations (\emph{20}, \emph{63}). Because these measures apply at one site, they do not by themselves ensure control at later deployments, and a fuller remedy is for emission standards to follow the equipment. The 2026 federal turbine standards already provide for this route by excluding portable turbines certified to future nonroad emission standards, and a proposed nonroad combustion turbine rule entered White House review in October 2026 (\emph{20}, \emph{85}). If mobile turbines had to meet the emission level achievable with SCR wherever they operate, controls would be specified at initial deployment, as already required for the permanent plant at the same site, and the same standard would apply on both sides of a state line.

The conditions that created this problem extend beyond Southaven. The less stringent federal standards for temporary turbines apply nationwide (\emph{20}). Mobile turbines can be leased and relocated, and data center developers plan to pair large amounts of new capacity with on-site generation (\emph{6}, \emph{66}). If a fleet like this one operated at several other publicly documented on-site generation projects, including projects 125 to 200 km from a state line, the in-state share of damage would be at most about 52\% under InMAP and below 30\% under AP4 (table S10D). At the location, operating intensity, and population of this site, we estimate that each gigawatt of turbines without SCR causes about 3.4 premature deaths and about \$38 million in damages per year (table S10A). Mobility also changes the pace of cross-border pollution. Mobile units can arrive within months and leave within about a year, faster than updates to emission inventories (\emph{83}, \emph{86}) or interstate petitions, such as Delaware's and Maryland's Section 126 petitions under the federal Clean Air Act, which took about two years from filing to a final decision (\emph{87}, \emph{88}). Remedies that act only after units arrive can therefore come too late. If emission standards follow the equipment and satellite verification supplements them, on-site generation can expand while shifting far smaller health damages onto nearby and downwind residents.

\FloatBarrier
\section*{References and Notes}
\begin{scirefs}
\item[1.] D. Chen, Z. Zhou, Y. Cai, J. Qin, A. L. Katchova, L. Chen, Artificial intelligence data centers could reach one percent of global electricity demand by 2030. \emph{Commun. Sustain.} \textbf{1}, 147 (2026). doi:10.1038/s44458-026-00152-5
\item[2.] X. Chen, X. Wang, A. Colacelli, M. Lee, L. Xie, Electricity demand and grid impacts of AI data centers: Challenges and prospects. \emph{Nexus} \textbf{3}, 100162 (2026). doi:10.1016/j.ynexs.2026.100162
\item[3.] S. J. Smith, A. Hubbard, A. Newkirk, M. Ganeshalingam, B. Holecek, D. A. Sartor, M. Mills, A. Shehabi, ``United States data center energy usage report: 2025 update'' (LBNL-2001758, Lawrence Berkeley National Laboratory, June 2026); \url{https://doi.org/10.71468/P1RP4F}.
\item[4.] T. Xiao, F. Fuso Nerini, H. D. Matthews, M. Tavoni, F. You, Environmental impact and net-zero pathways for sustainable artificial intelligence servers in the USA. \emph{Nat. Sustain.} \textbf{8}, 1541--1553 (2025). doi:10.1038/s41893-025-01681-y
\item[5.] P. Colangelo, A. K. Coskun, J. Megrue, C. Roberts, S. Sengupta, V. Sivaram, E. Tiao, A. Vijaykar, C. Williams, D. C. Wilson, B. Records, Z. MacFarland, D. Dreiling, N. Morey, A. Ratnayake, B. Vairamohan, AI data centres as grid-interactive assets. \emph{Nat. Energy} \textbf{11}, 254--261 (2026). doi:10.1038/s41560-025-01927-1
\item[6.] International Energy Agency, ``Key questions on energy and AI'' (World Energy Outlook special report, International Energy Agency, April 2026); \url{https://iea.blob.core.windows.net/assets/3179f7f8-01f6-4dd6-bffa-c9f7b73f1dc9/KeyQuestionsonEnergyandAI.pdf}.
\item[7.] J. Martos, ``U.S. gas power proposals tied to data centers nearly double in six months'' (Global Energy Monitor, 25 August 2026); \url{https://globalenergymonitor.org/research/us-gas-power-proposals-tied-data-centers-nearly-double-six-months}.
\item[8.] P. Achakulwisut, M. Brauer, P. Hystad, S. C. Anenberg, Global, national, and urban burdens of paediatric asthma incidence attributable to ambient NO\textsubscript{2} pollution: Estimates from global datasets. \emph{Lancet Planet. Health} \textbf{3}, e166--e178 (2019). doi:10.1016/S2542-5196(19)30046-4
\item[9.] S. F. Camilleri, G. H. Kerr, S. C. Anenberg, D. E. Horton, All-cause NO\textsubscript{2}-attributable mortality burden and associated racial and ethnic disparities in the United States. \emph{Environ. Sci. Technol. Lett.} \textbf{10}, 1159--1164 (2023). doi:10.1021/acs.estlett.3c00500
\item[10.] E. A. Gilmore, L. B. Lave, P. J. Adams, The costs, air quality, and human health effects of meeting peak electricity demand with installed backup generators. \emph{Environ. Sci. Technol.} \textbf{40}, 6887--6893 (2006). doi:10.1021/es061151q
\item[11.] Y. Han, Z. Wu, P. Li, A. Wierman, S. Ren, Health-informed computing: Estimating and addressing the public health impact of data centers. arXiv:2412.06288v4 {[}cs.CY{]} (2026).
\item[12.] N. Z. Muller, ``Measuring the impact of data centers in the United States economy: Monetary damage from air pollution and greenhouse gas emissions'' (NBER Working Paper 35100, National Bureau of Economic Research, 2026); \url{https://doi.org/10.3386/w35100}.
\item[13.] Y. Tao, P. Gao, Global data center expansion and human health: A call for empirical research. \emph{Eco Environ. Health} \textbf{4}, 100157 (2025). doi:10.1016/j.eehl.2025.100157
\item[14.] D. Raychaudhuri, V. Volcovici, ``Pollution from Musk's unpermitted xAI power project hits hardest in Black communities'' (Reuters, 14 July 2026); \url{https://www.reuters.com/world/americas/pollution-musks-unpermitted-xai-power-project-hits-hardest-black-communities-2026-07-14/}.
\item[15.] A. Gaffney, ``The data center backlash comes for SpaceXAI's Colossus'' (Scientific American, 9 September 2026); \url{https://www.scientificamerican.com/article/spacexais-colossus-data-center-is-fueling-a-massive-backlash/}.
\item[16.] Mississippi Commission on Environmental Quality, ``Agreed order no. 7739-26: Mississippi Commission on Environmental Quality v. MZX Tech LLC'' (30 July 2026); \url{https://opcgis.deq.state.ms.us/enonline/get_doc.aspx?dt=order&id=1862315}.
\item[17.] Mississippi Commission on Environmental Quality, ``Amendment to agreed order no. 7739-26: Mississippi Commission on Environmental Quality v. MZX Tech LLC'' (11 September 2026) {[}publicly posted copy{]}; \url{https://rachelandthecity.com/the-xai-files/records/who-told-epa/mdeq-agreed-order-7739-26-amended-2026-09-11.pdf}.
\item[18.] U.S. Energy Information Administration, Preliminary monthly electric generator inventory, August 2026 (Form EIA-860M) (24 September 2026); \url{https://www.eia.gov/electricity/data/eia860m/} {[}accessed 5 October 2026{]}.
\item[19.] Mississippi Department of Environmental Quality, ``Determination on portable gas-fired combustion turbines'' (letter to xAI, 29 July 2025) {[}publicly posted copy{]}; \url{https://rachelandthecity.com/the-xai-files/records/upcoming-hearing/mdeq-mobile-source-determination-2025-07-29.pdf}.
\item[20.] U.S. Environmental Protection Agency, New source performance standards review for stationary combustion turbines and stationary gas turbines, final rule. \emph{Fed. Regist.} \textbf{91}, 1910--2005 (2026); \url{https://www.govinfo.gov/content/pkg/FR-2026-01-15/pdf/2026-00677.pdf}.
\item[21.] Solaris Energy Infrastructure, ``Quarterly report on Form 10-Q for the quarter ended June 30, 2026'' (U.S. Securities and Exchange Commission, 2026); \url{https://www.sec.gov/Archives/edgar/data/1697500/000162828026054349/sei-20260630.htm}.
\item[22.] J. B. Bushnell, C. D. Wolfram, Enforcement of vintage differentiated regulations: The case of new source review. \emph{J. Environ. Econ. Manage.} \textbf{64}, 137--152 (2012). doi:10.1016/j.jeem.2012.01.006
\item[23.] R. N. Stavins, Vintage-differentiated environmental regulation. \emph{Stanford Environ. Law J.} \textbf{25}, 29--66 (2006).
\item[24.] E. Helland, A. B. Whitford, Pollution incidence and political jurisdiction: Evidence from the TRI. \emph{J. Environ. Econ. Manage.} \textbf{46}, 403--424 (2003). doi:10.1016/S0095-0696(03)00033-0
\item[25.] M. E. Kahn, Domestic pollution havens: Evidence from cancer deaths in border counties. \emph{J. Urban Econ.} \textbf{56}, 51--69 (2004). doi:10.1016/j.jue.2004.02.001
\item[26.] M. Lipscomb, A. M. Mobarak, Decentralization and pollution spillovers: Evidence from the re-drawing of county borders in Brazil. \emph{Rev. Econ. Stud.} \textbf{84}, 464--502 (2017). doi:10.1093/restud/rdw023
\item[27.] J. E. Monogan III, D. M. Konisky, N. D. Woods, Gone with the wind: Federalism and the strategic location of air polluters. \emph{Am. J. Pol. Sci.} \textbf{61}, 257--270 (2017). doi:10.1111/ajps.12278
\item[28.] J. M. Morehouse, E. Rubin, Downwind and out: The strategic dispersion of power plants and their pollution. \emph{J. Assoc. Environ. Resour. Econ.} \textbf{13}, 831--867 (2026). doi:10.1086/740146
\item[29.] H. Sigman, Transboundary spillovers and decentralization of environmental policies. \emph{J. Environ. Econ. Manage.} \textbf{50}, 82--101 (2005). doi:10.1016/j.jeem.2004.10.001
\item[30.] U.S. Environmental Protection Agency, Air Quality System: Pre-generated data files (2026); \url{https://aqs.epa.gov/aqsweb/airdata/download_files.html} {[}accessed 7 October 2026{]}.
\item[31.] S. Beirle, C. Borger, S. Dörner, A. Li, Z. Hu, F. Liu, Y. Wang, T. Wagner, Pinpointing nitrogen oxide emissions from space. \emph{Sci. Adv.} \textbf{5}, eaax9800 (2019). doi:10.1126/sciadv.aax9800
\item[32.] D. L. Goldberg, Z. Lu, D. G. Streets, B. de Foy, D. Griffin, C. A. McLinden, L. N. Lamsal, N. A. Krotkov, H. Eskes, Enhanced capabilities of TROPOMI NO\textsubscript{2}: Estimating NO\textsubscript{x} from North American cities and power plants. \emph{Environ. Sci. Technol.} \textbf{53}, 12594--12601 (2019). doi:10.1021/acs.est.9b04488
\item[33.] K. Sun, J. A. Saju, C. R. Nowlan, G. González Abad, X. Liu, Hourly nitrogen oxides emissions estimated from TEMPO and comparison with facility-level monitoring data. \emph{J. Geophys. Res. Atmos.} \textbf{130}, e2025JD044565 (2025). doi:10.1029/2025JD044565
\item[34.] P. Zoogman, X. Liu, R. M. Suleiman, W. F. Pennington, D. E. Flittner, J. A. Al-Saadi, B. B. Hilton, D. K. Nicks, M. J. Newchurch, J. L. Carr, S. J. Janz, M. R. Andraschko, A. Arola, B. D. Baker, B. P. Canova, C. Chan Miller, R. C. Cohen, J. E. Davis, M. E. Dussault, D. Edwards, J. Fishman, A. Ghulam, G. González Abad, M. Grutter, J. Herman, J. Houck, D. Jacob, J. Joiner, B. Kerridge, J. Kim, N. Krotkov, L. Lamsal, C. Li, A. Lindfors, R. Martin, C. McElroy, C. McLinden, V. Natraj, D. Neil, C. Nowlan, E. O'Sullivan, P. Palmer, R. Pierce, M. Pippin, A. Saiz-Lopez, R. Spurr, J. Szykman, O. Torres, J. Veefkind, B. Veihelmann, H. Wang, J. Wang, K. Chance, Tropospheric Emissions: Monitoring of Pollution (TEMPO). \emph{J. Quant. Spectrosc. Radiat. Transf.} \textbf{186}, 17--39 (2017). doi:10.1016/j.jqsrt.2016.05.008
\item[35.] J. P. Veefkind, I. Aben, K. McMullan, H. Förster, J. de Vries, G. Otter, J. Claas, H. J. Eskes, J. F. de Haan, Q. Kleipool, M. van Weele, O. Hasekamp, R. Hoogeveen, J. Landgraf, R. Snel, P. Tol, P. Ingmann, R. Voors, B. Kruizinga, R. Vink, H. Visser, P. F. Levelt, TROPOMI on the ESA Sentinel-5 Precursor: A GMES mission for global observations of the atmospheric composition for climate, air quality and ozone layer applications. \emph{Remote Sens. Environ.} \textbf{120}, 70--83 (2012). doi:10.1016/j.rse.2011.09.027
\item[36.] S. Beirle, K. F. Boersma, U. Platt, M. G. Lawrence, T. Wagner, Megacity emissions and lifetimes of nitrogen oxides probed from space. \emph{Science} \textbf{333}, 1737--1739 (2011). doi:10.1126/science.1207824
\item[37.] U. Simonsohn, J. P. Simmons, L. D. Nelson, Specification curve analysis. \emph{Nat. Hum. Behav.} \textbf{4}, 1208--1214 (2020). doi:10.1038/s41562-020-0912-z
\item[38.] S. Steegen, F. Tuerlinckx, A. Gelman, W. Vanpaemel, Increasing transparency through a multiverse analysis. \emph{Perspect. Psychol. Sci.} \textbf{11}, 702--712 (2016). doi:10.1177/1745691616658637
\item[39.] S. Beirle, C. Borger, S. Dörner, H. Eskes, V. Kumar, A. de Laat, T. Wagner, Catalog of NO\textsubscript{x} emissions from point sources as derived from the divergence of the NO\textsubscript{2} flux for TROPOMI. \emph{Earth Syst. Sci. Data} \textbf{13}, 2995--3012 (2021). doi:10.5194/essd-13-2995-2021
\item[40.] National Interagency Fire Center, WFIGS Interagency Perimeters Year-to-Date (2026); \url{https://services3.arcgis.com/T4QMspbfLg3qTGWY/arcgis/rest/services/WFIGS_Interagency_Perimeters_YearToDate/FeatureServer/0} {[}accessed 4 October 2026{]}.
\item[41.] A. Abadie, A. Diamond, J. Hainmueller, Synthetic control methods for comparative case studies: Estimating the effect of California's tobacco control program. \emph{J. Am. Stat. Assoc.} \textbf{105}, 493--505 (2010). doi:10.1198/jasa.2009.ap08746
\item[42.] National Oceanic and Atmospheric Administration, Hazard Mapping System fire and smoke product; \url{https://ospo.noaa.gov/products/land/hms.html} {[}accessed 11 October 2026{]}.
\item[43.] Texas Gas Transmission, ``Index of customers for the fourth quarter of 2026'' (Accession No. 20261001-5154, Federal Energy Regulatory Commission, 1 October 2026); \url{https://elibrary.ferc.gov/eLibrary/filelist/20261001-5154}.
\item[44.] U.S. Environmental Protection Agency, Clean Air Markets Program Data: Hourly emissions (2026); \url{https://campd.epa.gov/data/bulk-data-files} {[}accessed 5 October 2026{]}.
\item[45.] L. P. Clark, D. B. Millet, J. D. Marshall, National patterns in environmental injustice and inequality: Outdoor NO\textsubscript{2} air pollution in the United States. \emph{PLOS ONE} \textbf{9}, e94431 (2014). doi:10.1371/journal.pone.0094431
\item[46.] G. H. Kerr, D. L. Goldberg, S. C. Anenberg, COVID-19 pandemic reveals persistent disparities in nitrogen dioxide pollution. \emph{Proc. Natl. Acad. Sci. U.S.A.} \textbf{118}, e2022409118 (2021). doi:10.1073/pnas.2022409118
\item[47.] Y. Wang, P. Liu, J. Schwartz, E. Castro, W. Wang, H. Chang, N. Scovronick, L. Shi, Disparities in ambient nitrogen dioxide pollution in the United States. \emph{Proc. Natl. Acad. Sci. U.S.A.} \textbf{120}, e2208450120 (2023). doi:10.1073/pnas.2208450120
\item[48.] U.S. Census Bureau, 2020--2024 American Community Survey 5-year estimates, detailed tables B01001, B03002, B17001, B19013, and C17002 (2026); \url{https://api.census.gov/data/2024/acs/acs5.html} {[}accessed 21 October 2026{]}.
\item[49.] M. Fowlie, E. Rubin, R. Walker, Bringing satellite-based air quality estimates down to earth. \emph{AEA Pap. Proc.} \textbf{109}, 283--288 (2019). doi:10.1257/pandp.20191064
\item[50.] National Primary and Secondary Ambient Air Quality Standards for Oxides of Nitrogen (With Nitrogen Dioxide as the Indicator), 40 C.F.R. § 50.11 (2025); \url{https://www.govinfo.gov/content/pkg/CFR-2025-title40-vol2/pdf/CFR-2025-title40-vol2-sec50-11.pdf}.
\item[51.] Mississippi Department of Environmental Quality, ``MZX Tech LLC: Prevention of significant deterioration permit'' (11 March 2026); \url{https://opc.deq.state.ms.us/get_doc.aspx?dt=finalp&id=1834338}.
\item[52.] Montrose Air Quality Services, ``Calpine Green Leaf 1: 2021 TM2500 compliance report'' (2021); \url{https://www.energy.gov/sites/default/files/2022-02/06.\%20CAISO\%20202\%28c\%29\%20Covered\%20Resource\%20-\%20Greenleaf\%20Unit\%201_Yuba\%20City\%2C\%20CA\%20\%28Exhibit\%20E-1\%20to\%206\%2C\%20H-2\%20\%26\%205\%29_2.pdf}.
\item[53.] E. A. Gilmore, J. Heo, N. Z. Muller, C. W. Tessum, J. D. Hill, J. D. Marshall, P. J. Adams, An inter-comparison of the social costs of air quality from reduced-complexity models. \emph{Environ. Res. Lett.} \textbf{14}, 074016 (2019). doi:10.1088/1748-9326/ab1ab5
\item[54.] N. Z. Muller, AP4 integrated assessment model, Carnegie Mellon University; \url{https://nickmuller.tepper.cmu.edu/media/APModel/AP4\%20Model.zip} {[}accessed 4 October 2026{]}.
\item[55.] C. W. Tessum, J. D. Hill, J. D. Marshall, InMAP: A model for air pollution interventions. \emph{PLOS ONE} \textbf{12}, e0176131 (2017). doi:10.1371/journal.pone.0176131
\item[56.] P. Tschofen, I. L. Azevedo, N. Z. Muller, Fine particulate matter damages and value added in the US economy. \emph{Proc. Natl. Acad. Sci. U.S.A.} \textbf{116}, 19857--19862 (2019). doi:10.1073/pnas.1905030116
\item[57.] C. A. Pope III, J. S. Lefler, M. Ezzati, J. D. Higbee, J. D. Marshall, S. Y. Kim, M. Bechle, K. S. Gilliat, S. E. Vernon, A. L. Robinson, R. T. Burnett, Mortality risk and fine particulate air pollution in a large, representative cohort of U.S. adults. \emph{Environ. Health Perspect.} \textbf{127}, 077007 (2019). doi:10.1289/EHP4438
\item[58.] D. Krewski, M. Jerrett, R. T. Burnett, R. Ma, E. Hughes, Y. Shi, M. C. Turner, C. A. Pope III, G. Thurston, E. E. Calle, M. J. Thun, ``Extended follow-up and spatial analysis of the American Cancer Society study linking particulate air pollution and mortality'' (Research Report 140, Health Effects Institute, 2009); \url{https://www.healtheffects.org/system/files/Krewski140.pdf}.
\item[59.] H. Khreis, C. Kelly, J. Tate, R. Parslow, K. Lucas, M. Nieuwenhuijsen, Exposure to traffic-related air pollution and risk of development of childhood asthma: A systematic review and meta-analysis. \emph{Environ. Int.} \textbf{100}, 1--31 (2017). doi:10.1016/j.envint.2016.11.012
\item[60.] U.S. Environmental Protection Agency, AVoided Emissions and geneRation Tool (AVERT), 2023 data year (2024); \url{https://www.epa.gov/avert}.
\item[61.] Air Hygiene International, ``Emission compliance tests for Units 1--5, 7, and 8, Longhorn Data Center, Abilene, Texas'' (seven stack-test reports, TCEQ Records Online document IDs 9835200, 9852352, 9858896, 9858897, 10069366, 10072417, and 10150158, Texas Commission on Environmental Quality, 2026); \url{https://records.tceq.texas.gov/cs/idcplg?IdcService=TCEQ_EXTERNAL_SEARCH_GET_FILE&dID=9835200&Rendition=Web&searchType=External}.
\item[62.] Pinyon Environmental, ``Modification application: Air quality standard permit for electric generating units, Abilene DC 1, LLC, Longhorn Data Center, registration 177263'' (application submitted to the Texas Commission on Environmental Quality, 10 January 2025); \url{https://records.tceq.texas.gov/cs/idcplg?IdcService=TCEQ_EXTERNAL_SEARCH_GET_FILE&Rendition=Web&dID=8600639}.
\item[63.] U.S. Environmental Protection Agency, New source performance standards review for stationary combustion turbines and stationary gas turbines, correction. \emph{Fed. Regist.} \textbf{91}, 43561--43569 (2026); \url{https://www.govinfo.gov/content/pkg/FR-2026-07-16/pdf/2026-14371.pdf}.
\item[64.] Shelby County Health Department, ``Construction permit no. 01156-01PC: CTC Property LLC'' (air pollution control permit, 2 July 2025) {[}publicly posted copy{]}; \url{https://rachelandthecity.com/the-xai-files/records/county-and-epa/shelby-county-air-permit-01156-01pc-2025-07-02.pdf}.
\item[65.] U.S. Environmental Protection Agency, ``Air Pollution Control Cost Manual: Selective catalytic reduction'' (U.S. Environmental Protection Agency, 2019); \url{https://www.epa.gov/economic-and-cost-analysis-air-pollution-regulations/cost-reports-and-guidance-air-pollution}.
\item[66.] Solaris Energy Infrastructure, ``Annual report on Form 10-K for the fiscal year ended December 31, 2025'' (U.S. Securities and Exchange Commission, 2026); \url{https://www.sec.gov/Archives/edgar/data/1697500/000162828026012501/sei-20251231.htm}.
\item[67.] Solaris Energy Infrastructure, ``Solaris Energy Infrastructure announces fourth quarter 2024 financial and operational update and power solutions growth capital developments'' (press release, 4 December 2024); \url{https://ir.solaris-energy.com/news/2024/12-04-2024-232917240}.
\item[68.] Solaris Energy Infrastructure, ``First quarter 2025 earnings call prepared remarks'' (29 April 2025); \url{https://ir.solaris-energy.com/~/media/Files/S/Solaris-IR/documents/events/sei-q1-2025-prepared-remarks.pdf}.
\item[69.] U.S. Environmental Protection Agency, ``NOx mitigation measures: Selective catalytic reduction and combustion controls for combustion turbines'' (technical support document, Docket ID No. EPA-HQ-OAR-2024-0419-0242, Office of Air and Radiation, 2026); \url{https://www.regulations.gov/document/EPA-HQ-OAR-2024-0419-0242}.
\item[70.] U.S. Environmental Protection Agency, ``Economic impact analysis for the new source performance standards review for stationary combustion turbines: Final rule'' (EPA-452/R-26-002, Office of Clean Air Programs, 2026); \url{https://www.epa.gov/system/files/documents/2026-01/combustion_turbines_eia_final_2026-01.pdf}.
\item[71.] U.S. Environmental Protection Agency, ``Regulatory impact analysis for the new source performance standards review for stationary combustion turbines'' (EPA-452/R-24-016, Office of Air Quality Planning and Standards, November 2024); \url{https://downloads.regulations.gov/EPA-HQ-OAR-2024-0419-0005/content.pdf}.
\item[72.] Denham Capital, ``Nexif Energy reaches financial close for the 154MW Snapper Point power plant'' (press release, 4 May 2021); \url{https://www.denhamcapital.com/assets/2021/05/Nexif-Energy-Reaches-Financial-Close-for-the-154MW-Snapper-Point-Plant-04.05.21.pdf}.
\item[73.] Essential Services Commission of South Australia, ``Electricity generation licence variation and transfer: Iberdrola Australia Energy Markets'' (licensing decision, 29 August 2022); \url{https://www.escosa.sa.gov.au/projects-and-publications/projects/electricity/electricity-generation-licence-variation-and-transfer-iberdrola-australia-energy-markets/iaem}.
\item[74.] N. Filatoff, ``SA flicks a switch for back-up generation'' (GE News, 2 August 2017); \url{https://www.ge.com/news/reports/ge-tm2500-for-south-australia}.
\item[75.] Iberdrola Australia, ``Bolivar Power Station operational environmental management plan'' (Iberdrola Australia, May 2023); \url{https://www.iberdrola.com.au/assets/Documents/Firming-Asset-Documents/Bolivar-OEMP-Combined.pdf}.
\item[76.] California Energy Commission, ``Order no. 2026-0408-03g: Order approving the amendment to merge the RSPAPS license (21-TGP-01) into the existing Roseville Energy Park license (03-AFC-1) and permanent closure of the CEC RSPAPS docket'' (Orders and resolutions of the 8 April 2026 business meeting, Docket No. 26-BUSMTG-01, TN 269446, California Energy Commission, 2026); \url{https://efiling.energy.ca.gov/GetDocument.aspx?DocumentContentId=106537&tn=269446}.
\item[77.] California Energy Commission, ``Roseville Energy Park (03-AFC-01C): CEC staff analysis of petition to amend the final commission decision'' (TN 268922, California Energy Commission, 4 March 2026); \url{https://efiling.energy.ca.gov/GetDocument.aspx?tn=268922}.
\item[78.] I. C. Dedoussi, S. D. Eastham, E. Monier, S. R. H. Barrett, Premature mortality related to United States cross-state air pollution. \emph{Nature} \textbf{578}, 261--265 (2020). doi:10.1038/s41586-020-1983-8
\item[79.] M. P. S. Thind, C. W. Tessum, I. L. Azevedo, J. D. Marshall, Fine particulate air pollution from electricity generation in the US: Health impacts by race, income, and geography. \emph{Environ. Sci. Technol.} \textbf{53}, 14010--14019 (2019). doi:10.1021/acs.est.9b02527
\item[80.] Clean Air Act, 42 U.S.C. § 7410(a)(2)(D) (2023); \url{https://www.govinfo.gov/content/pkg/USCODE-2023-title42/html/USCODE-2023-title42-chap85-subchapI-partA-sec7410.htm}.
\item[81.] Mississippi Department of Environmental Quality, ``Response to comments, MZX Tech LLC, Permit No. 0680-00119'' (p. 23, 2026) {[}publicly posted copy{]}; \url{https://rachelandthecity.com/the-xai-files/records/upcoming-hearing/mdeq-response-to-comments-mzx-p23.pdf}.
\item[82.] Mississippi Commission on Environmental Quality, ``Minutes of the Commission meeting'' (executed minutes, 22 January 2026); \url{https://www.mdeq.ms.gov/wp-content/uploads/2026/03/EXECUTED-Commission-Meeting-Minutes-1_22_2026.pdf}.
\item[83.] Mississippi Department of Environmental Quality, ``DeSoto County, Mississippi second 10-year maintenance plan for the 2008 ozone NAAQS'' (public notice draft and emission inventory workbooks, 20 November 2025); \url{https://www.mdeq.ms.gov/wp-content/uploads/2025/11/Mississippi-Second-2008-Ozone-NAAQS-Maintenance-Plan-DRAFT.pdf}.
\item[84.] E. M. Krieger, J. A. Casey, S. B. C. Shonkoff, A framework for siting and dispatch of emerging energy resources to realize environmental and health benefits: Case study on peaker power plant displacement. \emph{Energy Policy} \textbf{96}, 302--313 (2016). doi:10.1016/j.enpol.2016.05.049
\item[85.] Office of Information and Regulatory Affairs, ``Nonroad combustion turbines rule (RIN 2060-AX08)'' (proposed rule received for review under Executive Order 12866, 15 October 2026); \url{https://www.reginfo.gov/public/do/eoDetails?rrid=1540963}.
\item[86.] U.S. Environmental Protection Agency, National Emissions Inventory (NEI); \url{https://www.epa.gov/air-emissions-inventories/national-emissions-inventory-nei} {[}accessed 28 October 2026{]}.
\item[87.] A. Underwood, R. Marcantonio, D. Wood, P. Crippa, Interstate air pollution governance in the United States: Exploring Clean Air Act section 126. \emph{Environ. Manage.} \textbf{74}, 401--413 (2024). doi:10.1007/s00267-024-02002-3
\item[88.] U.S. Environmental Protection Agency, Response to Clean Air Act section 126(b) petitions from Delaware and Maryland, notice of final action on petition. \emph{Fed. Regist.} \textbf{83}, 50444--50473 (2018); \url{https://www.govinfo.gov/content/pkg/FR-2018-10-05/pdf/2018-20854.pdf}.
\item[89.] W. W. Qiang, B. Huang, X. Liu, H. F. Lee, Data for: Temporary gas turbines powering a US AI data center created a cross-state air pollution hotspot, version 1.0.0, Zenodo (2026); \url{https://doi.org/10.5281/zenodo.23762139}.
\item[90.] W. W. Qiang, B. Huang, X. Liu, H. F. Lee, Building-allocated 2020 population grid around Southaven, Mississippi, version 1.0.0, Zenodo (2026); \url{https://doi.org/10.5281/zenodo.23877598}.
\item[91.] W. W. Qiang, B. Huang, X. Liu, H. F. Lee, Code for: Temporary gas turbines powering a US AI data center created a cross-state air pollution hotspot, version 1.0.0, Zenodo (2026); \url{https://doi.org/10.5281/zenodo.23877611}.
\item[92.] Texas Gas Transmission, Operationally available capacity (informational postings, GasQuest); \url{https://www.gasquest.com/capacity/operationally-available} {[}accessed 27 October 2026{]}.
\item[93.] X. Liu, TEMPO NO2 tropospheric and stratospheric columns V04 (PROVISIONAL), NASA Langley Atmospheric Science Data Center (2026); \url{https://doi.org/10.5067/is-40e/tempo/no2_l2.004}.
\item[94.] X. Liu, TEMPO gridded NO2 tropospheric and stratospheric columns V04 (PROVISIONAL), NASA Langley Atmospheric Science Data Center (2026); \url{https://doi.org/10.5067/is-40e/tempo/no2_l3.004}.
\item[95.] Copernicus Sentinel-5P (processed by ESA), TROPOMI Level 2 nitrogen dioxide total column products, Version 02, European Space Agency (2021); \url{https://doi.org/10.5270/S5P-9bnp8q8}.
\item[96.] D. L. Goldberg, J. Johnson, T. Shah, O. Nawaz, G. H. Kerr, L. Chen, D. Huber, L. Judd, Contrasting the TEMPO and TROPOMI views of tropospheric NO\textsubscript{2}: Implications for diurnal variability and model evaluation. \emph{J. Geophys. Res. Atmos.} \textbf{131}, e2026JD046905 (2026). doi:10.1029/2026JD046905
\item[97.] D. C. Dowell, C. R. Alexander, E. P. James, S. S. Weygandt, S. G. Benjamin, G. S. Manikin, B. T. Blake, J. M. Brown, J. B. Olson, M. Hu, T. G. Smirnova, T. Ladwig, J. S. Kenyon, R. Ahmadov, D. D. Turner, J. D. Duda, T. I. Alcott, The High-Resolution Rapid Refresh (HRRR): An hourly updating convection-allowing forecast model. Part I: Motivation and system description. \emph{Weather Forecast.} \textbf{37}, 1371--1395 (2022). doi:10.1175/waf-d-21-0151.1
\item[98.] National Oceanic and Atmospheric Administration, High-Resolution Rapid Refresh (HRRR), NOAA Open Data Dissemination; \url{https://registry.opendata.aws/noaa-hrrr-pds/} {[}accessed 16 October 2026{]}.
\item[99.] Copernicus Climate Change Service, ERA5 hourly data on single levels from 1940 to present, Climate Data Store (2018); \url{https://doi.org/10.24381/cds.adbb2d47}.
\item[100.] H. Hersbach, B. Bell, P. Berrisford, S. Hirahara, A. Horányi, J. Muñoz-Sabater, J. Nicolas, C. Peubey, R. Radu, D. Schepers, A. Simmons, C. Soci, S. Abdalla, X. Abellan, G. Balsamo, P. Bechtold, G. Biavati, J. Bidlot, M. Bonavita, G. De Chiara, P. Dahlgren, D. Dee, M. Diamantakis, R. Dragani, J. Flemming, R. Forbes, M. Fuentes, A. Geer, L. Haimberger, S. Healy, R. J. Hogan, E. Hólm, M. Janisková, S. Keeley, P. Laloyaux, P. Lopez, C. Lupu, G. Radnoti, P. de Rosnay, I. Rozum, F. Vamborg, S. Villaume, J.-N. Thépaut, The ERA5 global reanalysis. \emph{Q. J. R. Meteorol. Soc.} \textbf{146}, 1999--2049 (2020). doi:10.1002/qj.3803
\item[101.] M. Ghahremanloo, C. Nowlan, G. González Abad, C. Garraffo, X. Liu, H. Wang, B. Henderson, E. Baumann, L. Valin, J. A. Geddes, X. Zhao, Comprehensive analysis of bias in TEMPO NO\textsubscript{2} column densities through Pandora observations. \emph{J. Geophys. Res. Atmos.} \textbf{130}, e2025JD044150 (2025). doi:10.1029/2025JD044150
\item[102.] U.S. Census Bureau, 2024 TIGER/Line Shapefiles, primary and secondary roads (2024); \url{https://www2.census.gov/geo/tiger/TIGER2024/PRISECROADS/} {[}accessed 25 September 2026{]}.
\item[103.] Arkansas Department of Transportation, Average daily traffic points; \url{https://gis.ardot.gov/referenced/rest/services/SIR_TIS/Average_Daily_Traffic_Points/MapServer/0} {[}accessed 23 September 2026{]}.
\item[104.] Mississippi Department of Transportation, Traffic counts, 2016--2025 annual average daily traffic; \url{https://mdot.ms.gov/applications/trafficcounters/files/2016-2025\%20AADT.zip} {[}accessed 23 September 2026{]}.
\item[105.] Tennessee Department of Transportation, Traffic count points, annual average daily traffic; \url{https://services8.arcgis.com/tYbyA2CnUcjfSnwp/ArcGIS/rest/services/Traffic_Points/FeatureServer/0} {[}accessed 23 September 2026{]}.
\item[106.] U.S. Geological Survey, Annual National Land Cover Database (NLCD) Collection 1 Products, U.S. Geological Survey (2024); \url{https://doi.org/10.5066/P94UXNTS}.
\item[107.] N. Doudchenko, G. W. Imbens, ``Balancing, regression, difference-in-differences and synthetic control methods: A synthesis'' (NBER Working Paper 22791, National Bureau of Economic Research, 2016); \url{https://doi.org/10.3386/w22791}.
\item[108.] G. González Abad, C. Nowlan, H. Wang, H. Chong, J. Houck, X. Liu, K. Chance, ``Tropospheric Emissions: Monitoring of Pollution (TEMPO) project trace gas and cloud Level 2 and 3 data products: User guide'' (version 2.1, NASA Atmospheric Science Data Center, 2026); \url{https://doi.org/10.5067/doc/tempo/L2andL3UserGuide/V2.1}.
\item[109.] V. Chernozhukov, D. Chetverikov, M. Demirer, E. Duflo, C. Hansen, W. Newey, J. Robins, Double/debiased machine learning for treatment and structural parameters. \emph{Econom. J.} \textbf{21}, C1--C68 (2018). doi:10.1111/ectj.12097
\item[110.] H. R. Künsch, The jackknife and the bootstrap for general stationary observations. \emph{Ann. Stat.} \textbf{17}, 1217--1241 (1989). doi:10.1214/aos/1176347265
\item[111.] D. N. Politis, J. P. Romano, ``A circular block-resampling procedure for stationary data'' in \emph{Exploring the Limits of Bootstrap}, R. LePage, L. Billard, Eds. (Wiley, 1992), pp. 263--270.
\item[112.] D. B. Rubin, The Bayesian bootstrap. \emph{Ann. Stat.} \textbf{9}, 130--134 (1981). doi:10.1214/aos/1176345338
\item[113.] U.S. Energy Information Administration, Form EIA-860 detailed data with previous form data (EIA-860A/860B); \url{https://www.eia.gov/electricity/data/eia860/} {[}accessed 15 October 2026{]}.
\item[114.] U.S. Energy Information Administration, Preliminary monthly electric generator inventory, July 2026 (Form EIA-860M) (2026); \url{https://www.eia.gov/electricity/data/eia860m/} {[}accessed 16 October 2026{]}.
\item[115.] E. Ben-Michael, A. Feller, J. Rothstein, The augmented synthetic control method. \emph{J. Am. Stat. Assoc.} \textbf{116}, 1789--1803 (2021). doi:10.1080/01621459.2021.1929245
\item[116.] D. Arkhangelsky, S. Athey, D. A. Hirshberg, G. W. Imbens, S. Wager, Synthetic difference-in-differences. \emph{Am. Econ. Rev.} \textbf{111}, 4088--4118 (2021). doi:10.1257/aer.20190159
\item[117.] A. Abadie, A. Diamond, J. Hainmueller, Comparative politics and the synthetic control method. \emph{Am. J. Pol. Sci.} \textbf{59}, 495--510 (2015). doi:10.1111/ajps.12116
\item[118.] J. G. MacKinnon, H. White, Some heteroskedasticity-consistent covariance matrix estimators with improved finite sample properties. \emph{J. Econom.} \textbf{29}, 305--325 (1985). doi:10.1016/0304-4076(85)90158-7
\item[119.] J. S. Scire, D. G. Strimaitis, R. J. Yamartino, ``Model formulation and user's guide for the CALPUFF dispersion model'' (Report No. A025-2, California Air Resources Board, 1990); \url{https://ww2.arb.ca.gov/sites/default/files/classic/research/apr/past/a025-2a.pdf}.
\item[120.] U.S. Census Bureau, LEHD Origin-Destination Employment Statistics (LODES), version 8, workplace area characteristics, 2023; \url{https://lehd.ces.census.gov/data/} {[}accessed 23 September 2026{]}.
\item[121.] U.S. Census Bureau, 2020 Census redistricting data (P.L. 94-171) summary file (2021); \url{https://www.census.gov/programs-surveys/decennial-census/about/rdo/summary-files.html} {[}accessed 16 October 2026{]}.
\item[122.] U.S. Census Bureau, 2020 Census Demographic and Housing Characteristics File (DHC) (2023); \url{https://www.census.gov/data/tables/2023/dec/2020-census-dhc.html} {[}accessed 16 October 2026{]}.
\item[123.] J. Mennis, Generating surface models of population using dasymetric mapping. \emph{Prof. Geogr.} \textbf{55}, 31--42 (2003). doi:10.1111/0033-0124.10042
\item[124.] Microsoft, USBuildingFootprints (usbuildings-v2 files), GitHub; \url{https://github.com/microsoft/USBuildingFootprints} {[}accessed 15 October 2026{]}.
\item[125.] J. Dewitz, National Land Cover Database (NLCD) 2021 Products (ver. 2.0, June 2026), U.S. Geological Survey (2023); \url{https://doi.org/10.5066/p9jz7ao3}.
\item[126.] Trinity Consultants, ``Unit operating hours for June 2026, units 1--60, submitted to MDEQ'' (10 July 2026) {[}publicly posted copy{]}; \url{https://rachelandthecity.com/the-xai-files/records/upcoming-hearing/mzx-runtime-table-2026-07-10.pdf}.
\item[127.] Trinity Consultants, ``Emissions information for portable turbines, MZX Tech LLC, Southaven, Mississippi'' (15 May 2026; filed as Attachment 34, ECF No. 51-36, Case No. 3:26-cv-00074-DMB-JMV, N.D. Miss., 12 June 2026), CourtListener RECAP Archive; \url{https://storage.courtlistener.com/recap/gov.uscourts.msnd.52261/gov.uscourts.msnd.52261.51.36.pdf}.
\item[128.] U.S. Environmental Protection Agency, ``Stationary gas turbines'' (AP-42, 5th ed., vol. I, chap. 3, sect. 3.1, U.S. Environmental Protection Agency, April 2000); \url{https://www.epa.gov/sites/default/files/2020-10/documents/c03s01.pdf}.
\item[129.] U.S. Energy Information Administration, ``Electric power annual, Table 8.2: Average tested heat rates by prime mover and energy source, 2015--2025''; \url{https://www.eia.gov/electricity/annual/html/epa_08_02.html} {[}accessed 27 October 2026{]}.
\item[130.] National Energy Technology Laboratory, ``Cost and performance baseline for fossil energy plants, volume 5: Natural gas electricity generating units for flexible operation'' (DOE/NETL-2023/3855, National Energy Technology Laboratory, 5 May 2023); \url{https://netl.doe.gov/projects/files/CostandPerformanceBaselineforFossilEnergyPlantsVol5NatGasElectricityGenUnitsFlexOperation_050523.pdf}.
\item[131.] A. L. Goodkind, C. W. Tessum, J. S. Coggins, J. D. Hill, J. D. Marshall, Fine-scale damage estimates of particulate matter air pollution reveal opportunities for location-specific mitigation of emissions. \emph{Proc. Natl. Acad. Sci. U.S.A.} \textbf{116}, 8775--8780 (2019). doi:10.1073/pnas.1816102116
\item[132.] A. L. Goodkind, C. W. Tessum, J. S. Coggins, J. D. Hill, J. D. Marshall, InMAP source--receptor matrix (ISRM) dataset, version 1.2.1, Zenodo (2019); \url{https://doi.org/10.5281/zenodo.2589760}.
\item[133.] C. Tessum, ``Using EarthSciLab for calculations with the InMAP Source Receptor Matrix (ISRM)'' (InMAP blog, 9 September 2026); \url{https://inmap.run/blog/2026/09/09/isrm-esm}.
\item[134.] G. A. Briggs, ``Plume rise predictions'' in \emph{Lectures on Air Pollution and Environmental Impact Analyses}, D. A. Haugen, Ed. (American Meteorological Society, 1975), pp. 59--111.
\item[135.] Copernicus Climate Change Service, ERA5 hourly data on pressure levels from 1940 to present, Climate Data Store (2018); \url{https://doi.org/10.24381/cds.bd0915c6}.
\item[136.] Centers for Disease Control and Prevention, National Center for Health Statistics, Underlying cause of death, 2018--2024, single race (CDC WONDER database D158); \url{https://wonder.cdc.gov/ucd-icd10-expanded.html} {[}accessed 15 October 2026{]}.
\item[137.] X. Wu, D. Braun, J. Schwartz, M.-A. Kioumourtzoglou, F. Dominici, Evaluating the impact of long-term exposure to fine particulate matter on mortality among the elderly. \emph{Sci. Adv.} \textbf{6}, eaba5692 (2020). doi:10.1126/sciadv.aba5692
\item[138.] J. Heo, P. J. Adams, H. O. Gao, Reduced-form modeling of public health impacts of inorganic PM\textsubscript{2.5} and precursor emissions. \emph{Atmos. Environ.} \textbf{137}, 80--89 (2016). doi:10.1016/j.atmosenv.2016.04.026
\item[139.] U.S. Environmental Protection Agency, ``Technical support document: Estimating the benefit per ton of reducing directly-emitted PM2.5, PM2.5 precursors and ozone precursors from 21 sectors'' (June 2024); \url{https://www.epa.gov/system/files/documents/2024-06/source-apportionment-tsd-2024.pdf}.
\item[140.] U.S. Environmental Protection Agency, ``AVERT user manual'' (version 4.3, April 2024); \url{https://www.epa.gov/avert/avert-user-manual}.
\item[141.] Texas Commission on Environmental Quality, ``Air quality permit nos. 181009, PSDTX1670, and GHGPSDTX254: Fermi Equipment Holdco, LLC, Fermi America Project Matador, Carson County, Texas'' (25 February 2026); \url{https://records.tceq.texas.gov/cs/idcplg?IdcService=TCEQ_EXTERNAL_SEARCH_GET_FILE&Rendition=Web&dID=9562391&searchType=External}.
\item[142.] Williams Companies, ``Socrates power generation facilities, Licking County, Ohio'' (map, 6 March 2025); \url{https://www.williams.com/wp-content/uploads/sites/14/2025/03/Socrates-Power-Generation-Facilites-Map.pdf}.
\item[143.] Office of Management and Budget, ``Circular A-4'' (17 September 2003); \url{https://obamawhitehouse.archives.gov/omb/circulars_a004_a-4}.
\item[144.] Executive Office of the President, Executive Order 14192 of January 31, 2025, Unleashing prosperity through deregulation. \emph{Fed. Regist.} \textbf{90}, 9065--9067 (2025); \url{https://www.govinfo.gov/content/pkg/FR-2025-02-06/pdf/2025-02345.pdf}.
\item[145.] Bureau of Economic Analysis, NIPA Table 1.1.9, Implicit price deflators for gross domestic product, vintage of 26 June 2025 (archived in EPA cost workbook attachment 4, 2025); \url{https://downloads.regulations.gov/EPA-HQ-OAR-2024-0419-0242/attachment_4.xlsx}.
\item[146.] U.S. Environmental Protection Agency, ``Method 19: Determination of sulfur dioxide removal efficiency and particulate matter, sulfur dioxide, and nitrogen oxides emission rates''; \url{https://www.epa.gov/emc/method-19-sulfur-dioxide-removal-and-particulate-sulfur-dioxide-and-nitrogen-oxides-electric}.
\item[147.] Entergy Louisiana, ``Entergy Louisiana breaks ground on key substation to power data center in Richland Parish'' (press release, 27 June 2025); \url{https://www.entergy.com/news/entergy-louisiana-breaks-ground-on-key-substation-to-power-data-center-in-richland-parish}.
\item[148.] Louisiana Economic Development, ``Louisiana earns `Deal of the Year' award for historic \$10 billion Meta AI data center project in northeast Louisiana'' (press release, 8 January 2025); \url{https://www.opportunitylouisiana.gov/news/louisiana-earns-deal-of-the-year-award-for-historic-10-billion-meta-ai-data-center-project-in-northeast-louisiana}.
\item[149.] U.S. Environmental Protection Agency, ``Available and emerging technologies for reducing greenhouse gas emissions from combustion turbine electric generating units'' (Office of Air and Radiation, 21 April 2022); \url{https://nepis.epa.gov/Exe/ZyPURL.cgi?Dockey=P101CLPD.TXT}.
\item[150.] California Energy Commission, ``CEC staff's recommendation to license Greenleaf 1 temporary power generators (21-TPG-02)'' (TN 239617, California Energy Commission, 8 September 2021); \url{https://efiling.energy.ca.gov/GetDocument.aspx?tn=239617}.
\item[151.] Kiewit Engineering Group, ``Calpine Greenleaf 1 emergency generator: Facility closure plan'' (Document 20057371-PLN-001, Revision H, TN 261308, California Energy Commission, 27 January 2025); \url{https://efiling.energy.ca.gov/GetDocument.aspx?tn=261308}.
\item[152.] Mississippi Department of Environmental Quality, 11 Miss. Admin. Code Pt. 2, Ch. 2, R. 2.13.D, Mississippi air pollution control regulations; \url{https://www.mdeq.ms.gov/air/air-regulations/} {[}accessed 21 October 2026{]}.
\item[153.] Roseville Electric Utility, ``Roseville Energy Park temporary generators quarterly report: Second quarter 2023'' (Docket No. 21-TPG-01, TN 251362, California Energy Commission, 28 July 2023); \url{https://efiling.energy.ca.gov/GetDocument.aspx?tn=251362}.
\item[154.] H. S. Bloom, Minimum detectable effects: A simple way to report the statistical power of experimental designs. \emph{Eval. Rev.} \textbf{19}, 547--556 (1995). doi:10.1177/0193841X9501900504
\item[155.] California Energy Commission, ``CEC staff's recommendation to license Roseville State Power Augmentation Power Site (21-TPG-01)'' (TN 239616, California Energy Commission, 8 September 2021); \url{https://efiling.energy.ca.gov/GetDocument.aspx?tn=239616}.
\item[156.] R. Gasparini, ``Declaration of Dr. Roberto Gasparini'' (with exhibits, ECF No. 60-4, Case No. 3:26-cv-00074-DMB-JMV, N.D. Miss., 15 June 2026), CourtListener RECAP Archive; \url{https://storage.courtlistener.com/recap/gov.uscourts.msnd.52261/gov.uscourts.msnd.52261.60.4.pdf}.
\end{scirefs}

\textbf{Acknowledgments:} Generative AI tools were used to polish the language of the manuscript and to improve the quality of the code. All text and code were finally proofread and tested by the authors, who take full responsibility for the content of this work. No AI tool was used to generate results or conclusions.

\textbf{Data, code, and materials availability:} All data used in this study are publicly available from the sources listed in the supplementary materials. Data S1 contains the tabulated data underlying the figures. The processed data (\emph{89}), the population grid behind Figs. 1A and 3A, which contains information from Microsoft US Building Footprints made available under the Open Database License (\emph{90}), and the code that reproduces the reported statistics (\emph{91}) are archived at Zenodo. This study did not generate new materials.

\noindent\textbf{Supplementary Materials}\\ Materials and Methods, Supplementary Text, Figs.~S1 to S4, and Tables S1 to S13 follow the main text in this document. Data S1 is provided as an ancillary file (\texttt{Data\_S1.xlsx}, listed under Ancillary files on the arXiv abstract page).

\clearpage
\setcounter{page}{1}
\renewcommand{\thepage}{S\arabic{page}}
\begin{center}
{\Large Supplementary Materials for}\\[6pt]
{\Large\bfseries Temporary gas turbines powering a US AI data center created a cross-state air pollution hotspot}\\[8pt]
\paperauthors\\[4pt]
{\small\paperaffils}
\end{center}

\noindent\textbf{This PDF file includes:}\\
Materials and Methods\\
Supplementary Text\\
Figs. S1 to S4\\
Tables S1 to S13\\
References for the supplementary materials (92 to 156) are listed with the main text.

\noindent\textbf{Other Supplementary Materials for this manuscript include the following:}\\
Data S1 (ancillary file \texttt{Data\_S1.xlsx})

\section*{Materials and Methods}

The analysis had five parts. First, satellite observations measured the increase in tropospheric nitrogen dioxide (NO\textsubscript{2}) columns near the fleet after the turbines entered service. Second, we tested whether this increase was anomalous and whether it came from the fleet, using national comparisons, analyses of wind direction, distance, wind speed, and fleet capacity, and a screening of other emission sources. Third, overlaying the field of increases with population, community characteristics, and monitor locations identified the affected population and the monitoring network's coverage. Fourth, we converted emissions estimated from unit operating records into health damages and distributed them across states using reduced-complexity air quality models. Finally, we compared the benefits and costs of selective catalytic reduction (SCR) systems over different evaluation horizons, both for SCR on new turbines and for retrofits of existing units. The sections below describe the study site and data sources and then each part in turn.

\subsection*{Study site and data sources}

The turbine fleet studied here is located at 2875 Stanton Road, Southaven, Mississippi, about 1.6 km south of the Tennessee state line. The study covers this power generation site only and excludes the buildings of the Colossus 2 data center. Colossus 1, in Memphis next to the Tennessee Valley Authority (TVA) Allen plant, is a separate site. The unit roster compiled from public records contains 69 regulatory identifiers, 68 of which correspond to delivered units. According to Appendix A of the amended agreed order, 14 of the delivered units are Solar SMT-130 units equipped with SCR and oxidation catalysts, and the remaining units control nitrogen oxides (NO\emph{\textsubscript{x}}) with demineralized water injection. The water-injected models include several generations of the TM2500, the FT8 MOBILEPAC, the M35, and the JT35.

To establish the composition of the fleet and the regulatory record of each unit, we relied mainly on the agreed order issued by the Mississippi Commission on Environmental Quality on 30 July 2026 and the amended agreed order issued on 11 September 2026, whose Appendix A lists every unit (\emph{16}, \emph{17}). We also consulted the 29 July 2025 letter from the Mississippi Department of Environmental Quality (MDEQ) (\emph{19}), the biweekly reports required by the agreed order, and the inventory corrections recorded in the amended agreed order. For the permanent power plant at the same site, we used the Prevention of Significant Deterioration (PSD) permit issued on 11 March 2026 and the related response to comments (\emph{51}, \emph{81}). Company information comes from filings by Solaris Energy Infrastructure with the U.S. Securities and Exchange Commission (\emph{21}, \emph{66}), and information on gas contracts comes from the index of customers that Texas Gas Transmission filed with the Federal Energy Regulatory Commission (\emph{43}). Daily scheduled quantities at the site's delivery point come from the pipeline's public informational postings (Data S1) (\emph{92}). The reference list marks documents obtained as publicly posted copies. Unit status records are current as of 3 October 2026.

From these records, we documented the arrival, shutdown, disconnection, removal, and authorization deadline dates for each unit to separate actual equipment changes from regulatory deadlines. Monthly on-site capacity was calculated from confirmed arrival and removal dates. Where these dates were uncertain, we report lower and upper bounds on capacity (Data S1). The supplementary text describes the state and federal rules for temporary authorization and the deadlines of individual units.

\subsection*{Satellite and meteorological data}

The main analysis used Level-2 tropospheric NO\textsubscript{2} columns from the Tropospheric Emissions: Monitoring of Pollution (TEMPO) instrument (V04 collection) (\emph{93}) at native pixel resolution for the periods stated for each analysis, mainly April to September of 2024, 2025, and 2026. All columns are in units of 10\textsuperscript{15} molecules cm\textsuperscript{\ensuremath{-}2}. We retained only pixels of reliable quality, defined as a main quality flag of 0 and a finite column with an absolute value below 1,000. The operational retrieval converts slant columns to vertical columns with air mass factors that depend on a priori NO\textsubscript{2} profiles from a global chemical transport model, which do not include this new source. Where the true profile holds more NO\textsubscript{2} near the surface than the a priori profile, the air mass factor is too large and the retrieved column too low, and operational retrievals have shown such low biases near urban centers (\emph{32}). The retrieved columns therefore tend to understate the near-field increase. Errors that vary smoothly in space, such as those in separating stratospheric and tropospheric columns, largely cancel in the near-field excess. Pixels were weighted by their area of overlap with the analysis grid. When several data files (granules) covered the same observation time, we merged them before any calculation. Processor versions can differ among scans, and Data S1 lists the version used for each paired scan at 17 to 20 UTC.

The source-agnostic peak search used the quality-filtered TEMPO Level-3 V04 product on a 0.02° grid distributed by Google Earth Engine (NASA/TEMPO/NO2\_L3\_V4\_QA) (\emph{94}). The national comparison and the synthetic-control analyses used the gridded offline product of the TROPOspheric Monitoring Instrument (TROPOMI) provided by Google Earth Engine (COPERNICUS/S5P/OFFL/L3\_NO2), and TROPOMI Level-2 columns (\emph{95}) provide a record independent of TEMPO from 2019 to 2026. TROPOMI observes once a day near 13:30 local solar time, with larger ground pixels and a different retrieval. TEMPO columns are marginally higher than TROPOMI columns over polluted regions, mainly because of different air mass factor assumptions (\emph{96}), so the two instruments are compared by the timing and ranking of the change rather than by its magnitude. Wind rotation used 10-m winds from the High-Resolution Rapid Refresh (HRRR) (\emph{97}, \emph{98}), averaging the eastward and northward components over the hour of the scan and the three preceding whole hours to represent transport before the scan. Wind-speed correlations used ERA5 10-m winds (\emph{99}, \emph{100}), matched to the UTC hour containing each scan. Weak-wind shares were calculated separately with HRRR and ERA5. Smoke screening used daily smoke polygons from the Hazard Mapping System (HMS) of the National Oceanic and Atmospheric Administration (NOAA) (\emph{42}).

The main observation window was 17 to 20 UTC (noon to 4 p.m. local daylight time in Memphis). We chose this window because it overlaps the TROPOMI overpass time and has small solar zenith angles, averaging 25° in spring and 36° in late summer, compared with 53° and 58° for the other daytime hours. TEMPO total NO\textsubscript{2} columns agree best with ground-based Pandora measurements at small solar zenith angles and are less reliable early in the morning and near sunset (\emph{101}). The spring (April to July) main sample contains 310 paired scans on 113 days, and the late-summer (August to September) sample contains 142 paired scans on 41 days (table S1).

\subsection*{Paired estimate of the near-field NO\textsubscript{2} increase}

Following satellite studies of localized sources (\emph{31}, \emph{36}), we used the near-field excess to measure how far NO\textsubscript{2} columns near the fleet exceeded the regional background. We defined the near-field excess as the mean tropospheric NO\textsubscript{2} column within 5 km of the fleet (hereafter the core area) minus the weighted mean column over a reference area. The reference area is a weighted combination of candidate grid cells 10 to 20 km from the fleet and represents the regional background (hereafter the weighted reference area, to distinguish it from the fixed reference ring described below). To control for differences in season, time of day, and viewing geometry, we paired each 2026 TEMPO scan with the 2024 and 2025 scans closest in minute on the same calendar date and in the same UTC hour (hereafter a paired scan), and we computed all quantities only within the area validly observed in all three years (hereafter the common valid footprint). The main analysis set no further limit on the time difference and did not reweight dates by day of the week, and fig. S1C varies both. We dropped a paired scan if the common valid footprint covered less than half of the core area or if the valid coverage of the weighted reference area (the weighted sum of the valid-area fractions of the reference cells) was below 0.5. In the reference mean, each cell was weighted by its weight times its valid-area fraction, normalized to sum to 1.

For each year \emph{y} (2024, 2025, or 2026), we computed the near-field excess \(E_{y}\) by averaging the per-scan excess within each date and then averaging all dates with equal weight, so that dates with many scans would not dominate. The paired estimator is

\begin{equation*}
D = E_{2026} - \frac{1}{2}\left( E_{2024} + E_{2025} \right)\tag{S1}
\end{equation*}

where \(D\) is the change in 2026 relative to the 2024--2025 baseline. In the spring main sample, the 2024--2025 mean column over the core area was 2.24. To test whether the predeployment baseline was stable, we applied the same method to April to July 2025 relative to 2024.

We chose the weights so that the weighted average of the candidate cells resembled the core area in land use, traffic, and predeployment NO\textsubscript{2} columns. The candidate cells are 0.02° grid cells (about 4 km\textsuperscript{2} each) centered 10 to 20 km from the fleet. To limit the influence of other sources, we excluded 32 cells within 5 km of the TVA Allen plant, the Valero refinery, or Colossus 1. We also excluded 8 cells west of the core area that lacked 2024 traffic data. This left 201 candidate cells.

The weights were estimated from predeployment data only, and the same weights were used in both seasons. We compared candidate cells with the core area on seven indicators, namely the 2024 density of major roads, the 2024 impervious fraction, the 2024 vegetated fraction, the distances to Memphis International Airport and to the Valero refinery, the logarithm of weighted annual average daily traffic within 5 km, and the NO\textsubscript{2} column averaged over the daily means of all daytime scans from April to July 2024. Road density uses the 2024 TIGER/Line primary and secondary roads (\emph{102}). Traffic is the mean 2024 annual average daily traffic at state count stations within 5 km, weighted by \(\exp(-d/2\ \mathrm{km})\) for distance \(d\) (\emph{103}--\emph{105}). The impervious and vegetated fractions come from the Annual National Land Cover Database (Annual NLCD) for 2024 (\emph{106}), with forest, shrubland, grassland, pasture, cropland, and wetland classes counted as vegetated. Each indicator was standardized by its standard deviation in the pooled sample of core and candidate cells. Following the penalized synthetic-control approach of Doudchenko and Imbens (\emph{107}), we then solved for the weights that bring the weighted indicators of the candidate cells as close as possible to the mean indicators of the core area

\begin{equation*}
\min_{w}\ \frac{1}{2}\left\lbrack \left\lVert X_{\mathrm{core}} - \sum_{j}w_{j}X_{j} \right\rVert^{2} + \lambda\sum_{j}^{}w_{j}^{2} \right\rbrack,\quad\sum_{j}^{}w_{j} = 1,\ 0 \leq w_{j} \leq 0.1\tag{S2}
\end{equation*}

where \(w_{j}\) is the weight of candidate cell \(j\) (\(j\) = 1, \ldots, 201), \(X_{j}\) is the vector of its seven standardized indicators, and \(X_{\mathrm{core}}\) is the vector of mean indicators for the core area. The ridge penalty coefficient λ is 1, which makes the weights unique and prevents them from concentrating on a few cells. The weights sum to 1 and are capped at 0.1, so at least 10 cells contribute. The cap does not bind in the main estimate, whose largest weight is 0.08, and values of λ from 0.1 to 10 give changes of 1.10 to 1.13 in spring and 2.66 to 2.76 in late summer.

\subsubsection*{Field of increases and monthly series}

To map the NO\textsubscript{2} increase and delineate the hotspot, we computed the paired difference cell by cell on a 1-km grid following Eq. S1. TEMPO native pixels are about 2.1 km by 4.4 km (\emph{34}), so the 1-km field oversamples the native resolution, and the half-decay distance and the hotspot boundary include this instrumental smoothing. For each paired scan, we first computed the mean column of each 1-km equal-area cell (EPSG 5070 projection) within the common valid footprint. We then subtracted the mean of the 2024 and 2025 columns from the 2026 column in each cell and removed the corresponding change over the weighted reference area in the same paired scan, which gave the increase in each cell relative to the regional background.

Cell results were aggregated by date and then averaged across dates with equal weight, as for \(E_{y}\). This gave separate fields of increases for spring and late summer. The April to September field combines the two seasons weighted by calendar days, 122 for spring and 61 for late summer. Cells with fewer than 15 valid observation dates were excluded. For Fig. 1A only, we refined the grid spacing to 0.1 km by cubic interpolation and applied Gaussian smoothing at a scale of 0.3 km.

To follow the monthly change in the increase as the fleet expanded, we compared the near-field excess in each month from August 2025, when the first turbines arrived, to September 2026 with the same month of 2024, a year without turbines (Fig. 2A). Scans were paired by calendar date and UTC hour as in the main analysis, and calculations were restricted to the area validly observed in both years.

\subsubsection*{Sensitivity tests and same-scan placebos}

We tested the sensitivity of the main estimate to the reference area, observation window, data processor version, smoke screening, and weather adjustment (table S2). We re-estimated the change with three alternative reference areas. (1) Instead of estimating weights, we used the area-weighted mean over the complete 10- to 20-km annulus as a fixed reference ring. (2) Reference cells downwind of Colossus 1 under HRRR wind directions were excluded. (3) The distances to Memphis International Airport and the Valero refinery were omitted from the matching indicators when the reference weights were estimated (table S2E). Fleet NO\textsubscript{2} carried downwind into the 10- to 20-km reference cells would lower the near-field excess, so the specification curve (fig. S1C) also uses a 30- to 60-km reference ring farther from the fleet. The observation window of 17 to 20 UTC was also extended to all daytime hours. The processor-version test included only paired scans processed with version 4.8.4 in all three years, which removes the effect of version differences. The smoke-free subsample test kept only paired scans in which the core area did not overlap the smoke polygons flagged by the NOAA HMS at the scan times in any of the three years. The main quality flag imposes no cloud-fraction threshold (\emph{108}). Requiring an effective cloud fraction below 0.2 or 0.3 in all three years, with the reference weights of the main analysis, left 12 to 40 paired dates and gave increases of 0.90 to 1.03 in spring and 3.15 to 4.02 in late summer, all with intervals excluding zero (Data S1).

Weather adjustment took two forms (table S2A). The discrete approach divided scans into 48 strata by wind direction (16 sectors of 22.5°) and wind speed (below 2, 2 to 5, and above 5 m s\textsuperscript{\ensuremath{-}1}), retained strata observed in all three years, and reweighted each year's observations by the ratio of the 2026 stratum weights to that year's stratum weights so that their wind distribution matched 2026. We bounded weight ratios between 0.2 and 5 so that sparsely observed strata would not dominate, and removing the bounds changes the reweighted estimates by less than 0.01. We used HRRR and ERA5 winds separately. The continuous approach used ERA5 wind speed and its inverse, the first and second harmonics of wind direction, the logarithm of boundary-layer height, air temperature, relative humidity, and month and hour indicators. A multinomial logistic regression with threefold cross-fitting (\emph{109}) estimated the probability that each observation belonged to each year, with folds formed from 7-day blocks and training sets excluding observations within 3 days of the test dates. We dropped observations with a probability below 0.05 for any year and used inverse probability weights normalized within each year to standardize the three years to a common covariate distribution. Both approaches kept dates equally weighted and re-estimated the reference weights and the weather weights in every bootstrap replicate. The unweighted rows in the table use only the observations retained by each approach and therefore differ from the main estimate. Probability thresholds of 0.025 and 0.1, standardization with cross-fitted ridge outcome models, and augmented inverse probability weighting gave estimates of 0.76 to 1.20, with the widest intervals for the augmented estimator.

The specification curve in fig. S1C summarizes 300 point estimates. The 300 estimates fall into 20 strata that combine two seasons, two observation windows, and five reference approaches, namely reference weights estimated from 2024 data as in the main analysis, reference weights estimated from 2025 data, equal weights across the 201 candidate cells, a 10- to 20-km reference ring, and a 30- to 60-km reference ring. The remaining settings were core radius (5 or 10 km), minimum coverage (0.5 or 0.75), tolerance for differences in scan time across years (5, 15, or 30 min), baseline (the 2024--2025 mean, 2024, or 2025), and weekday weighting (original or standardized). Within each stratum, 15 combinations of these settings were drawn at random with a fixed seed. In each season, 15 of the 150 specifications were also drawn at random for confidence intervals, 30 in total. With a 5-min tolerance, the late-summer sample keeps a single paired date in 48 specifications, and the three intervals drawn among them have zero width. Of the other 27 intervals, 25 exclude zero.

We also performed same-scan placebo tests in other cities covered by the same TEMPO scans, to check whether the method would estimate an increase where no new source had appeared. Candidate cities had a 2020 population between 100,000 and 2 million and lay 100 to 800 km from the fleet. We then excluded any city with (1) a generating unit larger than 50 MW that entered service from 2024 to 2026 within 20 km, (2) a source or industrial corridor within 20 km whose annual NO\emph{\textsubscript{x}} emissions increased by more than 100 short tons from 2024 to 2025, or (3) more than 10 additional smoke days in 2026 compared with the previous two years. We fixed the final list of 15 cities before computing any NO\textsubscript{2} results. Each city was analyzed with the same procedure and the same scans as Southaven, estimating the paired change \(D\) with a 5-km core area and the fixed reference ring. Confidence intervals were computed from 1,999 circular block bootstrap replicates with blocks of 7 consecutive days (table S2C).

\subsubsection*{Statistical inference and detection threshold}

Confidence intervals for the main satellite analyses are based on 1,999 bootstrap replicates and were computed from the valid estimates among them. Unless stated otherwise, bootstrap intervals are 95\% percentile intervals, and we call a change significant when its 95\% interval excludes zero. The 30 intervals in the specification curve of fig. S1C instead use 499 replicates with the reference weights held fixed. Data S1 reports the number of valid estimates and \emph{P} values for the main estimated changes, two-sided except for the one-sided synthetic-control permutation tests. For bootstrap estimates, \(P = \min[1,\, 2(k + 1)/(B + 1)]\), where \(B\) is the number of valid replicates and \(k\) is the number of them on the smaller side of zero. For the paired estimator \(D\), the monthly changes, and the TROPOMI near-field excess at Southaven, which uses a fixed reference ring (Fig. 1B and table S3A), we used a circular block bootstrap (\emph{110}, \emph{111}) with blocks of 7 consecutive days, which preserves the weekly cycle of emissions and the correlation of weather across consecutive days. All scans from the same day entered the resampling together, and dates were weighted equally in aggregation.

For the cross-season field of increases and the hotspot population, we used a Bayesian bootstrap (\emph{112}) over calendar weeks. In each replicate, every calendar week received a random weight drawn from a standard exponential distribution and shared by all dates in that week. The weights were drawn once for all calendar weeks (Monday to Sunday) from April to September and then rescaled within each season to average 1 per calendar date, so the week spanning the end of July received the same random draw in both seasons.

In both resampling schemes, each replicate of an estimate that uses the weighted reference area, except in the specification curve, recomputed the NO\textsubscript{2} indicator from the April to July 2024 dates as resampled in that replicate, restandardized the indicators, and re-estimated the reference weights, so the confidence intervals include uncertainty in weight estimation. We computed contrasts such as the ratio of the late-summer to the spring increase and the year-to-year difference in wind-speed dependence from estimates within the same replicate, preserving their correlation. In each replicate, the 15-date requirement was evaluated from the resampled or weighted number of validly observed dates, and the confidence intervals for the hotspot population and the share of that population in Tennessee were computed from all 1,999 replicates.

To determine the earliest month in which satellite observations could detect the increase, we accumulated observations month by month starting in February 2026 and paired them with observations on the same dates and at the same UTC hours in 2024. With each additional month, we recomputed the increase and its confidence interval from all paired observations up to that month (the retrospective cumulative analysis) and defined the detection month as the first month in which the lower bound of the 95\% confidence interval exceeded zero.

We defined the detection threshold as the lowest emission rate at which the lower bound of the 95\% confidence interval of the increase would exceed zero under the observing conditions of the spring main sample. We held fixed the random variation reflected in the 1,999 bootstrap replicates of the spring main sample and assumed that the increase is proportional to the emission rate, which gives

\begin{equation*}
q^{*} = q_{0}\,\frac{{\widehat{D}}_{0} - L_{0}}{{\widehat{D}}_{0}}\tag{S3}
\end{equation*}

where \(q_{0}\) is the mean emission rate of the spring engineering inventory (215.6 kg per hour), \({\widehat{D}}_{0}\) = 1.103 is the corresponding estimate of the increase, and \(L_{0}\) = 0.584 is the lower bound of its 95\% confidence interval. This gives \(q^{*}\) ≈ 101 kg per hour. Requiring 80\% power, the criterion for the minimum detectable values in table S6, raises this rate to about 150 kg per hour, still below the inventory rate of 215.6 kg per hour.

\subsection*{Anomaly and attribution tests}

\subsubsection*{Comparison with urban locations nationwide and across the contiguous US}

To test whether the increase at Southaven was anomalous among US urban areas, we selected locations from the urban areas defined in the 2020 Census. Candidate grid points were spaced 10 km apart, and only points whose April to July background TROPOMI columns from 2020 to 2024 lay between the 10th and 99th percentiles were retained. We visited these points in a random order with a fixed seed and kept each one only if it lay at least 30 km from all previously kept locations. The selection used no location's 2026 NO\textsubscript{2} change. This yielded 655 comparison locations, or 656 including Southaven (table S3). For each location, we defined the near-field excess \(G\) as the mean column within a 5-km core area minus the mean column over a fixed reference ring 10 to 20 km away. For Southaven, the reference ring also excluded the area within 10 km of Colossus 1. We measured the change at each location relative to its own history with a stabilized anomaly score

\begin{equation*}
Z = \frac{G_{2026} - {\overline{G}}_{\mathrm{hist}}}{s_{\mathrm{hist}} + \sigma_{2026}}\tag{S4}
\end{equation*}

where \(G_{2026}\) is the TROPOMI near-field excess from 1 April to 31 July 2026, \({\overline{G}}_{\mathrm{hist}}\) and \(s_{\mathrm{hist}}\) are the mean and standard deviation of the near-field excess over the same period from 2020 to 2025 (six annual values), and \(\sigma_{2026}\) is the standard error of \(G_{2026}\), taken as the standard deviation of 1,999 circular 7-day block bootstrap estimates. Summing the two terms instead of adding them in quadrature makes the score conservative. The spring comparison required at least 30 valid observation days at each location. To place the 2026 score in the longer record, we computed the same score for every location in each year from 2020 to 2025, using the other five years from 2020 to 2025 as the history and requiring at least 30 valid observation days in the target year and in each historical year. Recomputing the 2026 scores with any five of the six historical years gave Southaven scores of 4.51 to 5.10, each ranking first among all location-years, and Southaven also ranked first with minimums of 20 or 40 valid observation days. For late summer (August to September), we ranked the locations by their change relative to the same months of 2024--2025 and, because this period is shorter, lowered the minimum to 10 valid observation days.

Like the point-source catalog of Beirle et al. (\emph{39}), the source-agnostic peak search does not prespecify the location of any emission source. It uses TEMPO observations to identify, across the contiguous US, local peaks in the increase that persisted over several months. We first reprojected TEMPO Level-3 columns at 0.02° to a 1-km grid, then grouped observations from 2024, 2025, and 2026 by calendar date and UTC hour, computing the increase only within areas validly observed in all three years. The search used the 17 to 20 UTC window and counted a pixel-day only when at least two UTC hours were observed in all three years. For each month, we averaged the results across dates with equal weight to obtain the monthly field of increases, then averaged the monthly fields across the selected months with equal weight.

Candidate search centers were the centers of grid cells spaced 5 km apart. For each center, we computed the increase in a 5-km core area relative to a reference ring and required valid coverage of at least 80\% in both areas. Peaks were then ranked from the largest to the smallest increase. When two peaks were closer than the specified minimum spacing, the smaller one was removed so that the same increase would not be counted twice. A peak was considered persistent if its increase was positive in at least three of the four months from April to July and no single month accounted for more than 60\% of the four-month total. This four-month rule was applied in every spring specification, including rankings that omitted May or used only June and July. In late summer, the increase had to be positive in both August and September.

To test whether the rank of the peak depended on parameter settings, we used specification curve analysis, repeating the search under many combinations of parameters and comparing ranks (\emph{37}, \emph{38}). We examined 60 combinations of four types of parameters, namely (1) the analysis months (April to July, April to July excluding May, and June to July), (2) the minimum spacing between peaks (15 or 20 km), (3) the minimum number of valid observation days per month (4 to 8), and (4) the distance of the reference ring from the search center (10 to 20 km or 30 to 60 km). The main specification, shown in Fig. 1C, uses April to July, a minimum spacing of 20 km, at least four valid observation days per month, and the 10- to 20-km reference ring. A combination was excluded from the ranking analysis if valid coverage was insufficient or if the Southaven peak could not be separated from a neighboring peak.

For the top-ranked peaks, we compared locations with fire perimeters from the National Interagency Fire Center (\emph{40}), HMS smoke polygons, and power plant locations to identify possible sources. These comparisons describe the peaks and leave the rankings unchanged. The spring specifications in which Southaven ranked second all excluded May or were limited to June and July, and in each of them the leading grid cell overlapped the perimeter of the Aspen Acres wildfire in Colorado, which was first reported on 29 June (\emph{40}). For late summer, we repeated the search over the contiguous US for August and September combined, August alone, and September alone, using a 10- to 20-km reference ring, a minimum peak spacing of 20 km, and at least five valid observation days common to all three years in each month.

We used monthly TROPOMI near-field excess to compare changes at Southaven and at comparison locations before and after deployment. From the 655 comparison locations, we excluded any location within 20 km of a combustion unit of at least 25 MW that Forms EIA-860 or EIA-860M of the U.S. Energy Information Administration (EIA) (\emph{113}, \emph{114}) record as entering service, retiring, or switching fuel between June 2018 and September 2026. We also excluded locations within 20 km of any power plant whose annual NO\emph{\textsubscript{x}} emissions changed by at least 100 short tons between consecutive years from 2017 to 2025 in the Clean Air Markets Program Data (CAMPD) (\emph{44}), which left 80 comparison locations. The predeployment period ran from June 2018 to July 2025, and the postdeployment period from August 2025 to September 2026. Because Southaven had no monthly observation in June 2018 or February 2021, the fits used the remaining 84 months. We estimated the postdeployment change with four methods, namely synthetic control (SCM) (\emph{41}), augmented synthetic control (ASCM) (\emph{115}), synthetic difference-in-differences (SDID) (\emph{116}), and equal-weight difference-in-differences (DID). ASCM uses ridge regression to correct the predeployment fit bias of synthetic control.

To judge whether the change at Southaven was unusual among these locations, we followed Abadie et al. (\emph{41}, \emph{117}) and repeated the estimation with Southaven and each of the 80 comparison locations in turn treated as the deployment site. For each location, we computed the root mean squared prediction error (RMSPE) between observed and predicted values before and after deployment, divided the postdeployment RMSPE by the predeployment RMSPE, and ranked the locations from the largest ratio to the smallest. The permutation \emph{P} value for Southaven is its rank divided by the total of 81 locations. We also ran a placebo-in-time test that treated August 2024 as the start of deployment, with June 2018 to July 2024 as the predeployment period and August 2024 to July 2025 as the postdeployment period, and repeated the analysis for the same 81 locations to see whether a similar change appeared before the actual deployment.

\subsubsection*{Tests of the source of the increase}

We first tested whether the spatial pattern of the increase matched emissions from the fleet carried by the wind. Centering on the fleet, we rotated each scan to align with the HRRR 10-m wind direction (\emph{32}) so all scans shared a common wind direction, then compared mean columns in the downwind and upwind boxes. The downwind box extends 0 to 15 km along the wind and the upwind box −15 to 0 km, and both extend 5 km on either side perpendicular to the wind (Fig. 2C). The main analysis used two settings, with each year rotated by its own winds and with all three years rotated by the 2026 winds at the corresponding times. As placebo tests, we also rotated all three years with the 2024 or the 2025 winds at the corresponding times (fig. S2A).

We next examined whether the increase declined with distance from the fleet. We aggregated the increase in concentric 1-km annuli centered on the fleet, with annulus midpoints from 0.5 to 29.5 km, and fitted the exponential decay model \(D(x) = b + A\exp( - x/\ell)\), where \(x\) is the distance from the fleet, \(b\) is the far-field level, \(A\) is the near-source increase relative to that level, and \(\ell\) sets the distance scale of the decay. The fit was weighted by the inverse of the bootstrap variance of each annulus, with the standard deviation floored at 0.03, and \(A\), \(\ell\), and \(b\) were bounded by {[}0, 50{]}, {[}0.1, 100{]}, and {[}−20, 20{]}. The distance at which the increase relative to the far-field level \(b\) falls by half, the half-decay distance, is \(\ell\ln 2\), and its confidence interval was obtained by refitting the model in the same block bootstrap replicates (Fig. 2E). The fitted far-field level is negative in both seasons (table S2B), meaning that distant annuli changed less than the weighted reference area. Measured against these distant annuli, the increase near the turbines is therefore larger than the near-field excess.

We also used wind-speed variation to test for a local source. An increase in near-source columns that is inversely related to wind speed is a common signature of a local source (\emph{36}). Using ERA5 wind speed, we computed the correlation between the near-field excess and inverse wind speed, \(1/\max(U,0.5\ \mathrm{m\,s}^{-1})\), separately for the baseline (2024 and 2025) and for 2026. Here \(U\) is wind speed, floored at 0.5 m s\textsuperscript{\ensuremath{-}1} to avoid very large inverse values. The calculation used individual scans with dates equally weighted, subtracted each year's mean from that year's data, and computed the difference between the 2026 and baseline correlations in each bootstrap replicate (Fig. 2D). Because the near-field excess already subtracts the reference column, regional accumulation of pollution common to both areas cancels in the difference. To quantify the contribution of weak winds, we set wind-speed thresholds of 2, 3, and 4 m s\textsuperscript{\ensuremath{-}1} and computed the cumulative shares of observation opportunities and of the net increase below each threshold, classifying each paired scan by its 2026 wind in the UTC hour containing it (fig. S2B and table S4). We also computed the increase in each season separately for scans with 2026 single-hour HRRR 10-m winds below 3 m s\textsuperscript{\ensuremath{-}1} and at or above 3 m s\textsuperscript{\ensuremath{-}1}, with dates weighted as in the all-wind estimate. We then compared the seasons after weighting both seasons' stratum estimates by the same distribution of wind speeds (fig. S2C).

In addition to these spatial and meteorological signatures, we examined whether the increase changed as fleet capacity grew. For the 14 months from August 2025 to September 2026, we estimated

\begin{equation*}
\Delta E_{m} = \alpha + \beta C_{m} + \gamma_{s(m)} + \varepsilon_{m}\tag{S5}
\end{equation*}

where \(\Delta E_{m}\) is the change in the near-field excess in month \(m\) relative to the same month of 2024, \(C_{m}\) is the mean daily on-site capacity of units without SCR in that month (in units of 100 MW), \(\alpha\) is the intercept, \(\gamma_{s(m)}\) is a fixed effect for the meteorological season of the month (with winter as the reference), \(\beta\) is the increase per additional 100 MW of capacity, and \(\varepsilon_{m}\) is the error term. We estimated the regression using ordinary least squares and report heteroskedasticity-robust (HC1) intervals with a \emph{t} reference distribution (\emph{118}), with 9 residual degrees of freedom, or 8 when the weather term described next is added, so that the intervals include the scatter of months around the fitted line. To test whether this association reflected changes in meteorology, we added to Eq. S5 the change in mean inverse wind speed, 1/max(\emph{U}, 0.5 m s\textsuperscript{\ensuremath{-}1}) from HRRR 10-m winds, between each month and the same month of 2024 over the same paired scans. Sensitivity analyses instead used the change in the share of scans with winds below 3 m s\textsuperscript{\ensuremath{-}1}, the change in the logarithm of ERA5 boundary-layer height, or both the wind and boundary-layer terms. Capacity counts units with uncertain departure dates until their latest possible departure date. Each bootstrap replicate re-estimated the reference weights, the monthly changes, the weather covariates, and the coefficients. These intervals, hereafter monthly sampling intervals, reflect only the sampling uncertainty of the monthly changes. Without weather terms they give 0.16 (0.12 to 0.22) in Fig. 2B and 0.16 (0.12 to 0.20) in fig. S2D, which keeps only replicates with valid estimates for all 14 months. The HC1 intervals are 0.05 to 0.26 without weather terms and 0.01 to 0.29 with the inverse wind speed term, and a wild bootstrap of HC2-rescaled residuals over all 16,384 sign combinations reproduces them to two decimal places as basic bootstrap intervals. Combining the monthly sampling draws with the wild bootstrap, which counts monthly sampling error twice because the residuals already contain it, gives wider percentile intervals of 0.05 to 0.27 and −0.01 to 0.31.

We also extracted the contracted capacity at the site's delivery point and its effective dates from the Texas Gas Transmission index of customers (\emph{43}), as timing evidence independent of the satellite observations.

To test whether other sources could explain the observed increase, we estimated the contributions of the TVA Southaven and TVA Allen plants to the increase in satellite columns from the hourly emissions that they report to the U.S. Environmental Protection Agency (EPA) (\emph{44}). TVA Southaven lies about 1.5 km from the fleet, within the core area, and TVA Allen about 14 km to the northwest. The calculation used a time-varying Gaussian puff model that accounts for chemical conversion (\emph{119}), sampled over the footprint of native satellite pixels. Inputs included HRRR 80-m winds over the preceding 6 hours. The NO\emph{\textsubscript{x}} lifetime was set to 4 hours (\emph{36}), the NO\textsubscript{2}/NO\emph{\textsubscript{x}} ratio was set to rise with a time scale of 15 min from 0.1 toward 0.7, close to the value of about 0.76 implied by the plume analysis of Beirle et al. (\emph{31}), the initial horizontal spread of each puff was 200 m, a puff was released every 150 s, and trajectories were integrated with 75-s steps. This analysis used a separately assembled paired sample of 101 scans on 42 days from April to June (fig. S2E). We also ran area-weighted regressions on the 1,198 cells of the 1-km grid that lie within 20 km of the fleet and have at least 95\% coverage for every covariate to estimate the increases from 0 to 5 km and 5 to 10 km from the fleet relative to the area 10 to 20 km away (table S5). The GIS-adjusted regressions controlled for six gridded covariates, namely a predeployment background indicator (the 90th percentile of the June 2025 column anomalies in each cell), major-road density, annual average daily traffic, impervious surface, vegetation, and transportation employment density, using the indicators described above and 2023 jobs in transportation and warehousing (\emph{120}). Road density, traffic, and transportation employment density entered the regressions as log(1 + \emph{x}), and all controls were standardized with area weights. Intervals for these regressions were obtained by refitting them to the fields of increases of the 1,999 circular block bootstrap replicates, holding the covariates and area weights fixed.

We also used two observed negative controls to test whether construction activity could explain the increase. The first is May to July 2025 at this site, when construction was under way before the turbines arrived. The second is a large data center campus under construction in Richland Parish, Louisiana (32.505°N, 91.636°W), analyzed for April to July of 2025 and 2026. Both were compared with the same months of 2024 using the paired estimator of the main analysis, with reference weights estimated for each site from 2024 data, to examine whether a similar satellite signal appears where construction is under way (fig. S2E and table S6). Their confidence intervals come from 1,999 resamples of whole calendar weeks drawn with replacement, with the reference weights re-estimated in each replicate. The other two contrasts in table S6 compared April to July 2026 with 2024 at Abilene (32.508°N, 99.777°W) and September to November 2024 with 2023 at Colossus 1 (35.060°N, 90.156°W). Both used reference weights estimated from baseline-year NO\textsubscript{2} columns and intervals from 1,999 circular 7-day block bootstrap replicates, with the weights re-estimated in each replicate. The Colossus 1 contrast used the 273 scans for which the TVA Allen puff simulation was available, and it does not subtract the simulated Allen contribution. For the diurnal comparison, on dates with valid observations in both time windows, we compared the date-weighted mean paired change at 18 to 21 UTC with that at 13 to 16 UTC, with intervals from the same 7-day block bootstrap replicates for both windows (fig. S2F).

\subsection*{Affected population and monitor coverage}

Total population, race, and ethnicity come from 2020 Census P.L. 94-171 block data (\emph{121}). We obtained the population under 18 by subtracting the population aged 18 and older from the total. The population aged 65 and older comes from table P12 of the 2020 Census Demographic and Housing Characteristics (DHC) file (\emph{122}), as the sum of male cells 20 to 25 and female cells 44 to 49. To locate people within blocks more accurately, we allocated block population to buildings in proportion to their footprint area, a form of dasymetric mapping (\emph{123}), and then overlaid the allocation on the analysis grid to compute the population of each cell. Buildings come from Microsoft usbuildings-v2 (\emph{124}). We kept only buildings larger than 40 m\textsuperscript{2} that lie on developed land in the 2021 National Land Cover Database (NLCD 2021) (\emph{125}) and removed Shelby County and DeSoto County parcels occupied by industrial facilities. When a block had no eligible building, its population was allocated to NLCD developed land, and when no developed land was available, to the whole block.

We defined the hotspot as the area where the April to September increase in NO\textsubscript{2} columns was at least 0.5 × 10\textsuperscript{15} molecules cm\textsuperscript{\ensuremath{-}2}, and we used 0.25 and 1.0 as alternative thresholds in sensitivity analyses (table S7). Each bootstrap replicate redelineated the hotspot, summed the population within it, and aggregated that population by state, so that the population estimates reflect the uncertainty of the hotspot boundary. The poverty rate in the hotspot uses block group data from table C17002 of the 2020--2024 American Community Survey (ACS) (\emph{48}), weighted by the 2020 population of each block group allocated to the hotspot. County comparisons use ACS county tables B03002 and C17002 for the same period.

We also tested the association between community composition and the increase with population-weighted regressions across block groups. The increase for each block group was averaged over grid cells with weights equal to the population allocated to each cell, and the regressions were weighted by the total population of each block group in the 2020--2024 ACS. The explanatory variables were the non-Hispanic Black share, the share below the poverty line (each per 10 percentage points), and median household income (per \$10,000, ACS table B19013) (\emph{48}), each entered in a separate regression. Each regression was estimated both without geographic controls and with controls for distance to the fleet (per 10 km), distance squared, the sine and cosine of direction, and their interactions with distance, to examine whether the association persisted after controlling for relative location (table S8). Each regression included the block groups that have population allocated to the grid and a finite increase and explanatory variable in every replicate, so sample sizes differ by variable. Intervals for these regressions were obtained by refitting them to the April to July fields of increases of the 1,999 circular block bootstrap replicates, and each replicate also redrew the block-group shares, median household incomes, and population weights from distributions based on their ACS margins of error, with the geographic controls and analysis samples held fixed.

To assess whether existing ground monitors could reflect the increase, we obtained monitor locations and data availability from the EPA Air Quality System (AQS) (\emph{30}). For the two nearest NO\textsubscript{2} monitors with data in 2025 or 2026, we estimated the increase in the satellite column at each monitor with the same scans as the main analysis, using a 5-km core area centered on each monitor and a 10- to 20-km fixed reference ring (table S2E). In AQS data through May 2026, the Shelby County, Tennessee, monitor (site 47-157-0100) reported no 2026 NO\textsubscript{2} data, and the only Mississippi NO\textsubscript{2} monitor reporting 2026 data (Jackson County, site 28-059-0006) was about 530 km from the fleet.

\subsection*{Emissions and health damages}

\subsubsection*{Fleet emission inventory}

We estimated the emissions of each of the 68 units on the roster as the product of rated capacity, load fraction, operating hours, heat rate, and emission factor, and summed them to obtain the engineering emission inventory (table S9). We obtained June 2026 operating hours from the unit-level operating-hours table the operator's consultant submitted to MDEQ on 10 July 2026 (\emph{126}), covering units 1 to 60 and totaling 21,439 unit-hours. Heat rates were set by model, and emission factors by model and control technology, from permit documents, manufacturer specifications compiled by the operator's consultant (\emph{127}), and EPA AP-42 (\emph{128}). For the M35, JT35, TM2500 Gen2, and TM2500 Gen7, which lack documented heat rates, we used 11.2 million British thermal units (MMBtu) per megawatt-hour (table S9D), which lies within the range of the capacity-weighted average tested heat rates that EIA reports for natural gas-fired combustion turbines from 2015 to 2025 (11.0 to 11.3 MMBtu per megawatt-hour) (\emph{129}). The 2024 fuel-weighted heat rate of natural gas-fired combustion turbines in EIA-923 data, which includes partial-load operation, is 11.18 MMBtu per megawatt-hour. Operating hours in April, May, and July were extrapolated by applying the combined June operating fraction of each model to each unit's time on site in that month. The central scenario used a load fraction of 0.8, and a range of 0.6 to 1.0 tested the effect of this assumption. Generation and delivered electricity were accounted for separately, with delivered electricity equal to generation less 3\% losses, slightly above the full-load auxiliary loads of 1.8 to 2.3\% for the aeroderivative simple-cycle reference plants of the National Energy Technology Laboratory (\emph{130}). Under these settings, mean NO\emph{\textsubscript{x}} emissions from April to July were about 216 kg per hour (as NO\textsubscript{2}), or 162 to 269 kg per hour for load fractions of 0.6 to 1.0 (tables S9 and S10).

We checked the engineering inventory independently against stack tests and gas scheduling records. Emission intensity in a stack test is the NO\emph{\textsubscript{x}} mass emission rate divided by electric output. For the two TM2500 units at the Greenleaf 1 plant in California, we used the mean of each unit's three test runs (tables 5-1 and 5-4 of the test report) (\emph{52}). For the seven units with SCR at the Abilene data center campus in Texas (\emph{61}), we summed the mass emission rates of 21 test runs and divided the sum by the sum of the corresponding outputs (fig. S3A). The fuel-based estimate used daily scheduled quantities at the site's Texas Gas Transmission delivery point from 13 July to 30 September 2026 as a proxy for activity. We converted scheduled quantities to fuel heat input on a higher heating value basis (1 dekatherm equals 1 MMBtu), allocated heat input according to the capacity and heat rate of the units on site that had June operating records, applied group-level emission factors to estimate daily NO\emph{\textsubscript{x}} emissions, and divided each day's emissions by 24 to obtain mean hourly emission rates for comparison with the engineering inventory (supplementary text). For units without SCR, the factors were the Greenleaf 1 stack-test intensities of 0.355 and 0.384 kg per generated megawatt-hour combined with heat rates of 9, 11, or 13 MMBtu per megawatt-hour, and for units with SCR, the factor corresponded to 2 parts per million (ppm, dry at 15\% oxygen).

\subsubsection*{Health damage estimation}

We used two reduced-complexity air quality models, the Intervention Model for Air Pollution (InMAP) (\emph{55}) and AP4 (\emph{54}), to convert the increase in particulate nitrate concentrations caused by NO\emph{\textsubscript{x}} emissions into health damages. Particulate nitrate is a component of fine particulate matter (PM\textsubscript{2.5}) formed from NO\emph{\textsubscript{x}} through atmospheric chemistry. Both models resolve damages by receptor location, allowing us to account for impacts borne by each state. Combining the annual mean concentration increase, population, and baseline mortality in each receptor cell, we calculated annual premature deaths as

\begin{equation*}
\Delta M = \sum_{i}^{}{\sum_{c}^{}P_{ic}}\, y_{c}\,\beta_{\mathrm{CR}}\,\Delta C_{i}\tag{S6}
\end{equation*}

where \(\Delta M\) is annual premature deaths, \(i\) is the receptor cell, and \(c\) is the county, with calculations performed on the intersections of cells and counties. \(P_{ic}\) is the population aged 30 and older in each intersection, \(y_{c}\) is the annual all-cause mortality rate for the same age group in the county, \(\Delta C_{i}\) is the annual mean increase in nitrate concentration in cell \(i\) (μg m\textsuperscript{\ensuremath{-}3}), and \(\beta_{\mathrm{CR}} = \ln(1.06)/10\) is the concentration--response coefficient, which corresponds to a relative risk of 1.06 per 10 μg m\textsuperscript{\ensuremath{-}3} (\emph{58}). We used this coefficient for the main estimate because it is the base case in an earlier application of the InMAP source--receptor matrix (\emph{131}), which keeps the estimates comparable with that work. Because the concentration increases are small, the log-linear concentration--response function is used in its linear form.

The concentration increase \(\Delta C_{i}\) depends on emissions and on their atmospheric transport and transformation. The main analysis used the InMAP source--receptor matrix (ISRM 1.2.1) (\emph{131}, \emph{132}), multiplied the source--receptor coefficients by the weights that allocate emissions to each release layer and by annual emissions, and summed the products to obtain the annual mean concentration increase in each receptor cell. We read the matrix rows directly from the original netCDF release and interpolated between layers with our own code, so neither defect reported in September 2026 (reversed interlayer weights in the srpredict tool and mislabeled layers in the version 1.2.1 Zarr release) affects these results (\emph{133}). To represent plume rise from the hot turbine exhaust, we allocated emissions to the matrix's three release layers, with midpoints of 29, 315, and 901 m, based on each turbine family's hourly effective release height (FT8 MOBILEPAC, SMT-130, and all other units). When the effective release height fell within the bounds of a layer, all emissions were assigned to that layer. When it fell in a gap between layers, emissions were split by linear interpolation between the midpoints of the adjacent layers, and when it lay outside the height range of the matrix, emissions were assigned to the nearest end layer.

The effective release height is the stack height plus plume rise. Plume rise was calculated with the Briggs buoyancy formulas as the smaller of the final rise under neutral conditions and the rise under stable conditions (\emph{134}). Inputs were the stack height, exhaust temperature, and mass flow rate of each turbine family, together with 17,544 hours of ERA5 meteorology from 2024 and 2025 (\emph{99}, \emph{135}). Atmospheric stability was estimated from the potential temperature gradient between 2 m and 850 hPa. Each unit was treated as a separate plume. The buoyancy flux was computed from the exhaust mass flow rate, the difference between the exhaust temperature and the 2-m air temperature, and surface pressure, and the wind speed at stack height was interpolated linearly in height between the ERA5 10-m and 100-m winds, with the 10-m wind used below 10 m and a minimum of 0.5 m s\textsuperscript{\ensuremath{-}1}. Hours without a positive potential temperature gradient used the neutral final rise. The stack parameters are listed in the note to table S10C. The resulting hourly layer weights were averaged over time for each turbine family and then combined with weights equal to each family's share of emissions to compute the annual mean concentration increase (table S10C). We refer to this setting, the hourly effective-release-height mixture, as the InMAP main setting.

\(P_{ic}\) is the 2020 total population of each cell--county intersection multiplied by the share of the county's population aged 30 and older in the DHC data, and \(y_{c}\) is the 2024 county all-cause mortality rate for the same age group from CDC WONDER (D158) (\emph{136}), with missing counties filled with the state rate. Two other concentration--response relationships served as sensitivity cases. The first is the function that Pope et al. (\emph{57}) estimated from a cohort of US adults (a relative risk of 1.12 per 10 μg m\textsuperscript{\ensuremath{-}3}), applied, as in the main estimate, to the population and mortality of people aged 30 and older, with only the coefficient replaced. The second is the function that Wu et al. (\emph{137}) estimated for Medicare beneficiaries aged 65 and older (a relative risk of 1.066 per 10 μg m\textsuperscript{\ensuremath{-}3} from their Cox model), applied to the population and mortality of that age group (table S10).

We also set up an InMAP ground-level release case and used AP4 ground-level release results for comparison. AP4 uses its own 2017 baseline of National Emissions Inventory emissions, population, and mortality. Because AP4 does not include this fleet as an elevated source, we used the AP4 coefficients of the nearby TVA Southaven combined-cycle plant as an elevated-release proxy (table S10A). EASIUR (Estimating Air pollution Social Impact Using Regression) (\emph{138}) results serve only as a sensitivity comparison (tables S10 to S13) and are not used for state-level accounting because the available product provides only damages per metric ton of emissions and cannot attribute damages to receptor locations. In addition, EASIUR source cells are 36 km wide, whereas ISRM cells in urban areas are about 1 km wide.

After obtaining deaths in each receptor cell, we summed them by state and counted the states that account for more than 1\% of total deaths to describe the cross-state distribution of damages. For the distance distribution, we accumulated deaths from the nearest to the farthest receptor cell by geodesic distance from the site to each cell center and report the distance at which the cumulative share first reaches 50\%. To express damages per unit of capacity without SCR, we assumed that damages are proportional to emissions, multiplied fleet-wide damages by the NO\emph{\textsubscript{x}} emission share of units without SCR, and divided by the calendar-day-weighted mean on-site capacity of those units from April to July, 1.252 GW (table S10).

We converted changes in mortality risk into health damages, or into benefits of emission reductions for the economic analysis, valuing each death at \$11.3 million (2019 dollars), the EPA value for 2025 income levels (\emph{139}), without further adjustment for income growth. Annual damages are reported without discounting. In the benefit--cost analysis, health benefits were distributed over time and discounted with the mortality lag structure used by the EPA (\emph{139}), described by \(L(r)\) in the next section.

To compare health damages from on-site generation and from grid power, we used the AVoided Emissions and geneRation Tool (AVERT) (\emph{60}, \emph{140}) with its 2023 Tennessee regional profile to simulate changes in emissions at each power plant when load falls by 1 MW or 10 MW in every hour of the year. We multiplied the resulting reductions in each plant's NO\emph{\textsubscript{x}}, sulfur dioxide, and primary particulate matter emissions by the damage coefficients for its location in each health model and divided by the generation displaced in the AVERT simulation to obtain health damages per unit of electricity (table S11). InMAP used its highest release layer, and EASIUR used its 300-m release-height category. AP4 matched damage coefficients by plant identifier and used county coefficients when no match was available. Results are reported as ranges across the 1-MW and 10-MW cases. On-site primary particulate matter and sulfur dioxide emissions were estimated both with AP-42 emission factors and with emission factors borrowed from the permit documents of similar models, under two fuel sulfur scenarios.

We also built emission-control and operating scenarios on the same engineering inventory so the measures are comparable. In the SCR retrofit scenario, NO\emph{\textsubscript{x}} emissions from units without SCR fall by 90\% and other units are unchanged. In the uniform limit scenario, the fuel input of each turbine model is held constant, and emissions are recalculated with the fuel-based emission factor that corresponds to the 2-ppm limit (dry at 15\% oxygen) in the PSD permit for the permanent plant (\emph{51}). Deaths avoided in both scenarios were computed with the damages per unit of emissions from the same air quality model (Fig. 4A and table S12). In the operating scenarios, delivered electricity was held at its June 2026 level and units were dispatched in order of increasing NO\emph{\textsubscript{x}} emission intensity per unit of generation at load fractions of 0.6, 0.8, and 1.0, to assess how changing the use of units affects damages when the supply of electricity is unchanged (table S12).

To examine the in-state share of damages at other sites, we transferred this fleet's turbine-family emission weights and hourly release-height mixture to the source cells of five sites of four other on-site generation projects with public source coordinates, computed the share of InMAP damages borne by the host state, and compared these shares with AP4 ground-level release results (table S10D). Source locations came from the coordinates in the Abilene permit application (\emph{62}) and the Fermi America Project Matador air permit (\emph{141}), the centroids of the project boundaries published for the two Socrates sites (\emph{142}), and the facility location of Colossus 1 (Data S1).

\subsection*{Benefit--cost analysis of SCR}

We assessed the economic merit of SCR over different evaluation horizons and geographic scopes of benefits by comparing the present values of health benefits and costs. Benefits are the health improvements from NO\emph{\textsubscript{x}} reductions net of damages from ammonia slip, and costs include the initial investment and subsequent operation and maintenance. Accounting for the lagged realization of health benefits and the discounting of future amounts, the benefit--cost ratio is

\begin{equation*}
\mathrm{BCR}(H) = \frac{(B_{N} - sB_{A})\, L(r)\, A(r,H)}{kK + O\, A(r,H)}\tag{S7}
\end{equation*}

where \(\mathrm{BCR}(H)\) is the benefit--cost ratio for an evaluation horizon of \(H\) years and \(r\) is the annual discount rate. \(B_{N}\) is the annual health damage avoided by the NO\emph{\textsubscript{x}} reductions of the evaluated units, \(B_{A}\) is the annual health damage per 1 ppm of ammonia slip, and \(s\) is the ammonia-slip concentration in ppm, dry at 15\% oxygen. \(B_{N} - sB_{A}\) is therefore the annual net health benefit after ammonia damages. In the denominator, \(K\) is the initial investment for fitting SCR to the evaluated units as new turbines, \(k\) is an investment adjustment factor equal to 1 for new turbines and 1.25 for retrofits of existing units, and \(O\) is the annual operation and maintenance cost. The initial investment \(kK\) is paid once in year 0, and the present value of operation and maintenance costs is \(O\, A(r,H)\). The payback period is the solution of \(\mathrm{BCR}(H) = 1\), which the main text rounds up to whole years.

Here, \(A(r,H) = \lbrack 1 - (1 + r)^{- H}\rbrack/r\) is the annuity factor, which converts equal annual amounts over the horizon to present value, and \(L(r) = \sum_{t = 0}^{19}\omega_{t}(1 + r)^{- t}\) is the lag discount factor for health benefits. \(L(r)\) follows the mortality lag structure used by the EPA (\emph{139}) and discounts all health benefits from one year of emission reductions to that year. At discount rates of 7\% and 3\%, \emph{L}(\emph{r}) is 0.816 and 0.906 (table S13A). The weight \(\omega_{t}\) is the share of benefits realized in year \(t\), 30\% in the year of the reduction (\(t\) = 0), 12.5\% in each of years 1 to 4, and the remaining 20\% spread evenly over years 5 to 19. Benefits realized after the end of the horizon are therefore also counted. The analysis used discount rates of 7\% and 3\% from Circular A-4 (\emph{143}), the version reinstated in 2025 (\emph{144}), and counted health benefits both in all states and in Mississippi only.

In this benefit--cost analysis, the evaluated units are the 29 units without SCR that reported operating hours in June 2026 and had not retired by September 2026, with a combined capacity of 891.44 MW. SCR was assumed to remove 90\% of NO\emph{\textsubscript{x}} (\emph{65}), resulting in an annual reduction of 915.5 metric tons, and annual avoided damages were computed from the site's damages per unit of emissions and receptor distribution from April to July 2026. The evaluation horizons were the remaining period on site, 1 year, 2 years, 20 years, and 30 years, with 30 years as the main horizon, matching the 30-year useful life that the EPA used in its economic impact analysis (\emph{70}). We computed each unit's remaining period on site by dividing the number of days from 2 October 2026 to the deadline by 365.25. SCR on new turbines and retrofits of existing units were evaluated separately, with retrofit investment multiplied by 1.25 as specified in the EPA technical support document (\(k\) in Eq. S7) (\emph{69}). Operation and maintenance costs, including taxes and insurance, were kept at their values for new turbines.

The cost parameters in Eq. S7 follow the technical support document and workbooks that the EPA prepared for its 2026 turbine standards (docket EPA-HQ-OAR-2024-0419-0242) (\emph{69}). The initial investment is based on the National Energy Technology Laboratory's aeroderivative simple-cycle model plant (\emph{130}), as scaled by the EPA to a single LM6000 turbine (\emph{69}). We scaled investment other than continuous emissions monitoring systems (CEMS) by the ratio of design heat inputs raised to the power 0.6, added one NO\emph{\textsubscript{x}} CEMS per unit, and assumed a hot-exhaust SCR configuration that requires tempering air. Fixed costs include maintenance, taxes and insurance, CEMS maintenance, and lost capacity payments. Variable costs include administration, ammonia, catalyst, catalyst disposal, auxiliary power, and back-pressure losses. The capital estimate above gives the initial investment \(K\), and these fixed and variable costs give the annual operation and maintenance cost \(O\).

We converted catalyst prices and auxiliary power prices using the EPA workbook, and computed each unit's capacity factor from its annual generation in the engineering inventory. Property taxes are transfers rather than social costs (\emph{143}), but the workbook combines them with insurance, so we report results that either keep both items in full or remove both. All amounts were converted to 2019 dollars with the ratio of the 2019 and 2024 GDP price deflators, 103.979/125.23 (\emph{145}), taken from the 26 June 2025 vintage of National Income and Product Accounts Table 1.1.9, consistent with the EPA workbook.

Benefit accounting also requires health damages from ammonia slip. We set ammonia-slip concentrations of 0, 1.4, 2.8, and 3.5 ppm, all dry at 15\% oxygen, with 3.5 ppm as a case above both permit limits. The values of 1.4 and 2.8 ppm correspond to the limits of 5 ppm for the PGM-130 and Titan 350 turbines and 10 ppm for the 6000PE turbines (dry at 0\% oxygen) in the permit for the permanent power plant at the same site (\emph{51}). To use these scenarios in the ammonia damage term of Eq. S7, we first converted them to annual ammonia emissions with the annual fuel heat input

\begin{equation*}
E_{\mathrm{NH}_{3}} = 10^{- 6}\, s\,\frac{20.9}{20.9 - 15}\, F_{\mathrm{d}}\, F\,\rho_{\mathrm{NH}_{3}}\tag{S8}
\end{equation*}

where \(E_{\mathrm{NH}_{3}}\) is annual ammonia emissions, converted to metric tons after calculation, \(s\) is the ammonia-slip concentration (ppm, dry at 15\% oxygen), and 20.9/(20.9 − 15) converts the concentration from the 15\% to the 0\% oxygen basis. \(F_{\mathrm{d}}\) is the dry flue gas volume factor (F factor) for natural gas, 8,710 dry standard cubic feet per MMBtu (\emph{146}), \(F\) is the annual fuel heat input on a higher heating value basis (MMBtu), and \(\rho_{\mathrm{NH}_{3}}\) is the density of ammonia at standard conditions, its molar mass of 17 lb per lb-mol divided by the molar volume of 385.3 standard cubic feet per lb-mol. Permit limits dry at 0\% oxygen are converted to the 15\% oxygen basis by multiplying by (20.9 − 15)/20.9.

The ammonia damage per 1 ppm of slip was then obtained by converting the ammonia emitted at 1 ppm to health damages with the same air quality model and receptor distribution as the NO\emph{\textsubscript{x}} benefits. We also added ozone benefits as a supplement to the particulate matter benefits, using the benefit-per-ton approach of the EPA regulatory impact analysis for its 2024 proposed turbine standards (\emph{71}, \emph{139}) with the EPA 2024 benefit-per-ton coefficient for ozone from industrial boilers in the East South Central region for the 2025 emission year. This coefficient uses short-term mortality functions and a 7\% discount rate and includes the morbidity benefits counted by the EPA. Because reductions were assumed to be spread evenly over the year, 153/365 of them were assigned to the May to September ozone season. Because the coefficient already includes discounting for the corresponding health endpoints, it was not multiplied by the mortality lag factor of the particulate matter analysis (table S13).

\subsection*{Computation and data availability}

Satellite processing, reference weight estimation, and bootstrapping were done in Python 3.12.10, mainly with NumPy 2.4.6, pandas 2.3.3, SciPy 1.18.0, GeoPandas 1.1.4, Shapely 2.1.2, pyproj 3.7.2, and Rasterio 1.5.1. Reference weights were solved with a damped Newton dual method, with BFGS as a fallback, at a convergence tolerance of 10\textsuperscript{\ensuremath{-}5}. The health models used the InMAP source--receptor matrix ISRM 1.2.1 and a locally archived AP4 model workspace.

All data used in this study are public. They include NASA TEMPO (Atmospheric Science Data Center), TROPOMI (NASA Goddard Earth Sciences Data and Information Services Center and Google Earth Engine), ERA5 (Copernicus Climate Change Service, accessed through Google Earth Engine), HRRR (NOAA open data on Amazon Web Services), NOAA HMS, 2020 Census and ACS data from the U.S. Census Bureau, the EPA Air Quality System, Clean Air Markets Program Data and AVERT, EIA-860, EIA-860M and EIA-923, NLCD 2021, Microsoft building footprints (usbuildings-v2), TIGER/Line roads, state traffic counts, Annual NLCD, LODES employment, CDC WONDER mortality, NIFC fire perimeters, OpenStreetMap, GeoNames place names, Texas Gas Transmission informational postings, the InMAP source--receptor matrix, the AP4 and EASIUR models, and the regulatory documents and company filings described above. Data S1 contains the data underlying the figures, the paired TEMPO scans with their processor versions, coverage, and inclusion in the main estimate, unit-level operating records and emissions, daily scheduled gas deliveries, and a statistics sheet giving estimates with 95\% intervals, sample sizes, and \emph{P} values. Processed data and the code that reproduces the reported statistics are archived at Zenodo (\emph{89}--\emph{91}), and a later version of the code record will add the complete processing code.

\section*{Supplementary Text}

\subsection*{Detectability early in the deployment}

In the first eight months after deployment, on-site capacity stayed below 700 MW and the central emission rate in the engineering inventory, with the June 2026 operating fraction of each turbine model applied to the units on site in each month, was 19 to 120 kg per hour, near or below the detection threshold of about 101 kg per hour (Eq. S3), consistent with the small increase in that period (Fig. 2A).

In the retrospective cumulative analysis (materials and methods), the 95\% confidence interval first excluded zero when May observations were included, so the increase became detectable in the second month of the fleet's spring expansion. To test whether the procedure behaves the same way in other cities, we repeated this cumulative analysis for the 15 same-scan placebo cities with the fixed reference ring. At both the end of May and the end of September, only one city had a cumulative confidence interval that excluded zero, and its change was negative. In the seasonal estimates, Southaven ranked first among the 16 locations in both seasons, and its late-summer increase was 8.3 times the largest placebo change (table S2C). With paired scans of processor version 4.8.4 only, it again ranked first in both seasons (table S2D). In the late-summer seasonal estimates, the intervals of three placebo cities excluded zero, but their changes were at most 0.38, less than one-sixth of the lower confidence bound at Southaven with the same reference ring (2.65).

\subsection*{Comparison sites and nearby power plants}

Other sites provide comparative evidence on emission control and construction activity (table S6). A 2025 permit modification application for the Abilene campus lists 10 turbines, each fitted with SCR (\emph{62}), and the change there was 0.19 (−0.05 to 0.51), consistent with the low emissions of units with SCR. The Richland Parish campus was under construction from April to July 2025 (\emph{147}, \emph{148}), when the estimated change was 0.05 (−0.08 to 0.18), and the estimate for the same months of 2026 was 0.04 (−0.10 to 0.19).

Under the April to June transport conditions of the puff simulation (materials and methods), with the 2024 and 2025 emissions held at their reported values, 2026 emissions from TVA Southaven would have to be about 65 times their reported level, or 2026 emissions from both plants about 75 times, for the simulated change in near-field excess to reach the April to July main estimate of 1.10. With NO\emph{\textsubscript{x}} lifetimes of 2 to 8 hours or final NO\textsubscript{2}/NO\emph{\textsubscript{x}} ratios of 0.5 to 0.85, the required factors are 55 to 86 for TVA Southaven alone and 63 to 99 for both plants.

At Colossus 1, the September to November contrast between 2024 and 2023 was 0.25 (0.00 to 0.46, table S6). The site lies about 1.7 km from the TVA Allen plant, so this contrast includes changes at Allen and was not used for attribution.

\subsection*{Features of the late-summer increase}

In the TEMPO source-agnostic peak search, Southaven ranked first for August and September combined and for September alone, and second for August alone. The top peak in August lay near Plaskett, California, and overlapped official fire perimeters (\emph{40}) and HMS smoke polygons in the same period. Adding a cloud-fraction limit of 0.3 to the quality screening of the TROPOMI Level-3 product changed the late-summer increase at Southaven from 2.17 to 2.16 (1.21 to 3.21, Data S1).

Winds were weaker in late summer, and mean on-site capacity was higher (table S1). Under single-hour HRRR 10-m winds below 3 m s\textsuperscript{\ensuremath{-}1}, the increase was 2.23 (95\% confidence interval, 1.11 to 3.46) in spring and 3.45 (2.72 to 4.19) in late summer, compared with 0.45 (0.18 to 0.72) and 1.35 (0.35 to 2.34) at winds of at least 3 m s\textsuperscript{\ensuremath{-}1} (fig. S2C). In both seasons, the increase under weak winds exceeded that under stronger winds, by 1.78 (0.75 to 2.98) in spring and 2.10 (0.76 to 3.34) in late summer. Weighting both seasons by the spring distribution of wind speeds (36.7\% of observation opportunities below 3 m s\textsuperscript{\ensuremath{-}1}) gives 2.12 in late summer and 1.10 in spring, a difference of 1.02 (0.18 to 1.85), and the late-summer distribution gives a difference of 1.10 (0.18 to 2.09). Weaker winds thus account for about one-third of the seasonal difference, 0.57 (0.21 to 0.91), and about two-thirds remains after the wind distribution is held fixed. With all daytime hours, the two standardized seasonal differences were 1.16 (0.62 to 1.67) and 1.01 (0.43 to 1.59).

The association with capacity also held in alternative specifications. In the monthly capacity regression, the increase per 100 MW of capacity without SCR remained positive when August and September 2026 were excluded (0.08, HC1 interval 0.03 to 0.13, monthly sampling interval 0.02 to 0.16). The monthly sampling intervals in fig. S2D give 0.15 (0.08 to 0.23) with the monthly change in inverse wind speed held fixed and 0.22 (0.10 to 0.25), 0.16 (0.11 to 0.24), and 0.15 (0.09 to 0.26) when the weak-wind share, boundary-layer height, or both wind and boundary-layer terms were held fixed instead. The monthly change multiplied by the monthly mean wind speed, a flux proxy that adjusts for dilution by wind, also rose with capacity, by 0.44 (monthly sampling interval, 0.30 to 0.58) per 100 MW in units of 10\textsuperscript{15} molecules cm\textsuperscript{\ensuremath{-}2} m s\textsuperscript{\ensuremath{-}1}.

Two further diagnostics point to the fleet in late summer. With each year rotated by its own wind directions, the late-summer contrast between the downwind and upwind boxes was 3.58 (2.94 to 4.29, table S2B). On days with both morning and afternoon observations, the late-summer afternoon increase exceeded the morning increase by 1.78 (0.87 to 2.74, 41 days), whereas the spring difference was 0.31 (−0.25 to 1.03, 54 days, fig. S2F).

\subsection*{Independent checks on the emission inventory}

The fuel-based estimate of NO\emph{\textsubscript{x}} emissions from mid-July to September is 181 to 309 kg per hour, consistent with the inventory. Extrapolating the inventory to September with the June operating fraction of each turbine model gives mean emissions of 206 to 212 kg per hour for April to September, and replacing emissions after mid-July with the fuel-based estimate, which reflects scheduled gas deliveries, gives 204 to 250 kg per hour for the same period (table S9). The consultant's sheet lists heat rates of 8.35 MMBtu per megawatt-hour for the TM2500 Gen4, Gen6, and Gen8 and 9.376 for the FT8 MOBILEPAC without stating their heating-value basis (\emph{127}), whereas the emission factors are on a higher heating value basis. If these heat rates are on a lower heating value basis and are converted with the ratio of 1.108 for natural gas (\emph{149}), mean emissions for April to July rise from 216 to 228 kg per hour, and applying the Greenleaf 1 mean of 10.2 MMBtu per megawatt-hour (\emph{52}) to all TM2500 units gives 233 kg per hour. Under these two cases, respectively, the TM2500 Gen4, Gen6, and Gen8 rate becomes 0.39 and 0.43 kg per megawatt-hour, premature deaths rise to 4.6 and 4.7 per year, and damages rise to \$52 million and \$53 million. The share of deaths outside Mississippi (81.8\%) and the share avoided by SCR on the units without it (88\%) are unchanged. The 30-year benefit--cost ratio for SCR on new turbines (7\% discount rate, no ammonia slip) becomes 1.51 to 1.65 and 1.55 to 1.70, compared with 1.49 to 1.62 in the main estimate.

\subsection*{Sensitivity of the health damage estimates}

We tested the sensitivity of the damage estimates to release height, health damage model, concentration--response function, and the period of the emission estimate (table S10 and fig. S3B). With ground-level release, InMAP estimates 5.44 premature deaths per year, 75.7\% of them outside Mississippi. The AP4 ground-level release case estimates 6.0 premature deaths per year, about 97\% of them out of state. The EASIUR area-source case estimates 1.51 premature deaths per year, and the product does not resolve results by state (materials and methods).

With the coefficient of Wu et al. (\emph{137}) for people aged 65 and older, estimated premature deaths are 3.36 per year, and using emissions for April to September instead of April to July gives 4.13 to 5.05 per year (table S10B).

In the InMAP main setting, 16 states each bear more than 1\% of total deaths, and 73.8\% of deaths occur within 500 km of the site. With ground-level release at the other sites of table S10D, the in-state share estimated by InMAP is at most about 58\%.

\subsection*{Damages per unit of electricity from on-site generation and the grid}

Counting NO\emph{\textsubscript{x}} only, on-site generation causes \$9.95 in health damages per megawatt-hour delivered, compared with \$2.75 to \$2.76 for grid power, about 3.6 times as much (table S11A).

When primary PM\textsubscript{2.5} and sulfur dioxide are also counted, on-site generation causes \$15.1 to \$16.5 in damages per megawatt-hour, a ratio of 0.95 to 1.04 relative to grid power (table S11B and Data S1). Coal-fired units account for about 28\% of the generation displaced in the AVERT simulation but cause about 99\% of the grid's sulfur dioxide damages and about 76\% of its primary PM\textsubscript{2.5} damages, so including these two pollutants raises the grid's damages per unit of electricity substantially.

We then assessed the effect of SCR. If SCR reduces NO\emph{\textsubscript{x}} emissions from units without SCR by 90\%, the ratio of on-site to grid health damages falls to 0.30 to 0.75 in all 32 combinations of models and scenarios. When ammonia slip at the permanent plant's permit limits of 1.4 and 2.8 ppm is also counted, the ratio is 0.34 to 0.96 and remains below 1 in every case (table S11).

\subsection*{SCR on mobile turbines}

The 14 units with SCR in the fleet are all mobile SMT-130 turbine packages, also fitted with oxidation catalysts, and they are covered by the same agreed order as the other units. Seven of the first eight units, which arrived between 1 and 21 August 2025, were SMT-130 units with SCR (fig. S4A and Data S1). The agreed order does not require mobile turbines to use such controls but requires performance tests on all 14 units and the submission of temperature and pressure monitoring data in biweekly reports (\emph{16}). SCR has also been fitted to temporary TM2500 units elsewhere. California Energy Commission staff recommended licensing the two Greenleaf 1 units used in the stack-test comparison above to operate temporarily in 2021 under an emergency proclamation, noting that the units would be retrofitted with SCR and oxidation catalysts as soon as practicable (\emph{150}). A closure plan from January 2025 states that both units were by then fitted with SCR (\emph{151}).

\subsection*{Rules and deadlines for temporary authorization}

Two temporary authorization pathways are relevant here, namely the state mobile-source pathway and the federal temporary category. Both define their scope by how long equipment stays at a single site, but they serve different regulatory roles. The former determines whether a stationary-source permit is required, and the latter determines the applicable emission limits. This fleet operates under the state mobile-source pathway. Mississippi regulations (11 Miss. Admin. Code Pt. 2, Ch. 2, R. 2.13.D) (\emph{152}) exempt mobile sources from stationary-source construction and operating permits. On this basis, MDEQ determined in its 29 July 2025 letter that turbines mounted on movable chassis and planned to remain for less than 12 months were exempt from permitting as mobile sources (\emph{19}). The July 2026 agreed order then set an operating deadline for each unit (\emph{16}). Except for 13 SMT-130 units with SCR, all units had to stop operating within 12 months of arrival. The agreed order also prohibits adding mobile turbines after 15 July 2026. The September 2026 amended agreed order extended to May 2027 the deadlines for two units that arrived in August 2025, set an early end date of 27 August 2026 for two units that arrived in May 2026 and stopped operating on that date, and specified that this adjustment may not increase net emissions or the number of operating units (\emph{17}).

The draft DeSoto County ozone maintenance plan of November 2025 and its public emission inventory workbooks list only five existing facilities in the point source projections, derive the 2026 emission projections by interpolating between the 2017 and 2036 inventories, and do not include the fleet (\emph{83}).

The 2026 new source performance standards for turbines (40 CFR 60 Subpart KKKKa) (\emph{20}, \emph{63}) establish a temporary category for stationary turbines. It applies to turbines that are planned to remain, and do remain, for no more than 24 consecutive months within the same stationary source or within a group of contiguous stationary sources under common control. For new natural gas-fired turbines in this category with a base-load rated heat input of no more than 850 MMBtu per hour, the NO\emph{\textsubscript{x}} limit is 25 ppm (dry at 15\% oxygen). The EPA designated combustion controls as the best system of emission reduction for this category for two reasons. First, it cited the argument of commenters that the complexity of SCR would defeat the purpose of bringing in turbines quickly for short-term power. Second, the EPA's general judgment for small and medium turbines is that SCR becomes less cost-effective as unit size and annual utilization decline, and the EPA expects medium temporary turbines to have annual utilization below 45\% because of downtime for storage and transport (\emph{20}). To keep the clock from being reset by replacing or moving units, the standard also requires that time on site be counted cumulatively when a unit is replaced by one that serves the same or a similar function (§ 60.4331a(e)(2)), and that the clock not be reset when a unit moves within these bounds while still serving the same electric, mechanical, or thermal load (§ 60.4331a(e)(3)). The standards also exclude from the definition of a stationary combustion turbine any portable turbine certified to nonroad emission standards under title II of the Clean Air Act, a provision that becomes operative only if the EPA adopts such standards. The EPA also noted commenters' concerns about the air quality effects of many temporary turbines brought together and operated at a single location (\emph{20}).

Of the 68 delivered units, most have deadlines about one year after arrival. All 14 SMT-130 units have a deadline of 31 December 2026, up to about five months beyond the first anniversary of arrival for 13 of them. Based on the month-end roster of authorized units, the capacity-weighted NO\emph{\textsubscript{x}} emission rate of the units still authorized in January 2027 is 0.439 kg per megawatt-hour, about 16\% higher than the 0.378 kg per megawatt-hour in December 2026. Under the amended agreed order, the deadlines for most units without SCR fall between March and July 2027. From the arrival of the first units to the latest deadline, the span is about 23 months (fig. S4A).

Commercial contracts and equipment accounting cover much longer periods than the site authorization deadlines. The equipment lessor's two most recent contracts with other data centers have 10-year base terms with extension options of five and eight years, and the eight-year option (Hatchbo) was extended in July 2026 from five years (\emph{21}, \emph{66}). The operator also holds natural gas transportation rights through 2035, and a permanent power plant of 41 units with a total capacity of about 1,240 MW is under construction at the same site (\emph{16}, \emph{43}, \emph{51}). The lessor further discloses that the equipment can be redeployed after a contract ends, and its quarterly report uses a 25-year depreciation period for the turbines and depreciates turbine engine cores separately over an expected 30,000 fired hours (fig. S4B) (\emph{21}, \emph{66}). The lessor's annual report for fiscal year 2025 also discloses that Stateline Power, its joint venture with the site operator MZX Tech LLC, is expected to own about 900 MW of generating equipment and that the lessor holds exclusive rights, subject to certain conditions, to manage and redeploy Stateline's assets (\emph{66}).

Equipment has also been redeployed to other sites after earlier contracts ended. In 2017, the government of South Australia built an emergency power plant with four TM2500 units at Lonsdale and five at Elizabeth South (\emph{74}, \emph{75}). When the units were purchased, the plan was to regroup them, after their backup role over the summers of 2017 and 2018, into a gas-fired plant for the remainder of a forecast 25-year service life (\emph{74}). The four Lonsdale units, leased in 2019 for 25 years, stopped operating at the end of April 2022 and moved to Bolivar. The regulator approved the change of generation site in August 2022, and commissioning on fuel oil was completed in December of that year (\emph{73}, \emph{75}). The other five units, leased from the state government in August 2019 for 25 years, moved from Elizabeth South to the Snapper Point power station at Outer Harbor and began commercial operation there on 10 June 2022 (\emph{72}, \emph{75}).

Temporary units can also remain in place for long-term operation. In September 2021, under the governor's emergency proclamation, the California Energy Commission certified two TM2500 Gen4 units to operate temporarily at the Roseville Energy Park, with the certification originally set to end on 8 September 2026. In April 2026, the commission merged the units' temporary license into the plant's regular license (\emph{76}). The merged license requires the two units to operate SCR and carbon monoxide oxidation catalysts and, during normal operation (excluding startup and shutdown), to emit NO\emph{\textsubscript{x}} at no more than 2.5 ppm (dry at 15\% oxygen, 1-hour average). The operating permit issued by the local air district in October 2025 also limits each unit to 2,200 hours of operation per year (\emph{77}). With SCR in place, stack tests at normal maximum load in July 2022 measured NO\emph{\textsubscript{x}} concentrations of 1.82 and 2.11 ppm (dry at 15\% oxygen) (\emph{153}).

\subsection*{A framework for evaluating SCR over the period on site and the service life}

To distinguish the authorization period at a single site from the service life of the equipment, we built an accounting framework (fig. S4D) that compares the conclusions reached when emission controls are evaluated over the period on site and over the service life. If SCR moves with the generating equipment and continues in use after relocation, its one-time investment can be spread over generation at different sites. If the investment must be recovered within a given period, the cost per megawatt-hour of delivered electricity is

\begin{equation*}
C(H) = \frac{K}{Q(H)} + o\tag{S9}
\end{equation*}

where \(H\) is the cost recovery period, \(C(H)\) is the cost per megawatt-hour of delivered electricity over that period, \(K\) is the one-time investment in SCR, \(Q(H)\) is the discounted value of the electricity delivered over \(H\), and \(o\) is the operating cost per unit of electricity. \(K/Q(H)\) is therefore the investment cost allocated to each unit of electricity, and adding \(o\) gives the total cost per unit of electricity. Let the stay of the equipment at the current site be \(\tau\) years and its service life \(\Lambda\) years, with \(\Lambda \geq \tau\). If the investment is recovered within the current period on site only, the unit cost is \(C(\tau)\). If it is spread over the service life of the equipment, the unit cost is \(C(\Lambda)\).

Within this cost framework, we compared two scopes of benefit accounting. Let \(\delta\) be the health damage avoided per megawatt-hour delivered with SCR and \(\theta\) the share of benefits that accrues to the authorizing state. The social criterion uses the service life of the equipment as the evaluation horizon and counts benefits in all affected states, so SCR passes this criterion when \(\delta \geq C(\Lambda)\). The site-level criterion counts only benefits during the current period on site, and when it also counts only benefits within the authorizing state, it requires \(\theta\delta \geq C(\tau)\). Rewriting this condition as \(\delta \geq C(\tau)/\theta\) shows that the avoided damage per unit of electricity that SCR must reach rises from \(C(\Lambda)\) to \(C(\tau)/\theta\). The ratio of these two break-even levels is

\begin{equation*}
\frac{C(\tau)/\theta}{C(\Lambda)} = \frac{C(\tau)}{C(\Lambda)} \times \frac{1}{\theta}\tag{S10}
\end{equation*}

In Eq. S10, the first factor, \(C(\tau)/C(\Lambda)\), reflects the effect of shortening the cost recovery period, and the second factor, \(1/\theta\), reflects the effect of counting only benefits in the authorizing state. The main text states the same comparisons in total rather than per-megawatt-hour terms, calling the discounted damages that SCR avoids over the period on site the site threshold and those over the service life the social threshold. When \(C(\Lambda) \leq \delta < C(\tau)/\theta\), the same SCR system passes the social criterion but fails the site-level criterion. Both judgments concern the same equipment and the same technology and differ only in the evaluation horizon and the geographic scope of benefits. Because benefits after a move accrue near the next site, possibly in another state, a horizon limited to the current period on site also omits out-of-state benefits.

This framework rests on two simplifying assumptions. First, avoided health damages are proportional to generation, consistent with the proportional response of the reduced-complexity air quality models used in this study. Second, SCR can move with the equipment and continue in use after relocation. Actual relocation involves dismantling, transport, and recommissioning, which may shorten subsequent use. In that case, \(Q(\Lambda)\) can be replaced by the discounted value of the electricity expected to be delivered over the current stay and the shortened subsequent use. As long as this adjusted value still exceeds \(Q(\tau)\) for the current period on site, the unit cost of recovering the investment over the shorter period remains higher, and the gap between the two criteria created by the horizon persists.

For SCR on new turbines on the 29 evaluated units, costed with the EPA 2026 method and valued with the InMAP main setting at a 7\% discount rate without ammonia slip, the first factor of Eq. S10 is about 8.3 to 9.0 for a 1-year horizon against a 30-year service life (10.3 to 11.4 at 3\%), and the second factor is about 5.5 (table S13, B and F).

\subsection*{Sensitivity of the SCR benefit--cost ratio}

Taking a 7\% discount rate, a 30-year horizon, and the InMAP main setting as the baseline, we examined how ammonia slip, capital cost, and other analytic settings affect the benefit--cost ratio of SCR (table S13). With ammonia slip of 3.5 ppm, above both permit limits, the ratio is 1.13 to 1.23, and with ammonia slip of 2.8 ppm, capital costs can rise 33 to 48\% above the base estimate before net benefits fall to zero (table S13, B and C).

With the InMAP or AP4 ground-level release case, the ratios are 1.86 to 2.03 and 2.05 to 2.24, respectively. With the concentration--response coefficient of Pope et al. (\emph{57}), likewise applied to people aged 30 and older, the 30-year ratio without ammonia slip rises to 2.89 to 3.16, whereas the 1- and 2-year ratios remain below 1 (0.35 and 0.64 to 0.66). Valued instead with the InMAP deaths per metric ton that this fleet's emissions would cause at the other sites of table S10D (55 to 93\% of the Southaven value), the 30-year ratio without ammonia slip exceeds 1 at every site at a 3\% discount rate. At 7\%, it is 1.38 to 1.51 at Socrates, 1.22 to 1.34 at Colossus 1, 0.95 to 1.04 at Abilene, and 0.82 to 0.90 at Fermi Matador.

Over the remaining period on site only, the benefit--cost ratio for retrofits of existing units is 0.075 without ammonia slip and 0.061 with ammonia slip of 2.8 ppm. Over 30 years, without ammonia slip and assuming no replacement, the ratio is 1.29 to 1.40, and with replacement of the entire SCR system in year 15 and no salvage value, it is still 1.05 to 1.11 (table S13, D and F). Without ammonia slip, the payback period is 8.7 to 9.6 years for SCR on new turbines and 12.0 to 13.5 years for retrofits, shorter than the 15-year remaining useful life that the EPA assumed when costing SCR for modified and reconstructed turbines (\emph{69}).

\FloatBarrier\clearpage
\section*{Supplementary Figures}

\begin{center}\begin{minipage}{\textwidth}\centering
\includegraphics[width=\textwidth,height=0.62\textheight,keepaspectratio]{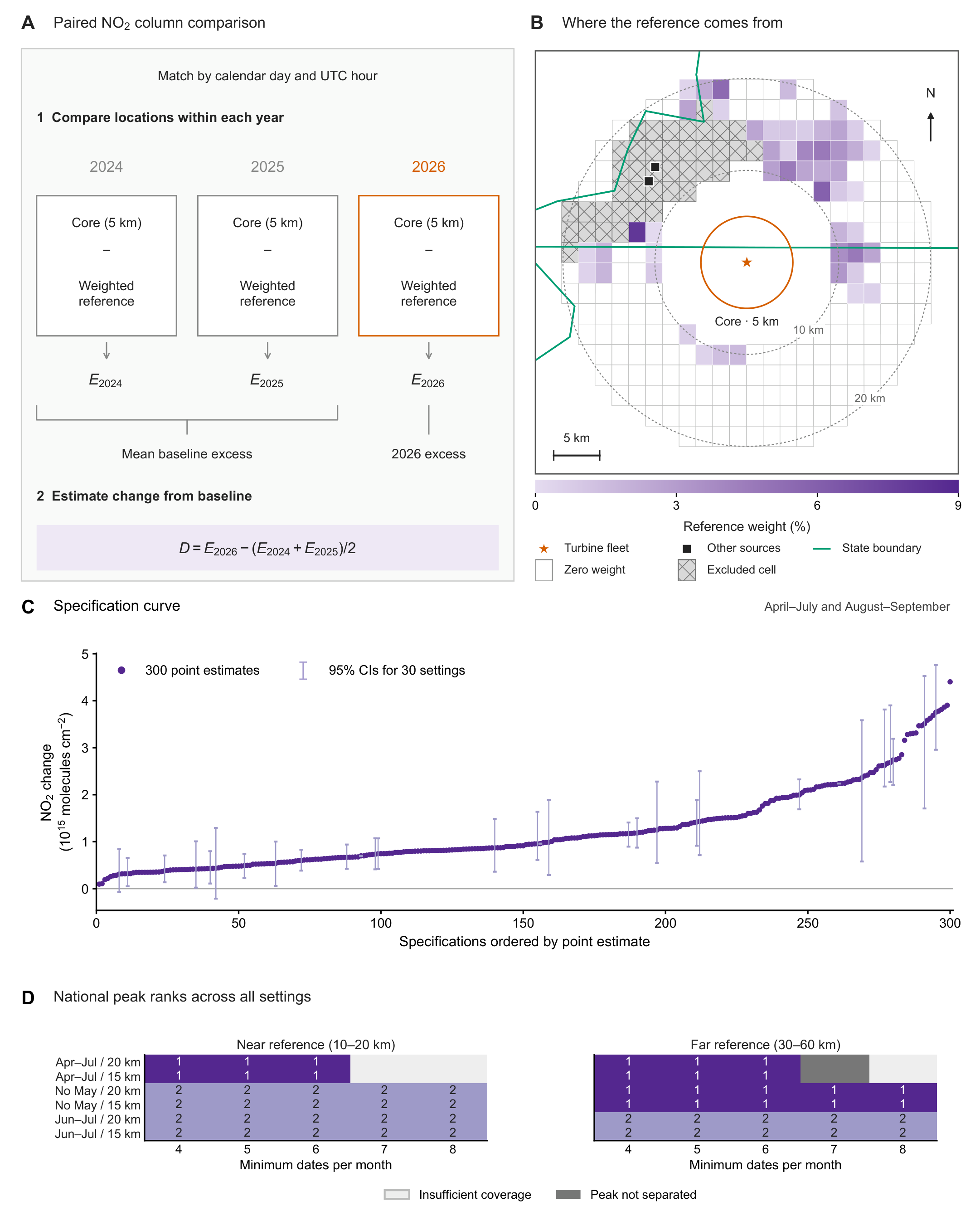}
\par\medskip
\begin{minipage}{\textwidth}\small \textbf{Fig. S1. Construction of the estimator and sensitivity to specifications.} (\textbf{A}) Schematic of the paired near-field estimator, which takes the difference between the core area and the weighted reference area within each year and compares 2026 with 2024 and 2025 on the same calendar day and UTC hour. (\textbf{B}) Geographic distribution of reference cells and their weights, distinguishing excluded cells from cells with zero weight. (\textbf{C}) The 300 point estimates, all positive, sorted by size, spanning seasons, observation windows, reference approaches, and other analytic settings, with 95\% confidence intervals drawn for 30 specifications. Of the 27 intervals based on more than one observation date, 25 exclude zero. (\textbf{D}) National rank of Southaven under 60 peak-search specifications, grouped by near and far reference rings. Rows show observation months and minimum peak spacing in kilometers, and columns show the minimum number of observation days per month. Southaven ranks first or second in all 52 rankable specifications, first in 22 of them, and first in all 12 rankable specifications that use the full April--July window. Changes in (C) are in units of 10\textsuperscript{15} molecules cm\textsuperscript{\ensuremath{-}2}.\end{minipage}
\end{minipage}\end{center}
\clearpage

\begin{center}\begin{minipage}{\textwidth}\centering
\includegraphics[width=\textwidth,height=0.62\textheight,keepaspectratio]{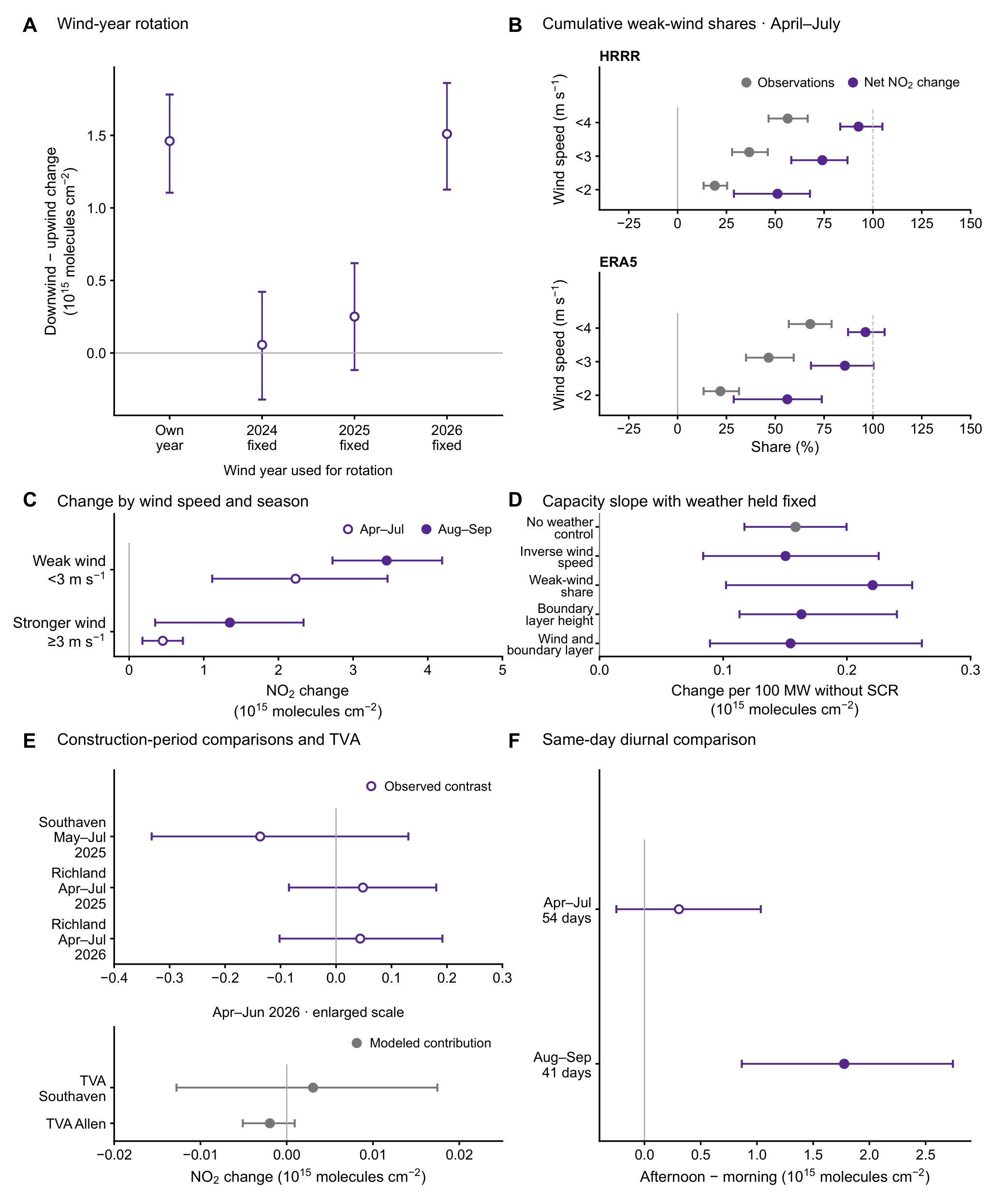}
\par\medskip
\begin{minipage}{\textwidth}\small \textbf{Fig. S2. Additional attribution diagnostics.} (\textbf{A}) Change in the downwind-minus-upwind contrast, April to July, when each year is rotated with its own winds or all three years with the 2024, 2025, or 2026 winds. (\textbf{B}) Cumulative weak-wind shares from HRRR and ERA5, April to July. Gray shows the date-weighted share of observations, and purple the share of the signed net NO\textsubscript{2} change, which can exceed 100\% when the change under stronger winds is negative. (\textbf{C}) Change under single-hour HRRR 10-m winds below 3 m s\textsuperscript{\ensuremath{-}1} and at or above 3 m s\textsuperscript{\ensuremath{-}1}, April to July (open circles) and August to September (filled circles). (\textbf{D}) Slope of the monthly change per 100 MW of capacity without SCR, without weather controls (gray circle) and with each weather term held fixed. Intervals reflect monthly sampling uncertainty. (\textbf{E}) Changes during construction at this site (May to July 2025) and at Richland (April to July 2025 and 2026), and simulated contributions of TVA Southaven and Allen (April to June 2026, 101 scans). (\textbf{F}) Afternoon (18 to 21 UTC) minus morning (13 to 16 UTC) contrast on the same days (54 days in April to July and 41 in August to September). Error bars are 95\% confidence intervals. Changes are in units of 10\textsuperscript{15} molecules cm\textsuperscript{\ensuremath{-}2}.\end{minipage}
\end{minipage}\end{center}
\clearpage

\begin{center}\begin{minipage}{\textwidth}\centering
\includegraphics[width=\textwidth,height=0.62\textheight,keepaspectratio]{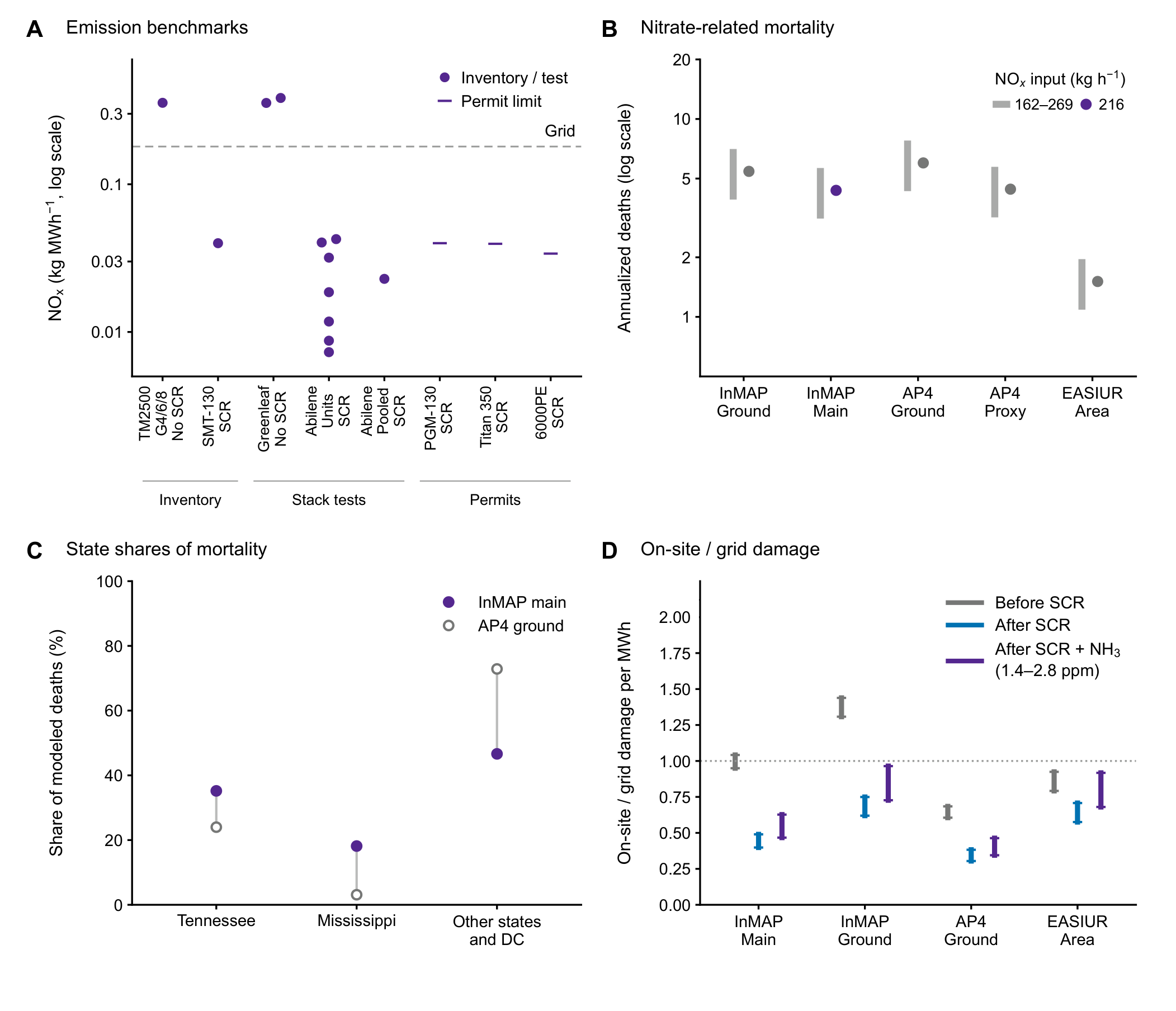}
\par\medskip
\begin{minipage}{\textwidth}\small \textbf{Fig. S3. Comparison of emission and health models.} (\textbf{A}) Emission intensity per unit of generation (log scale) from the engineering inventory, stack tests, and permit limits. The dashed line marks the marginal NO\emph{\textsubscript{x}} emission rate of the regional grid. (\textbf{B}) Annualized deaths from particulate nitrate under each model and release-height case. Gray bars show the range for NO\emph{\textsubscript{x}} emissions of 162 to 269 kg per hour, and filled circles show the central case of 216 kg per hour, with the InMAP main setting in purple. (\textbf{C}) State shares of deaths for the InMAP main setting and the AP4 ground-level case. (\textbf{D}) Ratio of on-site to grid damages per megawatt-hour, with grid damages expressed per megawatt-hour of generation displaced in the AVERT simulation, for the InMAP main setting, the InMAP and AP4 ground-level release cases, and the EASIUR area-source case, counting NO\emph{\textsubscript{x}}, primary particulate matter, and sulfur dioxide. Gray shows the ratio before SCR, blue after SCR, and purple after SCR with ammonia slip of 1.4 and 2.8 ppm.\end{minipage}
\end{minipage}\end{center}
\clearpage

\begin{center}\begin{minipage}{\textwidth}\centering
\includegraphics[width=\textwidth,height=0.62\textheight,keepaspectratio]{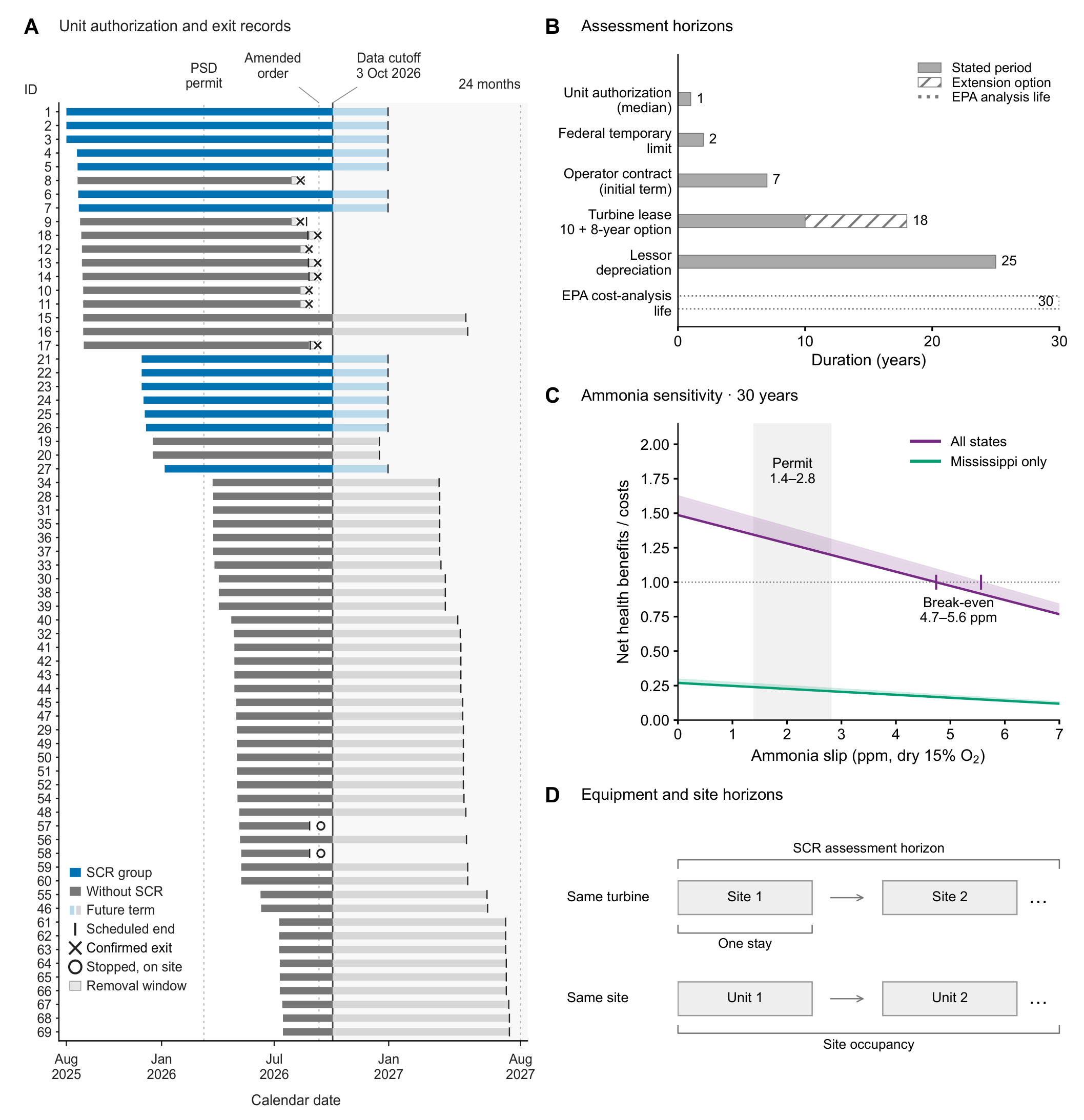}
\par\medskip
\begin{minipage}{\textwidth}\small \textbf{Fig. S4. Equipment, sites, and evaluation horizons.} (\textbf{A}) Authorization and exit records of 68 units as of 3 October 2026. Unit numbers follow Appendix A of the amended agreed order, which lists unit 53 as never delivered. Dark bars show time on site through 3 October 2026 (blue for units with SCR and gray for units without), light segments show the remaining authorized term up to each scheduled end, and removal windows, confirmed exits, and units stopped on site are marked separately. (\textbf{B}) Authorization, commercial, and cost-analysis horizons, with each row drawn from the initial term and extension options in the record. (\textbf{C}) Thirty-year benefit--cost ratio net of ammonia damages for SCR on new turbines, calculated for the 29 evaluated units at a 7\% discount rate, as a function of ammonia slip (dry at 15\% oxygen). Bands show the two tax and insurance treatments, and shading marks the permit limits of 1.4 to 2.8 ppm at the same site. Purple counts benefits in all states, and green counts benefits in Mississippi only. (\textbf{D}) Conceptual distinction between the same equipment moving across sites and successive units at the same site (not to scale).\end{minipage}
\end{minipage}\end{center}
\clearpage

\FloatBarrier\clearpage
\section*{Supplementary Tables}

\Needspace{14\baselineskip}\medskip\noindent \textbf{Table S1. Samples and descriptive statistics for the analysis windows.}\par\nopagebreak

Capacity and unit counts are means of the monthly upper-bound on-site values, which count units with uncertain departure dates until their latest possible departure date. Boundary-layer height, air temperature, and downward shortwave radiation are ERA5 values matched to the paired scans (17 to 20 UTC), and the weak-wind share uses the HRRR 10-m wind of the single UTC hour containing each scan.

{\small\def\LTcaptype{none}
\begin{longtable}[]{@{}
  >{\raggedright\arraybackslash}p{(\linewidth - 4\tabcolsep) * \real{0.4272}}
  >{\raggedright\arraybackslash}p{(\linewidth - 4\tabcolsep) * \real{0.2273}}
  >{\raggedright\arraybackslash}p{(\linewidth - 4\tabcolsep) * \real{0.3073}}@{}}
\toprule\noalign{}
\begin{minipage}[b]{\linewidth}\raggedright
Indicator
\end{minipage} & \begin{minipage}[b]{\linewidth}\raggedright
April--July 2026
\end{minipage} & \begin{minipage}[b]{\linewidth}\raggedright
August--September 2026
\end{minipage} \\
\midrule\noalign{}
\endfirsthead
\endhead
\bottomrule\noalign{}
\endlastfoot
On-site capacity (GW) & 1.473 & 1.719 \\
Without SCR (GW) & 1.249 & 1.495 \\
With SCR (GW) & 0.224 & 0.224 \\
Units on site & 54.1 & 63.1 \\
Paired scans & 310 & 142 \\
Paired dates & 113 & 41 \\
Share of observations with HRRR wind speed below 3 m s\textsuperscript{\ensuremath{-}1} (\%) & 36.7 & 64.0 \\
Boundary-layer height (m) & 1,309 & 1,566 \\
2-m air temperature (K) & 300.8 & 305.6 \\
Downward shortwave radiation (W m\textsuperscript{\ensuremath{-}2}) & 732 & 713 \\
\end{longtable}
}

\Needspace{14\baselineskip}\medskip\noindent \textbf{Table S2. Robustness of the change in near-field excess, attribution diagnostics, same-scan placebos, and tests at monitor locations.}\par\nopagebreak

Unless otherwise noted, units are 10\textsuperscript{15} molecules cm\textsuperscript{\ensuremath{-}2}, and brackets show 95\% confidence intervals. The main window is 17 to 20 UTC. Dashes mark specifications estimated for April--July only. The smoke-free rows keep only scans in which the core area was free of HMS smoke polygons in all three paired years. The same-processor row uses only paired scans processed with version 4.8.4 in all three years (143 scans on 55 days in spring and 88 scans on 26 days in late summer).

{\small\def\LTcaptype{none}
\begin{longtable}[]{@{}
  >{\raggedright\arraybackslash}p{(\linewidth - 8\tabcolsep) * \real{0.3175}}
  >{\raggedright\arraybackslash}p{(\linewidth - 8\tabcolsep) * \real{0.2482}}
  >{\raggedright\arraybackslash}p{(\linewidth - 8\tabcolsep) * \real{0.0798}}
  >{\raggedright\arraybackslash}p{(\linewidth - 8\tabcolsep) * \real{0.2383}}
  >{\raggedright\arraybackslash}p{(\linewidth - 8\tabcolsep) * \real{0.0798}}@{}}
\toprule\noalign{}
\multicolumn{5}{@{}>{\raggedright\arraybackslash}p{(\linewidth - 8\tabcolsep) * \real{0.9637} + 8\tabcolsep}@{}}{%
\begin{minipage}[b]{\linewidth}\raggedright
\emph{A. Main estimates and sensitivity}
\end{minipage}} \\
\midrule\noalign{}
\endfirsthead
\endhead
\bottomrule\noalign{}
\endlastfoot
Specification & April--July & \multicolumn{3}{>{\raggedright\arraybackslash}p{(\linewidth - 8\tabcolsep) * \real{0.3980} + 4\tabcolsep}@{}}{%
August--September} \\
Main estimate, 2024\nobreakdash--2025 baseline & 1.103 {[}0.584, 1.658{]} & \multicolumn{3}{>{\raggedright\arraybackslash}p{(\linewidth - 8\tabcolsep) * \real{0.3980} + 4\tabcolsep}@{}}{%
2.694 {[}2.201, 3.216{]}} \\
2024 baseline only & 1.058 {[}0.520, 1.633{]} & \multicolumn{3}{>{\raggedright\arraybackslash}p{(\linewidth - 8\tabcolsep) * \real{0.3980} + 4\tabcolsep}@{}}{%
2.680 {[}2.174, 3.239{]}} \\
Fixed 10\nobreakdash--20 km reference ring, area weighted & 1.298 {[}0.829, 1.837{]} & \multicolumn{3}{>{\raggedright\arraybackslash}p{(\linewidth - 8\tabcolsep) * \real{0.3980} + 4\tabcolsep}@{}}{%
3.211 {[}2.701, 3.719{]}} \\
Reference cells downwind of Colossus 1 (HRRR wind directions) excluded & 1.205 {[}0.695, 1.745{]} & \multicolumn{3}{>{\raggedright\arraybackslash}p{(\linewidth - 8\tabcolsep) * \real{0.3980} + 4\tabcolsep}@{}}{%
2.639 {[}2.079, 3.239{]}} \\
Same processor version only & 0.972 {[}0.240, 1.878{]} & \multicolumn{3}{>{\raggedright\arraybackslash}p{(\linewidth - 8\tabcolsep) * \real{0.3980} + 4\tabcolsep}@{}}{%
2.519 {[}1.346, 3.386{]}} \\
Predeployment, 2025 vs. 2024 & −0.088 {[}−0.267, 0.118{]} & \multicolumn{3}{>{\raggedright\arraybackslash}p{(\linewidth - 8\tabcolsep) * \real{0.3980} + 4\tabcolsep}@{}}{%
---} \\
HRRR discrete weather, unweighted & 0.991 {[}0.498, 1.404{]} & \multicolumn{3}{>{\raggedright\arraybackslash}p{(\linewidth - 8\tabcolsep) * \real{0.3980} + 4\tabcolsep}@{}}{%
2.677 {[}2.006, 3.563{]}} \\
HRRR discrete weather, reweighted & 0.982 {[}0.476, 1.377{]} & \multicolumn{3}{>{\raggedright\arraybackslash}p{(\linewidth - 8\tabcolsep) * \real{0.3980} + 4\tabcolsep}@{}}{%
2.709 {[}1.951, 3.617{]}} \\
ERA5 discrete weather, unweighted & 1.073 {[}0.503, 1.561{]} & \multicolumn{3}{>{\raggedright\arraybackslash}p{(\linewidth - 8\tabcolsep) * \real{0.3980} + 4\tabcolsep}@{}}{%
2.831 {[}1.958, 3.874{]}} \\
ERA5 discrete weather, reweighted & 1.085 {[}0.521, 1.618{]} & \multicolumn{3}{>{\raggedright\arraybackslash}p{(\linewidth - 8\tabcolsep) * \real{0.3980} + 4\tabcolsep}@{}}{%
2.839 {[}1.782, 3.921{]}} \\
Continuous weather, unweighted & 1.080 {[}0.442, 1.585{]} & \multicolumn{3}{>{\raggedright\arraybackslash}p{(\linewidth - 8\tabcolsep) * \real{0.3980} + 4\tabcolsep}@{}}{%
---} \\
Continuous weather, inverse probability weighted & 1.064 {[}0.364, 2.039{]} & \multicolumn{3}{>{\raggedright\arraybackslash}p{(\linewidth - 8\tabcolsep) * \real{0.3980} + 4\tabcolsep}@{}}{%
---} \\
Main estimate, scans/days & 310 / 113 & \multicolumn{3}{>{\raggedright\arraybackslash}p{(\linewidth - 8\tabcolsep) * \real{0.3980} + 4\tabcolsep}@{}}{%
142 / 41} \\
HMS smoke-free in all three paired years & 1.000 {[}0.475, 1.595{]} & \multicolumn{3}{>{\raggedright\arraybackslash}p{(\linewidth - 8\tabcolsep) * \real{0.3980} + 4\tabcolsep}@{}}{%
2.103 {[}1.573, 2.513{]}} \\
HMS smoke-free, scans/days & 192 / 88 & \multicolumn{3}{>{\raggedright\arraybackslash}p{(\linewidth - 8\tabcolsep) * \real{0.3980} + 4\tabcolsep}@{}}{%
63 / 30} \\
\multicolumn{5}{@{}>{\raggedright\arraybackslash}p{(\linewidth - 8\tabcolsep) * \real{0.9637} + 8\tabcolsep}@{}}{%
\emph{B. Attribution diagnostics in the main window and in all daytime hours}} \\*
Diagnostic and season & 17\nobreakdash--20 UTC & \multicolumn{3}{>{\raggedright\arraybackslash}p{(\linewidth - 8\tabcolsep) * \real{0.3980} + 4\tabcolsep}@{}}{%
All daytime hours} \\*
Change in near-field excess · April--July & 1.103 {[}0.584, 1.658{]} & \multicolumn{3}{>{\raggedright\arraybackslash}p{(\linewidth - 8\tabcolsep) * \real{0.3980} + 4\tabcolsep}@{}}{%
0.744 {[}0.423, 1.086{]}} \\
Change in near-field excess · August--September & 2.694 {[}2.201, 3.216{]} & \multicolumn{3}{>{\raggedright\arraybackslash}p{(\linewidth - 8\tabcolsep) * \real{0.3980} + 4\tabcolsep}@{}}{%
2.000 {[}1.578, 2.370{]}} \\
Downwind--upwind, own-year winds · April--July & 1.461 {[}1.105, 1.782{]} & \multicolumn{3}{>{\raggedright\arraybackslash}p{(\linewidth - 8\tabcolsep) * \real{0.3980} + 4\tabcolsep}@{}}{%
1.142 {[}0.912, 1.394{]}} \\
Downwind--upwind, own-year winds · August--September & 3.577 {[}2.935, 4.290{]} & \multicolumn{3}{>{\raggedright\arraybackslash}p{(\linewidth - 8\tabcolsep) * \real{0.3980} + 4\tabcolsep}@{}}{%
2.580 {[}2.152, 3.038{]}} \\
Downwind--upwind, 2026 winds · April--July & 1.510 {[}1.126, 1.861{]} & \multicolumn{3}{>{\raggedright\arraybackslash}p{(\linewidth - 8\tabcolsep) * \real{0.3980} + 4\tabcolsep}@{}}{%
1.165 {[}0.837, 1.471{]}} \\
Downwind--upwind, 2026 winds · August--September & 3.751 {[}3.279, 4.312{]} & \multicolumn{3}{>{\raggedright\arraybackslash}p{(\linewidth - 8\tabcolsep) * \real{0.3980} + 4\tabcolsep}@{}}{%
2.694 {[}2.324, 3.064{]}} \\
Half-decay distance (km) · April--July & 3.851 {[}2.925, 5.060{]} & \multicolumn{3}{>{\raggedright\arraybackslash}p{(\linewidth - 8\tabcolsep) * \real{0.3980} + 4\tabcolsep}@{}}{%
3.944 {[}2.973, 4.936{]}} \\
Half-decay distance (km) · August--September & 4.357 {[}3.777, 4.947{]} & \multicolumn{3}{>{\raggedright\arraybackslash}p{(\linewidth - 8\tabcolsep) * \real{0.3980} + 4\tabcolsep}@{}}{%
4.472 {[}3.676, 5.477{]}} \\
Far-field asymptote · April--July & −0.376 {[}−0.620, −0.150{]} & \multicolumn{3}{>{\raggedright\arraybackslash}p{(\linewidth - 8\tabcolsep) * \real{0.3980} + 4\tabcolsep}@{}}{%
−0.240 {[}−0.406, −0.089{]}} \\
Far-field asymptote · August--September & −1.095 {[}−1.492, −0.680{]} & \multicolumn{3}{>{\raggedright\arraybackslash}p{(\linewidth - 8\tabcolsep) * \real{0.3980} + 4\tabcolsep}@{}}{%
−0.827 {[}−1.231, −0.476{]}} \\
Baseline correlation with inverse wind speed · April--July & −0.008 {[}−0.105, 0.081{]} & \multicolumn{3}{>{\raggedright\arraybackslash}p{(\linewidth - 8\tabcolsep) * \real{0.3980} + 4\tabcolsep}@{}}{%
−0.032 {[}−0.094, 0.027{]}} \\
Baseline correlation with inverse wind speed · August--September & 0.005 {[}−0.149, 0.177{]} & \multicolumn{3}{>{\raggedright\arraybackslash}p{(\linewidth - 8\tabcolsep) * \real{0.3980} + 4\tabcolsep}@{}}{%
−0.040 {[}−0.162, 0.083{]}} \\
2026 correlation with inverse wind speed · April--July & 0.363 {[}0.163, 0.541{]} & \multicolumn{3}{>{\raggedright\arraybackslash}p{(\linewidth - 8\tabcolsep) * \real{0.3980} + 4\tabcolsep}@{}}{%
0.271 {[}0.152, 0.384{]}} \\
2026 correlation with inverse wind speed · August--September & 0.420 {[}0.208, 0.623{]} & \multicolumn{3}{>{\raggedright\arraybackslash}p{(\linewidth - 8\tabcolsep) * \real{0.3980} + 4\tabcolsep}@{}}{%
0.294 {[}0.131, 0.471{]}} \\
Difference in correlation with inverse wind speed · April--July & 0.371 {[}0.138, 0.596{]} & \multicolumn{3}{>{\raggedright\arraybackslash}p{(\linewidth - 8\tabcolsep) * \real{0.3980} + 4\tabcolsep}@{}}{%
0.304 {[}0.185, 0.425{]}} \\
Difference in correlation with inverse wind speed · August--September & 0.415 {[}0.191, 0.596{]} & \multicolumn{3}{>{\raggedright\arraybackslash}p{(\linewidth - 8\tabcolsep) * \real{0.3980} + 4\tabcolsep}@{}}{%
0.335 {[}0.132, 0.514{]}} \\
\multicolumn{5}{@{}>{\raggedright\arraybackslash}p{(\linewidth - 8\tabcolsep) * \real{0.9637} + 8\tabcolsep}@{}}{%
Half-decay distances are in kilometers, and correlations are dimensionless and centered separately within each year.} \\
\multicolumn{5}{@{}>{\raggedright\arraybackslash}p{(\linewidth - 8\tabcolsep) * \real{0.9637} + 8\tabcolsep}@{}}{%
\emph{C. Same-scan placebos, all processor versions}} \\*
Location & April--July & Rank & August--September & Rank \\*
Southaven & 1.296 {[}0.846, 1.846{]} & 1 & 3.200 {[}2.654, 3.697{]} & 1 \\
Cedar Rapids, IA & −0.192 {[}−0.324, −0.064{]} & 16 & −0.095 {[}−0.269, 0.085{]} & 16 \\
Columbia, MO & 0.056 {[}−0.066, 0.176{]} & 6 & 0.070 {[}−0.004, 0.149{]} & 8 \\
Davenport, IA--IL & 0.121 {[}−0.009, 0.290{]} & 3 & −0.025 {[}−0.173, 0.111{]} & 14 \\
Fayetteville--Springdale--Rogers, AR--MO & 0.004 {[}−0.135, 0.135{]} & 10 & 0.035 {[}−0.062, 0.131{]} & 9 \\
Fort Smith, AR--OK & 0.039 {[}−0.055, 0.141{]} & 7 & 0.079 {[}−0.018, 0.180{]} & 6 \\
Houma, LA & 0.029 {[}−0.067, 0.141{]} & 8 & 0.024 {[}−0.241, 0.287{]} & 12 \\
Iowa City, IA & 0.066 {[}−0.063, 0.200{]} & 5 & −0.041 {[}−0.160, 0.098{]} & 15 \\
Jackson, MS & 0.073 {[}−0.022, 0.169{]} & 4 & 0.162 {[}0.047, 0.274{]} & 3 \\
Lafayette, LA & 0.231 {[}−0.015, 0.619{]} & 2 & 0.035 {[}−0.108, 0.187{]} & 10 \\
Little Rock, AR & −0.042 {[}−0.206, 0.090{]} & 12 & 0.384 {[}0.002, 1.014{]} & 2 \\
Mandeville--Covington, LA & −0.018 {[}−0.122, 0.088{]} & 11 & 0.076 {[}−0.129, 0.253{]} & 7 \\
Monroe, LA & 0.020 {[}−0.107, 0.162{]} & 9 & −0.008 {[}−0.176, 0.182{]} & 13 \\
Shreveport, LA & −0.078 {[}−0.295, 0.175{]} & 15 & 0.027 {[}−0.138, 0.173{]} & 11 \\
Topeka, KS & −0.062 {[}−0.171, 0.056{]} & 14 & 0.118 {[}0.015, 0.220{]} & 4 \\
Tyler, TX & −0.056 {[}−0.164, 0.048{]} & 13 & 0.116 {[}−0.032, 0.281{]} & 5 \\
\multicolumn{5}{@{}>{\raggedright\arraybackslash}p{(\linewidth - 8\tabcolsep) * \real{0.9637} + 8\tabcolsep}@{}}{%
The 16 locations use the same scan list, a 5-km core area, and a 10\nobreakdash--20 km fixed reference ring, with dates weighted equally and 1,999 circular block bootstrap replicates with 7-day blocks. Ranks are computed among the 16 locations.} \\
\multicolumn{5}{@{}>{\raggedright\arraybackslash}p{(\linewidth - 8\tabcolsep) * \real{0.9637} + 8\tabcolsep}@{}}{%
\emph{D. Same-scan placebos, paired scans of version 4.8.4 only}} \\*
Location & April--July & Rank & August--September & Rank \\*
Southaven & 1.341 {[}0.586, 2.297{]} & 1 & 3.005 {[}1.971, 3.686{]} & 1 \\
Cedar Rapids, IA & −0.219 {[}−0.412, −0.080{]} & 16 & −0.063 {[}−0.228, 0.103{]} & 15 \\
Columbia, MO & 0.055 {[}−0.123, 0.253{]} & 3 & 0.054 {[}−0.005, 0.106{]} & 9 \\
Davenport, IA--IL & 0.024 {[}−0.106, 0.126{]} & 8 & 0.039 {[}−0.113, 0.211{]} & 11 \\
Fayetteville--Springdale--Rogers, AR--MO & −0.040 {[}−0.256, 0.132{]} & 11 & −0.008 {[}−0.106, 0.121{]} & 13 \\
Fort Smith, AR--OK & 0.018 {[}−0.147, 0.190{]} & 9 & 0.149 {[}0.067, 0.254{]} & 3 \\
Houma, LA & 0.055 {[}−0.107, 0.239{]} & 4 & −0.032 {[}−0.292, 0.251{]} & 14 \\
Iowa City, IA & 0.154 {[}−0.031, 0.346{]} & 2 & −0.131 {[}−0.246, 0.024{]} & 16 \\
Jackson, MS & 0.049 {[}−0.081, 0.164{]} & 5 & 0.129 {[}−0.020, 0.268{]} & 5 \\
Lafayette, LA & 0.043 {[}−0.159, 0.229{]} & 6 & 0.147 {[}−0.008, 0.319{]} & 4 \\
Little Rock, AR & −0.093 {[}−0.227, 0.036{]} & 14 & 0.449 {[}−0.136, 1.506{]} & 2 \\
Mandeville--Covington, LA & −0.123 {[}−0.295, 0.040{]} & 15 & 0.045 {[}−0.243, 0.375{]} & 10 \\
Monroe, LA & 0.043 {[}−0.149, 0.287{]} & 7 & 0.107 {[}−0.079, 0.372{]} & 7 \\
Shreveport, LA & −0.011 {[}−0.138, 0.116{]} & 10 & 0.034 {[}−0.140, 0.175{]} & 12 \\
Topeka, KS & −0.093 {[}−0.261, 0.074{]} & 13 & 0.117 {[}−0.039, 0.267{]} & 6 \\
Tyler, TX & −0.082 {[}−0.224, 0.039{]} & 12 & 0.057 {[}−0.127, 0.253{]} & 8 \\
\multicolumn{5}{@{}>{\raggedright\arraybackslash}p{(\linewidth - 8\tabcolsep) * \real{0.9637} + 8\tabcolsep}@{}}{%
\emph{E. Reference weights without distance indicators, and tests at monitor locations}} \\*
Specification & April--July & \multicolumn{3}{>{\raggedright\arraybackslash}p{(\linewidth - 8\tabcolsep) * \real{0.3980} + 4\tabcolsep}@{}}{%
August--September} \\*
Reference weights without airport and refinery distances & 1.124 {[}0.640, 1.678{]} & \multicolumn{3}{>{\raggedright\arraybackslash}p{(\linewidth - 8\tabcolsep) * \real{0.3980} + 4\tabcolsep}@{}}{%
2.829 {[}2.348, 3.395{]}} \\
Monitor 05-035-0005, Arkansas (28 km) & −0.095 {[}−0.227, 0.032{]} & \multicolumn{3}{>{\raggedright\arraybackslash}p{(\linewidth - 8\tabcolsep) * \real{0.3980} + 4\tabcolsep}@{}}{%
−0.190 {[}−0.470, 0.054{]}} \\
Monitor 47-157-0100, Tennessee (25 km) & 0.036 {[}−0.117, 0.191{]} & \multicolumn{3}{>{\raggedright\arraybackslash}p{(\linewidth - 8\tabcolsep) * \real{0.3980} + 4\tabcolsep}@{}}{%
−0.244 {[}−0.456, −0.023{]}} \\
\multicolumn{5}{@{}>{\raggedright\arraybackslash}p{(\linewidth - 8\tabcolsep) * \real{0.9637} + 8\tabcolsep}@{}}{%
The fixed-ring row of panel A uses 317 scans on 113 days, and panel C draws on the 310 scans of the main sample, of which each location keeps those in which it is validly observed. Monitor tests use a 5-km core area centered on each monitor and a 10- to 20-km fixed reference ring.} \\
\end{longtable}
}

\Needspace{14\baselineskip}\medskip\noindent \textbf{Table S3. Comparisons over time at Southaven and across US urban locations and the contiguous US.}\par\nopagebreak

Changes are in 10\textsuperscript{15} molecules cm\textsuperscript{\ensuremath{-}2}. The April--July national comparison requires at least 30 observation days at each location, and the August--September comparison at least 10. Synthetic-control comparison locations exclude those with power-sector changes within 20 km (materials and methods). SCM, synthetic control. ASCM, augmented synthetic control. SDID, synthetic difference-in-differences. DID, equal-weight difference-in-differences. A rank of 1 of 81 corresponds to a permutation \emph{P} value of 1/81. Brackets show 95\% confidence intervals.

{\small\def\LTcaptype{none}
\begin{longtable}[]{@{}
  >{\raggedright\arraybackslash}p{(\linewidth - 8\tabcolsep) * \real{0.2703}}
  >{\raggedright\arraybackslash}p{(\linewidth - 8\tabcolsep) * \real{0.2198}}
  >{\raggedright\arraybackslash}p{(\linewidth - 8\tabcolsep) * \real{0.1523}}
  >{\raggedright\arraybackslash}p{(\linewidth - 8\tabcolsep) * \real{0.1860}}
  >{\raggedright\arraybackslash}p{(\linewidth - 8\tabcolsep) * \real{0.1355}}@{}}
\toprule\noalign{}
\multicolumn{5}{@{}>{\raggedright\arraybackslash}p{(\linewidth - 8\tabcolsep) * \real{0.9638} + 8\tabcolsep}@{}}{%
\begin{minipage}[b]{\linewidth}\raggedright
\emph{A. TROPOMI near-field excess at Southaven}
\end{minipage}} \\
\midrule\noalign{}
\endfirsthead
\endhead
\bottomrule\noalign{}
\endlastfoot
Indicator & Value & \multicolumn{3}{>{\raggedright\arraybackslash}p{(\linewidth - 8\tabcolsep) * \real{0.4737} + 4\tabcolsep}@{}}{%
Note} \\
Annual values, April--July 2019\nobreakdash--2025 & 0.052\nobreakdash--0.213 & \multicolumn{3}{>{\raggedright\arraybackslash}p{(\linewidth - 8\tabcolsep) * \real{0.4737} + 4\tabcolsep}@{}}{%
Range, Level-2} \\
Change, April--July 2026 & 0.79 {[}0.54, 1.05{]} & \multicolumn{3}{>{\raggedright\arraybackslash}p{(\linewidth - 8\tabcolsep) * \real{0.4737} + 4\tabcolsep}@{}}{%
Level-2, relative to 2024\nobreakdash--2025} \\
Change on days paired with TEMPO, April--July 2026 & 0.63 & \multicolumn{3}{>{\raggedright\arraybackslash}p{(\linewidth - 8\tabcolsep) * \real{0.4737} + 4\tabcolsep}@{}}{%
Level-3 daily composite on the 14 days with paired TEMPO scans near the TROPOMI overpass (about 18:30 to 20:30 UTC)} \\
Change, August--September 2026 & 2.17 {[}1.22, 3.21{]} & \multicolumn{3}{>{\raggedright\arraybackslash}p{(\linewidth - 8\tabcolsep) * \real{0.4737} + 4\tabcolsep}@{}}{%
Level-3 national comparison, 32 observation days in 2026} \\
\multicolumn{5}{@{}>{\raggedright\arraybackslash}p{(\linewidth - 8\tabcolsep) * \real{0.9638} + 8\tabcolsep}@{}}{%
\emph{B. National comparison among 656 urban locations (TROPOMI Level-3 gridded product)}} \\*
Indicator & Value & \multicolumn{3}{>{\raggedright\arraybackslash}p{(\linewidth - 8\tabcolsep) * \real{0.4737} + 4\tabcolsep}@{}}{%
Note} \\*
Stabilized anomaly score, Southaven, April--July & 4.63 & \multicolumn{3}{>{\raggedright\arraybackslash}p{(\linewidth - 8\tabcolsep) * \real{0.4737} + 4\tabcolsep}@{}}{%
Eq. S4} \\
Highest score at the other 655 locations & 1.82 & \multicolumn{3}{>{\raggedright\arraybackslash}p{(\linewidth - 8\tabcolsep) * \real{0.4737} + 4\tabcolsep}@{}}{%
---} \\
Median change at the other 655 locations, April--July 2026 & 0.01 & \multicolumn{3}{>{\raggedright\arraybackslash}p{(\linewidth - 8\tabcolsep) * \real{0.4737} + 4\tabcolsep}@{}}{%
Relative to 2024\nobreakdash--2025} \\
Highest score at any location, 2020\nobreakdash--2025 & 2.68 & \multicolumn{3}{>{\raggedright\arraybackslash}p{(\linewidth - 8\tabcolsep) * \real{0.4737} + 4\tabcolsep}@{}}{%
Near Green Bay, Wisconsin, 2021, with each year scored against the other years} \\
Scores at Southaven, 2020\nobreakdash--2025 & −1.43 to 0.78 & \multicolumn{3}{>{\raggedright\arraybackslash}p{(\linewidth - 8\tabcolsep) * \real{0.4737} + 4\tabcolsep}@{}}{%
Before deployment} \\
Rank of the Southaven change, August--September & 1 of 656 & \multicolumn{3}{>{\raggedright\arraybackslash}p{(\linewidth - 8\tabcolsep) * \real{0.4737} + 4\tabcolsep}@{}}{%
Second-highest change 0.454} \\
\multicolumn{5}{@{}>{\raggedright\arraybackslash}p{(\linewidth - 8\tabcolsep) * \real{0.9638} + 8\tabcolsep}@{}}{%
\emph{C. TEMPO source-agnostic peak search}} \\*
Indicator & Value & \multicolumn{3}{>{\raggedright\arraybackslash}p{(\linewidth - 8\tabcolsep) * \real{0.4737} + 4\tabcolsep}@{}}{%
Note} \\*
Rank of Southaven, April--July & First in 22, second in 30 & \multicolumn{3}{>{\raggedright\arraybackslash}p{(\linewidth - 8\tabcolsep) * \real{0.4737} + 4\tabcolsep}@{}}{%
52 of 60 specifications rankable} \\
Specifications not rankable & 8 of 60 & \multicolumn{3}{>{\raggedright\arraybackslash}p{(\linewidth - 8\tabcolsep) * \real{0.4737} + 4\tabcolsep}@{}}{%
6 with insufficient coverage, 2 with peaks not separable} \\
Rank of Southaven, August--September & 1 & \multicolumn{3}{>{\raggedright\arraybackslash}p{(\linewidth - 8\tabcolsep) * \real{0.4737} + 4\tabcolsep}@{}}{%
Among 227,072 eligible cells} \\
Rank of the Southaven 2026 excess level (not the change), April--July & 9th among separated peaks (280,286 cells searched) & \multicolumn{3}{>{\raggedright\arraybackslash}p{(\linewidth - 8\tabcolsep) * \real{0.4737} + 4\tabcolsep}@{}}{%
Searched directly by level with 20-km peak separation, and the eight higher peaks are long-standing urban and industrial centers} \\
Rank of Southaven, August & 2 & \multicolumn{3}{>{\raggedright\arraybackslash}p{(\linewidth - 8\tabcolsep) * \real{0.4737} + 4\tabcolsep}@{}}{%
Among 227,074 eligible cells} \\
Rank of Southaven, September & 1 & \multicolumn{3}{>{\raggedright\arraybackslash}p{(\linewidth - 8\tabcolsep) * \real{0.4737} + 4\tabcolsep}@{}}{%
Among 311,974 eligible cells} \\
\multicolumn{5}{@{}>{\raggedright\arraybackslash}p{(\linewidth - 8\tabcolsep) * \real{0.9638} + 8\tabcolsep}@{}}{%
\emph{D. Synthetic control with 80 comparison locations (TROPOMI, monthly)}} \\*
Estimator & Rank after deployment & Estimated change & Rank in 2024 placebo & Placebo change \\*
SCM & 1/81 & 0.720 & 50/81 & 0.016 \\
ASCM & 1/81 & 0.718 & 38/81 & −0.014 \\
SDID & 1/81 & 0.719 & 51/81 & −0.014 \\
DID & 1/81 & 0.738 & 49/81 & 0.031 \\
\end{longtable}
}

\Needspace{14\baselineskip}\medskip\noindent \textbf{Table S4. Wind direction and weak winds.}\par\nopagebreak

Rotated contrasts are in 10\textsuperscript{15} molecules cm\textsuperscript{\ensuremath{-}2}, and weak-wind shares are percentages. Wind-speed thresholds are cumulative, and signed shares of the net increase can exceed 100\%. For both wind products, shares use the 10-m wind of the single UTC hour containing the scan. Brackets show 95\% confidence intervals.

{\small\def\LTcaptype{none}
\begin{longtable}[]{@{}
  >{\raggedright\arraybackslash}p{(\linewidth - 4\tabcolsep) * \real{0.3798}}
  >{\raggedright\arraybackslash}p{(\linewidth - 4\tabcolsep) * \real{0.2850}}
  >{\raggedright\arraybackslash}p{(\linewidth - 4\tabcolsep) * \real{0.2969}}@{}}
\toprule\noalign{}
\begin{minipage}[b]{\linewidth}\raggedright
Diagnostic
\end{minipage} & \begin{minipage}[b]{\linewidth}\raggedright
April--July
\end{minipage} & \begin{minipage}[b]{\linewidth}\raggedright
August--September
\end{minipage} \\
\midrule\noalign{}
\endfirsthead
\endhead
\bottomrule\noalign{}
\endlastfoot
Own-year wind directions & 1.461 {[}1.105, 1.782{]} & 3.577 {[}2.935, 4.290{]} \\
All years with 2026 wind directions & 1.510 {[}1.126, 1.861{]} & 3.751 {[}3.279, 4.312{]} \\
All years with 2024 wind directions & 0.056 {[}−0.321, 0.422{]} & −1.197 {[}−2.156, −0.354{]} \\
All years with 2025 wind directions & 0.251 {[}−0.118, 0.619{]} & −0.604 {[}−1.787, 0.507{]} \\
HRRR 10 m, \textless2 m s\textsuperscript{\ensuremath{-}1}, share of observations (\%) & 19.03 {[}13.32, 25.37{]} & 34.35 {[}19.54, 49.76{]} \\
HRRR 10 m, \textless2 m s\textsuperscript{\ensuremath{-}1}, share of net increase (\%) & 51.13 {[}28.82, 67.76{]} & 52.41 {[}33.09, 70.79{]} \\
HRRR 10 m, \textless3 m s\textsuperscript{\ensuremath{-}1}, share of observations (\%) & 36.65 {[}27.89, 46.17{]} & 64.02 {[}55.68, 74.20{]} \\
HRRR 10 m, \textless3 m s\textsuperscript{\ensuremath{-}1}, share of net increase (\%) & 74.09 {[}58.20, 86.98{]} & 81.98 {[}71.27, 94.86{]} \\
HRRR 10 m, \textless4 m s\textsuperscript{\ensuremath{-}1}, share of observations (\%) & 56.34 {[}46.56, 66.59{]} & 86.99 {[}81.11, 92.92{]} \\
HRRR 10 m, \textless4 m s\textsuperscript{\ensuremath{-}1}, share of net increase (\%) & 92.61 {[}83.25, 104.85{]} & 90.40 {[}82.46, 100.29{]} \\
ERA5 10 m, \textless2 m s\textsuperscript{\ensuremath{-}1}, share of observations (\%) & 21.90 {[}13.30, 31.42{]} & 49.59 {[}32.59, 64.38{]} \\
ERA5 10 m, \textless2 m s\textsuperscript{\ensuremath{-}1}, share of net increase (\%) & 56.17 {[}28.72, 73.78{]} & 68.41 {[}45.67, 88.13{]} \\
ERA5 10 m, \textless3 m s\textsuperscript{\ensuremath{-}1}, share of observations (\%) & 46.61 {[}35.04, 59.44{]} & 76.02 {[}62.03, 88.34{]} \\
ERA5 10 m, \textless3 m s\textsuperscript{\ensuremath{-}1}, share of net increase (\%) & 85.64 {[}68.27, 100.32{]} & 83.57 {[}70.22, 95.54{]} \\
ERA5 10 m, \textless4 m s\textsuperscript{\ensuremath{-}1}, share of observations (\%) & 67.92 {[}56.93, 78.84{]} & 96.34 {[}89.18, 100.00{]} \\
ERA5 10 m, \textless4 m s\textsuperscript{\ensuremath{-}1}, share of net increase (\%) & 96.17 {[}87.25, 105.98{]} & 97.58 {[}92.66, 100.00{]} \\
\end{longtable}
}

\Needspace{14\baselineskip}\medskip\noindent \textbf{Table S5. Other sources and spatial covariates.}\par\nopagebreak

Units are 10\textsuperscript{15} molecules cm\textsuperscript{\ensuremath{-}2}. TVA simulations use a separately assembled sample of 101 scans on 42 days from April to June, limited by archived transport and emission inputs, and give the 2026 change in near-field excess relative to the 2024\nobreakdash--2025 mean. Unadjusted and GIS-adjusted regressions use the same 1,198 cells. For 5 to 10 km relative to 10 to 20 km, the coefficients are 0.379 (0.217 to 0.546) unadjusted and 0.359 (0.199 to 0.525) adjusted. Brackets show 95\% confidence intervals.

{\small\def\LTcaptype{none}
\begin{longtable}[]{@{}
  >{\raggedright\arraybackslash}p{(\linewidth - 8\tabcolsep) * \real{0.3623}}
  >{\raggedright\arraybackslash}p{(\linewidth - 8\tabcolsep) * \real{0.2719}}
  >{\raggedright\arraybackslash}p{(\linewidth - 8\tabcolsep) * \real{0.1136}}
  >{\raggedright\arraybackslash}p{(\linewidth - 8\tabcolsep) * \real{0.1136}}
  >{\raggedright\arraybackslash}p{(\linewidth - 8\tabcolsep) * \real{0.1024}}@{}}
\toprule\noalign{}
\multicolumn{5}{@{}>{\raggedright\arraybackslash}p{(\linewidth - 8\tabcolsep) * \real{0.9638} + 8\tabcolsep}@{}}{%
\begin{minipage}[b]{\linewidth}\raggedright
\emph{A. Simulated power plant contributions and spatial covariates}
\end{minipage}} \\
\midrule\noalign{}
\endfirsthead
\endhead
\bottomrule\noalign{}
\endlastfoot
Comparison & Estimate & \multicolumn{3}{>{\raggedright\arraybackslash}p{(\linewidth - 8\tabcolsep) * \real{0.3297} + 4\tabcolsep}@{}}{%
Window and sample} \\
Simulated change in near-field excess, TVA Southaven & 0.003 {[}−0.013, 0.017{]} & \multicolumn{3}{>{\raggedright\arraybackslash}p{(\linewidth - 8\tabcolsep) * \real{0.3297} + 4\tabcolsep}@{}}{%
April--June, 101 scans/42 days} \\
Simulated change in near-field excess, TVA Allen & −0.002 {[}−0.005, 0.001{]} & \multicolumn{3}{>{\raggedright\arraybackslash}p{(\linewidth - 8\tabcolsep) * \real{0.3297} + 4\tabcolsep}@{}}{%
April--June, 101 scans/42 days} \\
Unadjusted, 0\nobreakdash--5 km relative to 10\nobreakdash--20 km & 1.254 {[}0.813, 1.747{]} & \multicolumn{3}{>{\raggedright\arraybackslash}p{(\linewidth - 8\tabcolsep) * \real{0.3297} + 4\tabcolsep}@{}}{%
April--July, 1,198 cells} \\
GIS-adjusted, 0\nobreakdash--5 km relative to 10\nobreakdash--20 km & 1.209 {[}0.744, 1.699{]} & \multicolumn{3}{>{\raggedright\arraybackslash}p{(\linewidth - 8\tabcolsep) * \real{0.3297} + 4\tabcolsep}@{}}{%
April--July, 1,198 cells} \\
\multicolumn{5}{@{}>{\raggedright\arraybackslash}p{(\linewidth - 8\tabcolsep) * \real{0.9638} + 8\tabcolsep}@{}}{%
\emph{B. NO\textsubscript{x} emissions of the two nearby power plants, April--June (CAMPD)}} \\*
Plant & 2024 (t) & 2025 (t) & 2026 (t) & Change \\*
TVA Allen & 30.102 & 40.963 & 40.314 & +13.46\% \\
TVA Southaven & 41.228 & 36.029 & 44.792 & +15.96\% \\
\multicolumn{5}{@{}>{\raggedright\arraybackslash}p{(\linewidth - 8\tabcolsep) * \real{0.9638} + 8\tabcolsep}@{}}{%
Sums of hourly EPA CAMPD emissions (metric tons, all hours of April--June). Changes are for 2026 relative to the mean of 2024 and 2025. Plant identifiers are 3393 (TVA Allen) and 55269 (TVA Southaven).} \\
\end{longtable}
}

\Needspace{14\baselineskip}\medskip\noindent \textbf{Table S6. Cross-site comparison and minimum detectable values.}\par\nopagebreak

Units are 10\textsuperscript{15} molecules cm\textsuperscript{\ensuremath{-}2}. Paired changes are defined as in Eq. S1. The minimum detectable value is 2.8 times the bootstrap standard deviation, for two-sided 5\% significance and 80\% power (\emph{154}). Reference weights for Richland were estimated from all quality-screened daytime observations from 1 April to 31 July 2024 and re-estimated in each bootstrap replicate, and the contrasts used paired scans at 17 to 20 UTC. Colossus 1 shares its near field with TVA Allen, and its contrast does not remove changes at Allen. Intervals for Richland and the Southaven construction row come from resampling whole calendar weeks, and the other rows use the circular 7-day block bootstrap.

{\small\def\LTcaptype{none}
\begin{longtable}[]{@{}
  >{\raggedright\arraybackslash}p{(\linewidth - 6\tabcolsep) * \real{0.3383}}
  >{\raggedright\arraybackslash}p{(\linewidth - 6\tabcolsep) * \real{0.2222}}
  >{\raggedright\arraybackslash}p{(\linewidth - 6\tabcolsep) * \real{0.1272}}
  >{\raggedright\arraybackslash}p{(\linewidth - 6\tabcolsep) * \real{0.2750}}@{}}
\toprule\noalign{}
\begin{minipage}[b]{\linewidth}\raggedright
Site and comparison
\end{minipage} & \begin{minipage}[b]{\linewidth}\raggedright
\emph{D} {[}95\% interval{]}
\end{minipage} & \begin{minipage}[b]{\linewidth}\raggedright
Scans/days
\end{minipage} & \begin{minipage}[b]{\linewidth}\raggedright
Minimum detectable value
\end{minipage} \\
\midrule\noalign{}
\endfirsthead
\endhead
\bottomrule\noalign{}
\endlastfoot
Southaven, 2026 vs. 2024\nobreakdash--2025, April--July & 1.10 {[}0.58, 1.66{]} & 310 / 113 & 0.77 \\
Abilene, 2026 vs. 2024, April--July & 0.19 {[}−0.05, 0.51{]} & 359 / 115 & 0.41 \\
Colossus 1, 2024 vs. 2023, September--November & 0.25 {[}0.00, 0.46{]} & 273 / 87 & 0.33 \\
Richland, 2025 vs. 2024, April--July & 0.05 {[}−0.08, 0.18{]} & 357 / 115 & 0.20 \\
Richland, 2026 vs. 2024, April--July & 0.04 {[}−0.10, 0.19{]} & 349 / 116 & 0.21 \\
Southaven construction before turbine arrival, 2025 vs. 2024, May--July & −0.14 {[}−0.33, 0.13{]} & 279 / 87 & 0.33 \\
\end{longtable}
}

\Needspace{14\baselineskip}\medskip\noindent \textbf{Table S7. Population beneath the April--September hotspot.}\par\nopagebreak

Population is the 2020 Census population allocated to buildings. Thresholds are April--September increases in units of 10\textsuperscript{15} molecules cm\textsuperscript{\ensuremath{-}2}. Poverty rates are ACS 2020\nobreakdash--2024 block group proportions weighted by the population allocated to the hotspot, and county values come from ACS 2020\nobreakdash--2024 county tables. Brackets show 95\% intervals. Intervals for the Tennessee share at each threshold and for the resident counts at the 0.5 threshold (panel B) come from 1,999 cross-season replicates weighted by calendar week, each re-estimating the reference weights, and only cells with at least 15 dates with valid observations are retained.

{\small\def\LTcaptype{none}
\begin{longtable}[]{@{}
  >{\raggedright\arraybackslash}p{(\linewidth - 8\tabcolsep) * \real{0.2547}}
  >{\raggedright\arraybackslash}p{(\linewidth - 8\tabcolsep) * \real{0.2071}}
  >{\raggedright\arraybackslash}p{(\linewidth - 8\tabcolsep) * \real{0.1673}}
  >{\raggedright\arraybackslash}p{(\linewidth - 8\tabcolsep) * \real{0.1911}}
  >{\raggedright\arraybackslash}p{(\linewidth - 8\tabcolsep) * \real{0.1435}}@{}}
\toprule\noalign{}
\multicolumn{5}{@{}>{\raggedright\arraybackslash}p{(\linewidth - 8\tabcolsep) * \real{0.9637} + 8\tabcolsep}@{}}{%
\begin{minipage}[b]{\linewidth}\raggedright
\emph{A. Residents and community characteristics by threshold and state}
\end{minipage}} \\
\midrule\noalign{}
\endfirsthead
\endhead
\bottomrule\noalign{}
\endlastfoot
Threshold/area & Residents & Tennessee share (\%) & Non-Hispanic Black (\%) & Poverty rate (\%) \\
Hotspot, ≥0.25 & 156,099 & 64.0 {[}56.4, 79.1{]} & 74.1 & 26.8 \\
Hotspot, ≥0.5 & 116,116 & 65.5 {[}52.4, 76.3{]} & 76.2 & 27.0 \\
Hotspot, ≥1.0 & 81,419 & 64.2 {[}44.7, 80.7{]} & 77.4 & 26.7 \\
Tennessee, ≥0.5 & 76,005 & --- & 92.7 & 31.3 \\
Mississippi, ≥0.5 & 40,111 & --- & 44.8 & 18.7 \\
Shelby County (Tennessee) & --- & --- & 52.3 & 17.8 \\
DeSoto County (Mississippi) & --- & --- & 32.6 & 9.7 \\
\multicolumn{5}{@{}>{\raggedright\arraybackslash}p{(\linewidth - 8\tabcolsep) * \real{0.9637} + 8\tabcolsep}@{}}{%
\emph{B. Residents by age group at the 0.5 threshold}} \\*
Indicator & Value & \multicolumn{3}{>{\raggedright\arraybackslash}p{(\linewidth - 8\tabcolsep) * \real{0.5020} + 4\tabcolsep}@{}}{%
Note} \\*
Residents, threshold 0.5 & 116,116 {[}102,954\nobreakdash--137,755{]} & \multicolumn{3}{>{\raggedright\arraybackslash}p{(\linewidth - 8\tabcolsep) * \real{0.5020} + 4\tabcolsep}@{}}{%
1,999 cross-season weighted replicates} \\
Residents aged 65 and older, threshold 0.5 & 18,026 {[}15,786\nobreakdash--21,099{]} & \multicolumn{3}{>{\raggedright\arraybackslash}p{(\linewidth - 8\tabcolsep) * \real{0.5020} + 4\tabcolsep}@{}}{%
2020 DHC table P12, 1,999 replicates} \\
Residents under 18, threshold 0.5 & 30,091 {[}26,731\nobreakdash--35,524{]} & \multicolumn{3}{>{\raggedright\arraybackslash}p{(\linewidth - 8\tabcolsep) * \real{0.5020} + 4\tabcolsep}@{}}{%
1,999 replicates} \\
\end{longtable}
}

\Needspace{14\baselineskip}\medskip\noindent \textbf{Table S8. Association between community composition and the April--July increase.}\par\nopagebreak

Population-weighted regressions across block groups for April--July. Coefficients are the increase (10\textsuperscript{15} molecules cm\textsuperscript{\ensuremath{-}2}) per 10 percentage points of population share or per \$10,000 of household income, and brackets show 95\% intervals. With distance and direction controls, all three intervals include zero.

{\small\def\LTcaptype{none}
\begin{longtable}[]{@{}
  >{\raggedright\arraybackslash}p{(\linewidth - 6\tabcolsep) * \real{0.3071}}
  >{\raggedright\arraybackslash}p{(\linewidth - 6\tabcolsep) * \real{0.2560}}
  >{\raggedright\arraybackslash}p{(\linewidth - 6\tabcolsep) * \real{0.2663}}
  >{\raggedright\arraybackslash}p{(\linewidth - 6\tabcolsep) * \real{0.1334}}@{}}
\toprule\noalign{}
\begin{minipage}[b]{\linewidth}\raggedright
Variable
\end{minipage} & \begin{minipage}[b]{\linewidth}\raggedright
Unadjusted
\end{minipage} & \begin{minipage}[b]{\linewidth}\raggedright
Controlling for distance and direction
\end{minipage} & \begin{minipage}[b]{\linewidth}\raggedright
Sample size
\end{minipage} \\
\midrule\noalign{}
\endfirsthead
\endhead
\bottomrule\noalign{}
\endlastfoot
Non-Hispanic Black share & 0.035 {[}0.013, 0.049{]} & −0.010 {[}−0.018, 0.001{]} & 587 \\
Share below the poverty line & 0.060 {[}0.018, 0.074{]} & 0.005 {[}−0.006, 0.013{]} & 582 \\
Median household income & −0.026 {[}−0.036, −0.009{]} & 0.002 {[}−0.004, 0.007{]} & 514 \\
\end{longtable}
}

\Needspace{14\baselineskip}\medskip\noindent \textbf{Table S9. Engineering emission inventory and fuel-based estimates.}\par\nopagebreak

Rows by model extrapolate the June operating fraction of each model to time on site in April--July with a load fraction of 0.8, and panel A intensities use generation as the denominator. June delivered electricity uses a delivery efficiency of 0.97. The fuel-based estimate uses scheduled gas quantities as a proxy for activity.

{\small\def\LTcaptype{none}
\begin{longtable}[]{@{}
  >{\raggedright\arraybackslash}p{(\linewidth - 10\tabcolsep) * \real{0.2004}}
  >{\raggedright\arraybackslash}p{(\linewidth - 10\tabcolsep) * \real{0.1629}}
  >{\raggedright\arraybackslash}p{(\linewidth - 10\tabcolsep) * \real{0.1629}}
  >{\raggedright\arraybackslash}p{(\linewidth - 10\tabcolsep) * \real{0.1254}}
  >{\raggedright\arraybackslash}p{(\linewidth - 10\tabcolsep) * \real{0.1629}}
  >{\raggedright\arraybackslash}p{(\linewidth - 10\tabcolsep) * \real{0.1504}}@{}}
\toprule\noalign{}
\multicolumn{6}{@{}>{\raggedright\arraybackslash}p{(\linewidth - 10\tabcolsep) * \real{0.9647} + 10\tabcolsep}@{}}{%
\begin{minipage}[b]{\linewidth}\raggedright
\emph{A. Inventory by model (April--July 2026)}
\end{minipage}} \\
\midrule\noalign{}
\endfirsthead
\endhead
\bottomrule\noalign{}
\endlastfoot
Model & Generation (GWh) & NO\emph{\textsubscript{x}} (metric tons) & Intensity (kg/MWh) & With SCR added (metric tons) & At 2 ppm (metric tons) \\
SMT-130 & 371.609 & 14.831 & 0.040 & 14.831 & 13.953 \\
TM2500 Gen6 & 378.744 & 134.842 & 0.356 & 13.484 & 10.569 \\
TM2500 Gen4 & 18.289 & 6.511 & 0.356 & 0.651 & 0.510 \\
TM2500 Gen8 & 434.668 & 154.753 & 0.356 & 15.475 & 12.130 \\
FT8 MOBILEPAC & 96.249 & 38.477 & 0.400 & 3.848 & 3.016 \\
M35 & 68.548 & 45.271 & 0.660 & 4.527 & 2.566 \\
TM2500 Gen7 & 205.382 & 135.640 & 0.660 & 13.564 & 7.688 \\
JT35/35060 & 113.288 & 74.819 & 0.660 & 7.482 & 4.241 \\
TM2500 Gen2 & 39.490 & 26.080 & 0.660 & 2.608 & 1.478 \\
All units, April--July & 1,726.265 & 631.225 & 0.366 & 76.470 & 56.151 \\
\multicolumn{6}{@{}>{\raggedright\arraybackslash}p{(\linewidth - 10\tabcolsep) * \real{0.9647} + 10\tabcolsep}@{}}{%
\emph{B. Operating records, June 2026}} \\*
Model & Units (operating) & Unit-hours & Delivered MWh & NO\emph{\textsubscript{x}} (metric tons) & kg/delivered MWh \\*
FT8 MOBILEPAC & 12 (8) & 1,036 & 23,636 & 9.741 & 0.412 \\
JT35/35060 & 3 (3) & 1,071 & 29,088 & 19.805 & 0.681 \\
M35 & 1 (1) & 602 & 16,350 & 11.132 & 0.681 \\
SMT-130 & 14 (14) & 7,139 & 88,638 & 3.647 & 0.041 \\
TM2500 Gen2 & 4 (3) & 747 & 13,043 & 8.880 & 0.681 \\
TM2500 Gen4 & 2 (2) & 326 & 5,692 & 2.089 & 0.367 \\
TM2500 Gen6 & 7 (7) & 4,093 & 98,423 & 36.125 & 0.367 \\
TM2500 Gen7 & 5 (4) & 2,165 & 52,061 & 35.446 & 0.681 \\
TM2500 Gen8 & 11 (10) & 4,260 & 117,581 & 43.157 & 0.367 \\
Total without SCR & 45 (38) & 14,300 & 355,875 & 166.375 & 0.468 \\
All units & 59 (52) & 21,439 & 444,512 & 170.022 & 0.382 \\
\multicolumn{6}{@{}>{\raggedright\arraybackslash}p{(\linewidth - 10\tabcolsep) * \real{0.9647} + 10\tabcolsep}@{}}{%
\emph{C. Emission checks by period}} \\*
Indicator & Value & \multicolumn{4}{>{\raggedright\arraybackslash}p{(\linewidth - 10\tabcolsep) * \real{0.6015} + 6\tabcolsep}@{}}{%
Note} \\*
April--July reconstruction & 215.6 kg/h & \multicolumn{4}{>{\raggedright\arraybackslash}p{(\linewidth - 10\tabcolsep) * \real{0.6015} + 6\tabcolsep}@{}}{%
June operating fraction extrapolated to time on site} \\
June fuel input sensitivity, load fraction 0.8 & 144,285\nobreakdash--193,203 MMBtu per day & \multicolumn{4}{>{\raggedright\arraybackslash}p{(\linewidth - 10\tabcolsep) * \real{0.6015} + 6\tabcolsep}@{}}{%
June operating records, with 9 to 13 MMBtu/MWh for units without SCR and 11.23 for units with SCR} \\
June only & 236 kg/h & \multicolumn{4}{>{\raggedright\arraybackslash}p{(\linewidth - 10\tabcolsep) * \real{0.6015} + 6\tabcolsep}@{}}{%
June operating records} \\
Fuel proxy, 13 July--30 September & 181\nobreakdash--309 kg/h & \multicolumn{4}{>{\raggedright\arraybackslash}p{(\linewidth - 10\tabcolsep) * \real{0.6015} + 6\tabcolsep}@{}}{%
Proxy from scheduled gas quantities, range of monthly means across factor and heat-rate scenarios} \\
Inventory extended to April--September & 205.8\nobreakdash--211.5 kg/h & \multicolumn{4}{>{\raggedright\arraybackslash}p{(\linewidth - 10\tabcolsep) * \real{0.6015} + 6\tabcolsep}@{}}{%
Bounds on on-site dates} \\
April--September, inventory and fuel-based estimate combined & 204.5\nobreakdash--250.4 kg/h & \multicolumn{4}{>{\raggedright\arraybackslash}p{(\linewidth - 10\tabcolsep) * \real{0.6015} + 6\tabcolsep}@{}}{%
Inventory for 1 April--12 July, fuel-based for 13 July--30 September} \\
Annual permit, permanent plant & 401.04 short tons (363.8 metric tons) & \multicolumn{4}{>{\raggedright\arraybackslash}p{(\linewidth - 10\tabcolsep) * \real{0.6015} + 6\tabcolsep}@{}}{%
41 units, a separate emission source} \\
\multicolumn{6}{@{}>{\raggedright\arraybackslash}p{(\linewidth - 10\tabcolsep) * \real{0.9647} + 10\tabcolsep}@{}}{%
\emph{D. Heat rate and NO\textsubscript{x} emission factor by model}} \\*
Model & Heat rate (MMBtu/MWh) & NO\emph{\textsubscript{x}} emission factor (lb/MMBtu) & \multicolumn{3}{>{\raggedright\arraybackslash}p{(\linewidth - 10\tabcolsep) * \real{0.4386} + 4\tabcolsep}@{}}{%
Basis} \\*
SMT-130 & 11.23 & 0.0078 & \multicolumn{3}{>{\raggedright\arraybackslash}p{(\linewidth - 10\tabcolsep) * \real{0.4386} + 4\tabcolsep}@{}}{%
PSD permit of the permanent plant (PGM-130)} \\
TM2500 Gen4, Gen6, Gen8 & 8.35 & 0.094 & \multicolumn{3}{>{\raggedright\arraybackslash}p{(\linewidth - 10\tabcolsep) * \real{0.4386} + 4\tabcolsep}@{}}{%
Manufacturer specifications (\emph{127})} \\
FT8 MOBILEPAC & 9.376 & 0.094 & \multicolumn{3}{>{\raggedright\arraybackslash}p{(\linewidth - 10\tabcolsep) * \real{0.4386} + 4\tabcolsep}@{}}{%
Manufacturer specifications (\emph{127})} \\
M35, JT35/35060, TM2500 Gen2, Gen7 & 11.2 & 0.13 & \multicolumn{3}{>{\raggedright\arraybackslash}p{(\linewidth - 10\tabcolsep) * \real{0.4386} + 4\tabcolsep}@{}}{%
EIA tested heat rate and AP-42 factor with water injection} \\
\multicolumn{6}{@{}>{\raggedright\arraybackslash}p{(\linewidth - 10\tabcolsep) * \real{0.9647} + 10\tabcolsep}@{}}{%
Emission factors are in pounds of NO\emph{\textsubscript{x}} (as NO\textsubscript{2}) per MMBtu of fuel input on a higher heating value basis.} \\
\end{longtable}
}

\Needspace{14\baselineskip}\medskip\noindent \textbf{Table S10. Health scenarios under different models, release heights, and sites.}\par\nopagebreak

Deaths are annualized deaths from particulate nitrate, and damages value each death at \$11.3 million (2019 dollars). The AP4 elevated-source proxy uses the AP4 coefficients of the nearby TVA Southaven combined-cycle plant. EASIUR provides no attribution to receptor locations and is therefore not used in panel D. Per-GW values multiply fleet deaths and damages by the NO\emph{\textsubscript{x}} share of units without SCR (97.7\%) and divide by their calendar-day-weighted mean capacity over the 122 days (1.252 GW), whereas table S1 averages the four monthly means (1.249 GW).

{\small\def\LTcaptype{none}
\begin{longtable}[]{@{}
  >{\raggedright\arraybackslash}p{(\linewidth - 10\tabcolsep) * \real{0.1953}}
  >{\raggedright\arraybackslash}p{(\linewidth - 10\tabcolsep) * \real{0.1587}}
  >{\raggedright\arraybackslash}p{(\linewidth - 10\tabcolsep) * \real{0.1587}}
  >{\raggedright\arraybackslash}p{(\linewidth - 10\tabcolsep) * \real{0.1587}}
  >{\raggedright\arraybackslash}p{(\linewidth - 10\tabcolsep) * \real{0.1526}}
  >{\raggedright\arraybackslash}p{(\linewidth - 10\tabcolsep) * \real{0.1406}}@{}}
\toprule\noalign{}
\multicolumn{6}{@{}>{\raggedright\arraybackslash}p{(\linewidth - 10\tabcolsep) * \real{0.9647} + 10\tabcolsep}@{}}{%
\begin{minipage}[b]{\linewidth}\raggedright
\emph{A. Deaths and damages by model and release height}
\end{minipage}} \\
\midrule\noalign{}
\endfirsthead
\endhead
\bottomrule\noalign{}
\endlastfoot
Model and release height & Deaths per year at 216 kg/h & Deaths per year at 162\nobreakdash--269 kg/h & \multicolumn{3}{>{\raggedright\arraybackslash}p{(\linewidth - 10\tabcolsep) * \real{0.4519} + 4\tabcolsep}@{}}{%
Damages at 216 kg/h (million dollars per year)} \\
InMAP main setting, hourly effective heights & 4.358 & 3.269\nobreakdash--5.428 & \multicolumn{3}{>{\raggedright\arraybackslash}p{(\linewidth - 10\tabcolsep) * \real{0.4519} + 4\tabcolsep}@{}}{%
49.2} \\
InMAP, ground level & 5.437 & 4.078\nobreakdash--6.772 & \multicolumn{3}{>{\raggedright\arraybackslash}p{(\linewidth - 10\tabcolsep) * \real{0.4519} + 4\tabcolsep}@{}}{%
61.4} \\
AP4, ground level & 5.999 & 4.500\nobreakdash--7.472 & \multicolumn{3}{>{\raggedright\arraybackslash}p{(\linewidth - 10\tabcolsep) * \real{0.4519} + 4\tabcolsep}@{}}{%
67.8} \\
AP4, elevated-source proxy & 4.420 & 3.315\nobreakdash--5.504 & \multicolumn{3}{>{\raggedright\arraybackslash}p{(\linewidth - 10\tabcolsep) * \real{0.4519} + 4\tabcolsep}@{}}{%
49.9} \\
EASIUR, area source & 1.510 & 1.132\nobreakdash--1.880 & \multicolumn{3}{>{\raggedright\arraybackslash}p{(\linewidth - 10\tabcolsep) * \real{0.4519} + 4\tabcolsep}@{}}{%
17.1} \\
InMAP main setting, per GW of capacity without SCR & 3.399 & --- & \multicolumn{3}{>{\raggedright\arraybackslash}p{(\linewidth - 10\tabcolsep) * \real{0.4519} + 4\tabcolsep}@{}}{%
38.4} \\
\multicolumn{6}{@{}>{\raggedright\arraybackslash}p{(\linewidth - 10\tabcolsep) * \real{0.9647} + 10\tabcolsep}@{}}{%
\emph{B. Other concentration--response functions and emission period (InMAP main setting)}} \\*
Case & Deaths per year & \multicolumn{4}{>{\raggedright\arraybackslash}p{(\linewidth - 10\tabcolsep) * \real{0.6107} + 6\tabcolsep}@{}}{%
Basis} \\*
Pope et al. (\emph{57}) coefficient & 8.48 & \multicolumn{4}{>{\raggedright\arraybackslash}p{(\linewidth - 10\tabcolsep) * \real{0.6107} + 6\tabcolsep}@{}}{%
Adults aged 30 and older} \\
Wu et al. (\emph{137}) coefficient & 3.36 & \multicolumn{4}{>{\raggedright\arraybackslash}p{(\linewidth - 10\tabcolsep) * \real{0.6107} + 6\tabcolsep}@{}}{%
Adults aged 65 and older} \\
Emissions for April--September & 4.13\nobreakdash--5.05 & \multicolumn{4}{>{\raggedright\arraybackslash}p{(\linewidth - 10\tabcolsep) * \real{0.6107} + 6\tabcolsep}@{}}{%
Inventory to 12 July, fuel-based estimate thereafter} \\
\multicolumn{6}{@{}>{\raggedright\arraybackslash}p{(\linewidth - 10\tabcolsep) * \real{0.9647} + 10\tabcolsep}@{}}{%
\emph{C. Release-layer weights in the InMAP main setting}} \\*
Turbine family & Share of fleet NO\emph{\textsubscript{x}}, April--July (\%) & 29-m layer & 315-m layer & \multicolumn{2}{>{\raggedright\arraybackslash}p{(\linewidth - 10\tabcolsep) * \real{0.2932} + 2\tabcolsep}@{}}{%
901-m layer} \\*
TM2500, M35, and JT35 (water injection) & 91.6 & 0.193 & 0.753 & \multicolumn{2}{>{\raggedright\arraybackslash}p{(\linewidth - 10\tabcolsep) * \real{0.2932} + 2\tabcolsep}@{}}{%
0.054} \\
FT8 MOBILEPAC (water injection) & 6.1 & 0.174 & 0.769 & \multicolumn{2}{>{\raggedright\arraybackslash}p{(\linewidth - 10\tabcolsep) * \real{0.2932} + 2\tabcolsep}@{}}{%
0.058} \\
SMT-130 (SCR) & 2.3 & 0.277 & 0.689 & \multicolumn{2}{>{\raggedright\arraybackslash}p{(\linewidth - 10\tabcolsep) * \real{0.2932} + 2\tabcolsep}@{}}{%
0.035} \\
\multicolumn{6}{@{}>{\raggedright\arraybackslash}p{(\linewidth - 10\tabcolsep) * \real{0.9647} + 10\tabcolsep}@{}}{%
Weights are the shares of 17,544 hours of 2024\nobreakdash--2025 ERA5 meteorology in which the effective release height falls in, or is interpolated to, each matrix layer (materials and methods). Stack height, exhaust temperature, and exhaust mass flow rate are 8.0 m, 801 K, and 80.8 kg s\textsuperscript{\ensuremath{-}1} for the TM2500, M35, and JT35, 10.0 m, 764 K, and 92 kg s\textsuperscript{\ensuremath{-}1} for the FT8 MOBILEPAC, and 10.7 m, 728 K, and 64.8 kg s\textsuperscript{\ensuremath{-}1} for the SMT-130. The first set is reported for TM2500 Gen4 units at the Roseville Energy Park at an ambient temperature of 38°C (\emph{155}). The exhaust conditions of the other two families and the SMT-130 stack height come from the FT8 MOBILEPAC specification sheet and from the design drawing of the SMT-130 SCR system, which includes tempering air (\emph{156}). The FT8 MOBILEPAC stack height is assumed.} \\
\multicolumn{6}{@{}>{\raggedright\arraybackslash}p{(\linewidth - 10\tabcolsep) * \real{0.9647} + 10\tabcolsep}@{}}{%
\emph{D. In-state share of deaths at other on-site generation sites}} \\*
Site & State & Distance to nearest state line (km) & InMAP, hourly heights (\%) & InMAP, ground level (\%) & AP4, ground level (\%) \\*
Southaven (this fleet) & Mississippi & 1.6 & 18.2 & 24.3 & 3.1 \\
Colossus 1 & Tennessee & 2.0 & 27.5 & 28.1 & 8.2 \\
Abilene & Texas & 198 & 26.0 & 29.3 & 26.3 \\
Socrates South & Ohio & 126 & 52.4 & 58.2 & 26.8 \\
Socrates North & Ohio & 131 & 52.4 & 58.2 & 26.8 \\
Fermi Matador & Texas & 132 & 12.7 & 13.9 & 12.5 \\
\multicolumn{6}{@{}>{\raggedright\arraybackslash}p{(\linewidth - 10\tabcolsep) * \real{0.9647} + 10\tabcolsep}@{}}{%
Shares are percentages of national deaths that occur in the host state. Other sites use this fleet's emissions and release-height mixture. Socrates South and North are two sites of one project and fall in the same InMAP source cell.} \\
\end{longtable}
}

\Needspace{14\baselineskip}\medskip\noindent \textbf{Table S11. Damages per megawatt-hour from on-site generation and the displaced grid.}\par\nopagebreak

The on-site side uses June operating hours, a load fraction of 0.8, and a delivery efficiency of 0.97. The grid side uses flat 1-MW and 10-MW load reductions in AVERT and is expressed per megawatt-hour of generation displaced. Ranges span the two grid cases and the on-site particulate and fuel sulfur assumptions. Ammonia slip of 1.4 and 2.8 ppm corresponds to the permanent plant's permit limits at the same site.

{\small\def\LTcaptype{none}
\begin{longtable}[]{@{}
  >{\raggedright\arraybackslash}p{(\linewidth - 6\tabcolsep) * \real{0.3275}}
  >{\raggedright\arraybackslash}p{(\linewidth - 6\tabcolsep) * \real{0.2663}}
  >{\raggedright\arraybackslash}p{(\linewidth - 6\tabcolsep) * \real{0.1437}}
  >{\raggedright\arraybackslash}p{(\linewidth - 6\tabcolsep) * \real{0.2254}}@{}}
\toprule\noalign{}
\multicolumn{4}{@{}>{\raggedright\arraybackslash}p{(\linewidth - 6\tabcolsep) * \real{0.9628} + 6\tabcolsep}@{}}{%
\begin{minipage}[b]{\linewidth}\raggedright
\emph{A. Damages per megawatt-hour and ratio of on-site to grid damages, NO\textsubscript{x} only}
\end{minipage}} \\
\midrule\noalign{}
\endfirsthead
\endhead
\bottomrule\noalign{}
\endlastfoot
Model or quantity & \multicolumn{3}{>{\raggedright\arraybackslash}p{(\linewidth - 6\tabcolsep) * \real{0.6353} + 4\tabcolsep}@{}}{%
NO\emph{\textsubscript{x}} only} \\
InMAP main setting, hourly heights & \multicolumn{3}{>{\raggedright\arraybackslash}p{(\linewidth - 6\tabcolsep) * \real{0.6353} + 4\tabcolsep}@{}}{%
3.613\nobreakdash--3.624} \\
InMAP ground level & \multicolumn{3}{>{\raggedright\arraybackslash}p{(\linewidth - 6\tabcolsep) * \real{0.6353} + 4\tabcolsep}@{}}{%
3.858\nobreakdash--5.136} \\
AP4 ground level & \multicolumn{3}{>{\raggedright\arraybackslash}p{(\linewidth - 6\tabcolsep) * \real{0.6353} + 4\tabcolsep}@{}}{%
2.335\nobreakdash--2.493} \\
EASIUR area source & \multicolumn{3}{>{\raggedright\arraybackslash}p{(\linewidth - 6\tabcolsep) * \real{0.6353} + 4\tabcolsep}@{}}{%
2.940\nobreakdash--3.136} \\
Main setting NO\emph{\textsubscript{x}} damages (\$/delivered MWh) & \multicolumn{3}{>{\raggedright\arraybackslash}p{(\linewidth - 6\tabcolsep) * \real{0.6353} + 4\tabcolsep}@{}}{%
9.955} \\
Grid NO\emph{\textsubscript{x}} damages (\$/displaced MWh) & \multicolumn{3}{>{\raggedright\arraybackslash}p{(\linewidth - 6\tabcolsep) * \real{0.6353} + 4\tabcolsep}@{}}{%
2.747\nobreakdash--2.755} \\
\multicolumn{4}{@{}>{\raggedright\arraybackslash}p{(\linewidth - 6\tabcolsep) * \real{0.9628} + 6\tabcolsep}@{}}{%
\emph{B. Ratio for NO\textsubscript{x}, primary PM\textsubscript{2.5}, and SO\textsubscript{2} before and after SCR (eight paired scenarios per model)}} \\*
Model & Before SCR & \multicolumn{2}{>{\raggedright\arraybackslash}p{(\linewidth - 6\tabcolsep) * \real{0.3691} + 2\tabcolsep}@{}}{%
After SCR (no ammonia)} \\*
InMAP main setting & 0.950\nobreakdash--1.042 & \multicolumn{2}{>{\raggedright\arraybackslash}p{(\linewidth - 6\tabcolsep) * \real{0.3691} + 2\tabcolsep}@{}}{%
0.398\nobreakdash--0.490} \\
InMAP ground level & 1.308\nobreakdash--1.439 & \multicolumn{2}{>{\raggedright\arraybackslash}p{(\linewidth - 6\tabcolsep) * \real{0.3691} + 2\tabcolsep}@{}}{%
0.620\nobreakdash--0.749} \\
AP4 ground level & 0.606\nobreakdash--0.684 & \multicolumn{2}{>{\raggedright\arraybackslash}p{(\linewidth - 6\tabcolsep) * \real{0.3691} + 2\tabcolsep}@{}}{%
0.304\nobreakdash--0.383} \\
EASIUR area source & 0.792\nobreakdash--0.924 & \multicolumn{2}{>{\raggedright\arraybackslash}p{(\linewidth - 6\tabcolsep) * \real{0.3691} + 2\tabcolsep}@{}}{%
0.575\nobreakdash--0.707} \\
\multicolumn{4}{@{}>{\raggedright\arraybackslash}p{(\linewidth - 6\tabcolsep) * \real{0.9628} + 6\tabcolsep}@{}}{%
\emph{C. Ratio with SCR and ammonia slip}} \\*
Model & Ammonia slip (ppm, dry at 15\% O\textsubscript{2}) & On-site/grid & Cases with ratio ≥ 1 \\*
InMAP main setting & 1.4 & 0.467\nobreakdash--0.558 & 0/8 \\
InMAP main setting & 2.8 & 0.536\nobreakdash--0.627 & 0/8 \\
InMAP main setting & 3.5 & 0.570\nobreakdash--0.661 & 0/8 \\
InMAP ground level & 1.4 & 0.727\nobreakdash--0.857 & 0/8 \\
InMAP ground level & 2.8 & 0.834\nobreakdash--0.964 & 0/8 \\
InMAP ground level & 3.5 & 0.888\nobreakdash--1.018 & 2/8 \\
AP4 ground level & 1.4 & 0.345\nobreakdash--0.423 & 0/8 \\
AP4 ground level & 2.8 & 0.385\nobreakdash--0.463 & 0/8 \\
AP4 ground level & 3.5 & 0.405\nobreakdash--0.484 & 0/8 \\
EASIUR area source & 1.4 & 0.680\nobreakdash--0.812 & 0/8 \\
EASIUR area source & 2.8 & 0.785\nobreakdash--0.917 & 0/8 \\
EASIUR area source & 3.5 & 0.837\nobreakdash--0.970 & 0/8 \\
\end{longtable}
}

\Needspace{14\baselineskip}\medskip\noindent \textbf{Table S12. Reductions in deaths under control and dispatch scenarios.}\par\nopagebreak

Reductions are relative to the engineering baseline for the same period and model. Dispatch scenarios hold June 2026 delivered electricity fixed. April to July engineering baselines use the inventory rate of 215.6 kg per hour and June baselines use the June rate of 236 kg per hour, so they differ slightly from table S10, which is evaluated at 216 kg per hour. Ammonia slip is not included.

{\small\def\LTcaptype{none}
\begin{longtable}[]{@{}
  >{\raggedright\arraybackslash}p{(\linewidth - 6\tabcolsep) * \real{0.3420}}
  >{\raggedright\arraybackslash}p{(\linewidth - 6\tabcolsep) * \real{0.2140}}
  >{\raggedright\arraybackslash}p{(\linewidth - 6\tabcolsep) * \real{0.1927}}
  >{\raggedright\arraybackslash}p{(\linewidth - 6\tabcolsep) * \real{0.2141}}@{}}
\toprule\noalign{}
\multicolumn{4}{@{}>{\raggedright\arraybackslash}p{(\linewidth - 6\tabcolsep) * \real{0.9627} + 6\tabcolsep}@{}}{%
\begin{minipage}[b]{\linewidth}\raggedright
\emph{A. Reduction in deaths (\%)}
\end{minipage}} \\
\midrule\noalign{}
\endfirsthead
\endhead
\bottomrule\noalign{}
\endlastfoot
Scenario and period & InMAP main setting & AP4 ground level & EASIUR area source \\
June 2026, SCR on all units without SCR (90\% removal) & 88.0 & 88.1 & 88.1 \\
June 2026, all units at 2 ppm & 91.2 & 91.3 & 91.3 \\
June 2026, lowest-emitting units dispatched first, load fraction 0.6 & 20.0 & 19.9 & 19.9 \\
June 2026, lowest-emitting units dispatched first, load fraction 0.8 & 28.0 & 28.0 & 28.0 \\
June 2026, lowest-emitting units dispatched first, load fraction 1.0 & 33.9 & 34.0 & 34.0 \\
April--July 2026, SCR on all units without SCR (90\% removal) & 87.8 & 87.9 & 87.9 \\
April--July 2026, all units at 2 ppm & 91.0 & 91.1 & 91.1 \\
\multicolumn{4}{@{}>{\raggedright\arraybackslash}p{(\linewidth - 6\tabcolsep) * \real{0.9627} + 6\tabcolsep}@{}}{%
\emph{B. Annualized deaths (deaths per year)}} \\*
Scenario and period & InMAP main setting & AP4 ground level & EASIUR area source \\*
June 2026, engineering baseline & 4.765 & 6.559 & 1.650 \\
June 2026, SCR on all units without SCR (90\% removal) & 0.571 & 0.783 & 0.197 \\
June 2026, all units at 2 ppm & 0.418 & 0.572 & 0.144 \\
June 2026, lowest-emitting units dispatched first, load fraction 0.6 & 3.813 & 5.254 & 1.322 \\
June 2026, lowest-emitting units dispatched first, load fraction 0.8 & 3.432 & 4.720 & 1.188 \\
June 2026, lowest-emitting units dispatched first, load fraction 1.0 & 3.148 & 4.327 & 1.089 \\
April--July 2026, engineering baseline & 4.350 & 5.988 & 1.507 \\
April--July 2026, SCR on all units without SCR (90\% removal) & 0.530 & 0.725 & 0.183 \\
April--July 2026, all units at 2 ppm & 0.390 & 0.533 & 0.134 \\
\end{longtable}
}

\Needspace{14\baselineskip}\medskip\noindent \textbf{Table S13. Benefits and costs of SCR for the 29 evaluated units.}\par\nopagebreak

Panels B, D, and E give ratios for 30 years (main evaluation) and 20 years. The two ends of each range keep or remove taxes and insurance. Ammonia slip is in ppm, dry at 15\% oxygen. Operating intensity and receptor distribution follow this site in April--July 2026. EASIUR provides no attribution to receptor states. All panels except E exclude ozone.

{\footnotesize\def\LTcaptype{none}
\begin{longtable}[]{@{}
  >{\raggedright\arraybackslash}p{(\linewidth - 10\tabcolsep) * \real{0.2729}}
  >{\raggedright\arraybackslash}p{(\linewidth - 10\tabcolsep) * \real{0.2217}}
  >{\raggedright\arraybackslash}p{(\linewidth - 10\tabcolsep) * \real{0.1282}}
  >{\raggedright\arraybackslash}p{(\linewidth - 10\tabcolsep) * \real{0.1196}}
  >{\raggedright\arraybackslash}p{(\linewidth - 10\tabcolsep) * \real{0.1112}}
  >{\raggedright\arraybackslash}p{(\linewidth - 10\tabcolsep) * \real{0.1112}}@{}}
\toprule\noalign{}
\multicolumn{6}{@{}>{\raggedright\arraybackslash}p{(\linewidth - 10\tabcolsep) * \real{0.9648} + 10\tabcolsep}@{}}{%
\begin{minipage}[b]{\linewidth}\raggedright
\emph{A. Inputs to Eq. S7 for the 29 evaluated units (InMAP main setting, 2019 dollars)}
\end{minipage}} \\
\midrule\noalign{}
\endfirsthead
\endhead
\bottomrule\noalign{}
\endlastfoot
Parameter & Value & \multicolumn{4}{>{\raggedright\arraybackslash}p{(\linewidth - 10\tabcolsep) * \real{0.4702} + 6\tabcolsep}@{}}{%
Note} \\
Initial investment for SCR on new turbines, \emph{K} & 96.7 million dollars & \multicolumn{4}{>{\raggedright\arraybackslash}p{(\linewidth - 10\tabcolsep) * \real{0.4702} + 6\tabcolsep}@{}}{%
Retrofits use 1.25\emph{K}} \\
Annual operation and maintenance, \emph{O} & 5.28 million dollars per year & \multicolumn{4}{>{\raggedright\arraybackslash}p{(\linewidth - 10\tabcolsep) * \real{0.4702} + 6\tabcolsep}@{}}{%
Includes taxes and insurance of 1.10 million dollars per year, removed at the upper end of each range} \\
NO\emph{\textsubscript{x}} removed & 915.5 metric tons per year & \multicolumn{4}{>{\raggedright\arraybackslash}p{(\linewidth - 10\tabcolsep) * \real{0.4702} + 6\tabcolsep}@{}}{%
90\% removal} \\
Annual avoided damage from NO\emph{\textsubscript{x}} reductions, all states & 23.8 million dollars per year & \multicolumn{4}{>{\raggedright\arraybackslash}p{(\linewidth - 10\tabcolsep) * \real{0.4702} + 6\tabcolsep}@{}}{%
} \\
Annual avoided damage from NO\emph{\textsubscript{x}} reductions, Mississippi only & 4.32 million dollars per year & \multicolumn{4}{>{\raggedright\arraybackslash}p{(\linewidth - 10\tabcolsep) * \real{0.4702} + 6\tabcolsep}@{}}{%
} \\
Annual ammonia damage per 1 ppm of slip, all states & 1.64 million dollars per year & \multicolumn{4}{>{\raggedright\arraybackslash}p{(\linewidth - 10\tabcolsep) * \real{0.4702} + 6\tabcolsep}@{}}{%
Dry at 15\% oxygen} \\
Annual ammonia damage per 1 ppm of slip, Mississippi only & 0.344 million dollars per year & \multicolumn{4}{>{\raggedright\arraybackslash}p{(\linewidth - 10\tabcolsep) * \real{0.4702} + 6\tabcolsep}@{}}{%
Dry at 15\% oxygen} \\
Lag discount factor, \emph{L}(\emph{r}) & 0.816 (7\%), 0.906 (3\%) & \multicolumn{4}{>{\raggedright\arraybackslash}p{(\linewidth - 10\tabcolsep) * \real{0.4702} + 6\tabcolsep}@{}}{%
} \\
Annuity factor over 30 years, \emph{A}(\emph{r}, 30) & 12.41 (7\%), 19.60 (3\%) & \multicolumn{4}{>{\raggedright\arraybackslash}p{(\linewidth - 10\tabcolsep) * \real{0.4702} + 6\tabcolsep}@{}}{%
} \\
\multicolumn{6}{@{}>{\raggedright\arraybackslash}p{(\linewidth - 10\tabcolsep) * \real{0.9648} + 10\tabcolsep}@{}}{%
At 7\% over 30 years without ammonia slip, these inputs give ratios of 1.49 with taxes and insurance and 1.62 without them.} \\
\multicolumn{6}{@{}>{\raggedright\arraybackslash}p{(\linewidth - 10\tabcolsep) * \real{0.9648} + 10\tabcolsep}@{}}{%
\emph{B. Benefit--cost ratios by model, geographic scope, discount rate, and ammonia slip}} \\*
Model & Benefits counted & Discount rate & Ammonia slip & 30 years & 20 years \\*
InMAP main setting & All states & 7\% & 0 & 1.487\nobreakdash--1.624 & 1.349\nobreakdash--1.461 \\
InMAP main setting & All states & 7\% & 1.4 & 1.343\nobreakdash--1.467 & 1.218\nobreakdash--1.320 \\
InMAP main setting & All states & 7\% & 2.8 & 1.199\nobreakdash--1.310 & 1.088\nobreakdash--1.178 \\
InMAP main setting & All states & 7\% & 3.5 & 1.127\nobreakdash--1.231 & 1.023\nobreakdash--1.108 \\
InMAP main setting & All states & 3\% & 0 & 2.113\nobreakdash--2.369 & 1.832\nobreakdash--2.021 \\
InMAP main setting & All states & 3\% & 1.4 & 1.908\nobreakdash--2.140 & 1.655\nobreakdash--1.826 \\
InMAP main setting & All states & 3\% & 2.8 & 1.704\nobreakdash--1.911 & 1.478\nobreakdash--1.631 \\
InMAP main setting & All states & 3\% & 3.5 & 1.602\nobreakdash--1.796 & 1.389\nobreakdash--1.533 \\
InMAP main setting & Mississippi only & 7\% & 0 & 0.270\nobreakdash--0.294 & 0.245\nobreakdash--0.265 \\
InMAP main setting & Mississippi only & 7\% & 1.4 & 0.239\nobreakdash--0.262 & 0.217\nobreakdash--0.235 \\
InMAP main setting & Mississippi only & 7\% & 2.8 & 0.209\nobreakdash--0.229 & 0.190\nobreakdash--0.206 \\
InMAP main setting & Mississippi only & 7\% & 3.5 & 0.194\nobreakdash--0.212 & 0.176\nobreakdash--0.191 \\
InMAP main setting & Mississippi only & 3\% & 0 & 0.383\nobreakdash--0.430 & 0.332\nobreakdash--0.367 \\
InMAP main setting & Mississippi only & 3\% & 1.4 & 0.340\nobreakdash--0.382 & 0.295\nobreakdash--0.326 \\
InMAP main setting & Mississippi only & 3\% & 2.8 & 0.298\nobreakdash--0.334 & 0.258\nobreakdash--0.285 \\
InMAP main setting & Mississippi only & 3\% & 3.5 & 0.276\nobreakdash--0.310 & 0.239\nobreakdash--0.264 \\
InMAP ground level & All states & 7\% & 0 & 1.856\nobreakdash--2.028 & 1.684\nobreakdash--1.824 \\
InMAP ground level & All states & 7\% & 1.4 & 1.628\nobreakdash--1.778 & 1.477\nobreakdash--1.600 \\
InMAP ground level & All states & 7\% & 2.8 & 1.400\nobreakdash--1.529 & 1.270\nobreakdash--1.376 \\
InMAP ground level & All states & 7\% & 3.5 & 1.286\nobreakdash--1.405 & 1.167\nobreakdash--1.264 \\
InMAP ground level & All states & 3\% & 0 & 2.638\nobreakdash--2.958 & 2.287\nobreakdash--2.524 \\
InMAP ground level & All states & 3\% & 1.4 & 2.314\nobreakdash--2.595 & 2.006\nobreakdash--2.214 \\
InMAP ground level & All states & 3\% & 2.8 & 1.990\nobreakdash--2.231 & 1.725\nobreakdash--1.904 \\
InMAP ground level & All states & 3\% & 3.5 & 1.827\nobreakdash--2.049 & 1.584\nobreakdash--1.748 \\
InMAP ground level & Mississippi only & 7\% & 0 & 0.452\nobreakdash--0.494 & 0.410\nobreakdash--0.444 \\
InMAP ground level & Mississippi only & 7\% & 1.4 & 0.388\nobreakdash--0.424 & 0.352\nobreakdash--0.381 \\
InMAP ground level & Mississippi only & 7\% & 2.8 & 0.324\nobreakdash--0.354 & 0.294\nobreakdash--0.319 \\
InMAP ground level & Mississippi only & 7\% & 3.5 & 0.292\nobreakdash--0.319 & 0.265\nobreakdash--0.287 \\
InMAP ground level & Mississippi only & 3\% & 0 & 0.642\nobreakdash--0.720 & 0.557\nobreakdash--0.614 \\
InMAP ground level & Mississippi only & 3\% & 1.4 & 0.551\nobreakdash--0.618 & 0.478\nobreakdash--0.528 \\
InMAP ground level & Mississippi only & 3\% & 2.8 & 0.461\nobreakdash--0.517 & 0.399\nobreakdash--0.441 \\
InMAP ground level & Mississippi only & 3\% & 3.5 & 0.415\nobreakdash--0.466 & 0.360\nobreakdash--0.397 \\
AP4 ground level & All states & 7\% & 0 & 2.048\nobreakdash--2.237 & 1.858\nobreakdash--2.013 \\
AP4 ground level & All states & 7\% & 1.4 & 1.832\nobreakdash--2.001 & 1.662\nobreakdash--1.800 \\
AP4 ground level & All states & 7\% & 2.8 & 1.615\nobreakdash--1.764 & 1.466\nobreakdash--1.587 \\
AP4 ground level & All states & 7\% & 3.5 & 1.507\nobreakdash--1.646 & 1.367\nobreakdash--1.481 \\
AP4 ground level & All states & 3\% & 0 & 2.911\nobreakdash--3.264 & 2.524\nobreakdash--2.785 \\
AP4 ground level & All states & 3\% & 1.4 & 2.603\nobreakdash--2.919 & 2.257\nobreakdash--2.491 \\
AP4 ground level & All states & 3\% & 2.8 & 2.296\nobreakdash--2.574 & 1.990\nobreakdash--2.196 \\
AP4 ground level & All states & 3\% & 3.5 & 2.142\nobreakdash--2.402 & 1.857\nobreakdash--2.049 \\
AP4 ground level & Mississippi only & 7\% & 0 & 0.064\nobreakdash--0.070 & 0.058\nobreakdash--0.063 \\
AP4 ground level & Mississippi only & 7\% & 1.4 & 0.038\nobreakdash--0.042 & 0.035\nobreakdash--0.038 \\
AP4 ground level & Mississippi only & 7\% & 2.8 & 0.013\nobreakdash--0.014 & 0.012\nobreakdash--0.013 \\
AP4 ground level & Mississippi only & 7\% & 3.5 & 0.000\nobreakdash--0.001 & 0.000\nobreakdash--0.000 \\
AP4 ground level & Mississippi only & 3\% & 0 & 0.091\nobreakdash--0.102 & 0.079\nobreakdash--0.087 \\
AP4 ground level & Mississippi only & 3\% & 1.4 & 0.055\nobreakdash--0.061 & 0.047\nobreakdash--0.052 \\
AP4 ground level & Mississippi only & 3\% & 2.8 & 0.019\nobreakdash--0.021 & 0.016\nobreakdash--0.018 \\
AP4 ground level & Mississippi only & 3\% & 3.5 & 0.001\nobreakdash--0.001 & 0.001\nobreakdash--0.001 \\
EASIUR area source & All states & 7\% & 0 & 0.515\nobreakdash--0.563 & 0.468\nobreakdash--0.506 \\
EASIUR area source & All states & 7\% & 1.4 & 0.319\nobreakdash--0.348 & 0.289\nobreakdash--0.313 \\
EASIUR area source & All states & 7\% & 2.8 & 0.122\nobreakdash--0.133 & 0.111\nobreakdash--0.120 \\
EASIUR area source & All states & 7\% & 3.5 & 0.024\nobreakdash--0.026 & 0.021\nobreakdash--0.023 \\
EASIUR area source & All states & 3\% & 0 & 0.732\nobreakdash--0.821 & 0.635\nobreakdash--0.701 \\
EASIUR area source & All states & 3\% & 1.4 & 0.453\nobreakdash--0.508 & 0.393\nobreakdash--0.433 \\
EASIUR area source & All states & 3\% & 2.8 & 0.173\nobreakdash--0.194 & 0.150\nobreakdash--0.166 \\
EASIUR area source & All states & 3\% & 3.5 & 0.034\nobreakdash--0.038 & 0.029\nobreakdash--0.032 \\
\multicolumn{6}{@{}>{\raggedright\arraybackslash}p{(\linewidth - 10\tabcolsep) * \real{0.9648} + 10\tabcolsep}@{}}{%
\emph{C. Allowable capital increase and break-even ammonia slip over 30 years (InMAP main setting, all states, 7\%)}} \\*
Ammonia slip (ppm) & Allowable capital increase (\%) & \multicolumn{4}{>{\raggedright\arraybackslash}p{(\linewidth - 10\tabcolsep) * \real{0.4702} + 6\tabcolsep}@{}}{%
Break-even ammonia slip (ppm)} \\*
0 & 81.6\nobreakdash--95.8 & \multicolumn{4}{>{\raggedright\arraybackslash}p{(\linewidth - 10\tabcolsep) * \real{0.4702} + 6\tabcolsep}@{}}{%
4.74\nobreakdash--5.56} \\
1.4 & 57.5\nobreakdash--71.7 & \multicolumn{4}{>{\raggedright\arraybackslash}p{(\linewidth - 10\tabcolsep) * \real{0.4702} + 6\tabcolsep}@{}}{%
4.74\nobreakdash--5.56} \\
2.8 & 33.4\nobreakdash--47.6 & \multicolumn{4}{>{\raggedright\arraybackslash}p{(\linewidth - 10\tabcolsep) * \real{0.4702} + 6\tabcolsep}@{}}{%
4.74\nobreakdash--5.56} \\
3.5 & 21.3\nobreakdash--35.5 & \multicolumn{4}{>{\raggedright\arraybackslash}p{(\linewidth - 10\tabcolsep) * \real{0.4702} + 6\tabcolsep}@{}}{%
4.74\nobreakdash--5.56} \\
\multicolumn{6}{@{}>{\raggedright\arraybackslash}p{(\linewidth - 10\tabcolsep) * \real{0.9648} + 10\tabcolsep}@{}}{%
\emph{D. Retrofits (InMAP main setting, all states, 7\%, capital 1.25 times that of SCR on new turbines)}} \\*
Ammonia slip (ppm) & Midlife replacement & 30 years & \multicolumn{3}{>{\raggedright\arraybackslash}p{(\linewidth - 10\tabcolsep) * \real{0.3420} + 4\tabcolsep}@{}}{%
20 years} \\*
0 & None & 1.29\nobreakdash--1.40 & \multicolumn{3}{>{\raggedright\arraybackslash}p{(\linewidth - 10\tabcolsep) * \real{0.3420} + 4\tabcolsep}@{}}{%
1.16\nobreakdash--1.25} \\
0 & Full system in year 15 & 1.05\nobreakdash--1.11 & \multicolumn{3}{>{\raggedright\arraybackslash}p{(\linewidth - 10\tabcolsep) * \real{0.3420} + 4\tabcolsep}@{}}{%
0.93\nobreakdash--0.99} \\
1.4 & None & 1.17\nobreakdash--1.26 & \multicolumn{3}{>{\raggedright\arraybackslash}p{(\linewidth - 10\tabcolsep) * \real{0.3420} + 4\tabcolsep}@{}}{%
1.05\nobreakdash--1.13} \\
1.4 & Full system in year 15 & 0.95\nobreakdash--1.01 & \multicolumn{3}{>{\raggedright\arraybackslash}p{(\linewidth - 10\tabcolsep) * \real{0.3420} + 4\tabcolsep}@{}}{%
0.84\nobreakdash--0.89} \\
2.8 & None & 1.04\nobreakdash--1.13 & \multicolumn{3}{>{\raggedright\arraybackslash}p{(\linewidth - 10\tabcolsep) * \real{0.3420} + 4\tabcolsep}@{}}{%
0.94\nobreakdash--1.01} \\
2.8 & Full system in year 15 & 0.84\nobreakdash--0.90 & \multicolumn{3}{>{\raggedright\arraybackslash}p{(\linewidth - 10\tabcolsep) * \real{0.3420} + 4\tabcolsep}@{}}{%
0.75\nobreakdash--0.79} \\
3.5 & None & 0.98\nobreakdash--1.06 & \multicolumn{3}{>{\raggedright\arraybackslash}p{(\linewidth - 10\tabcolsep) * \real{0.3420} + 4\tabcolsep}@{}}{%
0.88\nobreakdash--0.95} \\
3.5 & Full system in year 15 & 0.79\nobreakdash--0.84 & \multicolumn{3}{>{\raggedright\arraybackslash}p{(\linewidth - 10\tabcolsep) * \real{0.3420} + 4\tabcolsep}@{}}{%
0.71\nobreakdash--0.75} \\
\multicolumn{6}{@{}>{\raggedright\arraybackslash}p{(\linewidth - 10\tabcolsep) * \real{0.9648} + 10\tabcolsep}@{}}{%
The case with full replacement in year 15 adds the full retrofit capital discounted to year 0. Routine catalyst consumption is included in operation and maintenance.} \\
\multicolumn{6}{@{}>{\raggedright\arraybackslash}p{(\linewidth - 10\tabcolsep) * \real{0.9648} + 10\tabcolsep}@{}}{%
\emph{E. Ozone co-benefits (InMAP main setting plus EPA short-term ozone benefit per ton, all states, 7\%, no ammonia slip)}} \\*
Scenario & 30 years & \multicolumn{4}{>{\raggedright\arraybackslash}p{(\linewidth - 10\tabcolsep) * \real{0.4702} + 6\tabcolsep}@{}}{%
20 years} \\*
InMAP particulate matter + EPA short-term ozone & 1.814\nobreakdash--1.981 & \multicolumn{4}{>{\raggedright\arraybackslash}p{(\linewidth - 10\tabcolsep) * \real{0.4702} + 6\tabcolsep}@{}}{%
1.645\nobreakdash--1.782} \\
\multicolumn{6}{@{}>{\raggedright\arraybackslash}p{(\linewidth - 10\tabcolsep) * \real{0.9648} + 10\tabcolsep}@{}}{%
EPA 2024 benefit-per-ton tables, 2025 emission year, East South Central industrial boiler coefficient. Because reductions are assumed to be spread evenly over the year, 153/365 of them are assigned to the ozone season.} \\
\multicolumn{6}{@{}>{\raggedright\arraybackslash}p{(\linewidth - 10\tabcolsep) * \real{0.9648} + 10\tabcolsep}@{}}{%
\emph{F. Remaining period on site, cost per metric ton, and payback (InMAP main setting, all states)}} \\*
Item & Value & \multicolumn{4}{>{\raggedright\arraybackslash}p{(\linewidth - 10\tabcolsep) * \real{0.4702} + 6\tabcolsep}@{}}{%
Conditions} \\*
Retrofit benefit--cost ratio, remaining period on site & 0.075 & \multicolumn{4}{>{\raggedright\arraybackslash}p{(\linewidth - 10\tabcolsep) * \real{0.4702} + 6\tabcolsep}@{}}{%
No ammonia slip, 7\%, starting 2 October 2026} \\
Retrofit benefit--cost ratio, remaining period on site & 0.061 & \multicolumn{4}{>{\raggedright\arraybackslash}p{(\linewidth - 10\tabcolsep) * \real{0.4702} + 6\tabcolsep}@{}}{%
2.8 ppm, 7\%, starting 2 October 2026} \\
Benefit--cost ratio, SCR on new turbines, 1 year & 0.179\nobreakdash--0.181 & \multicolumn{4}{>{\raggedright\arraybackslash}p{(\linewidth - 10\tabcolsep) * \real{0.4702} + 6\tabcolsep}@{}}{%
No ammonia slip, 7\%, all states} \\
Benefit--cost ratio, SCR on new turbines, 2 years & 0.331\nobreakdash--0.337 & \multicolumn{4}{>{\raggedright\arraybackslash}p{(\linewidth - 10\tabcolsep) * \real{0.4702} + 6\tabcolsep}@{}}{%
No ammonia slip, 7\%, all states} \\
Cost per metric ton, SCR on new turbines, 30 years & 13,069\nobreakdash--14,275 & \multicolumn{4}{>{\raggedright\arraybackslash}p{(\linewidth - 10\tabcolsep) * \real{0.4702} + 6\tabcolsep}@{}}{%
2019 dollars per metric ton of NO\emph{\textsubscript{x}}} \\
Cost per metric ton, SCR on new turbines, 20 years & 14,527\nobreakdash--15,734 & \multicolumn{4}{>{\raggedright\arraybackslash}p{(\linewidth - 10\tabcolsep) * \real{0.4702} + 6\tabcolsep}@{}}{%
2019 dollars per metric ton of NO\emph{\textsubscript{x}}} \\
Payback, SCR on new turbines & 8.67\nobreakdash--9.62 & \multicolumn{4}{>{\raggedright\arraybackslash}p{(\linewidth - 10\tabcolsep) * \real{0.4702} + 6\tabcolsep}@{}}{%
No ammonia slip, 7\%, years} \\
Payback, retrofit & 11.95\nobreakdash--13.46 & \multicolumn{4}{>{\raggedright\arraybackslash}p{(\linewidth - 10\tabcolsep) * \real{0.4702} + 6\tabcolsep}@{}}{%
No ammonia slip, 7\%, years} \\
Payback, retrofit & 19.67\nobreakdash--24.86 & \multicolumn{4}{>{\raggedright\arraybackslash}p{(\linewidth - 10\tabcolsep) * \real{0.4702} + 6\tabcolsep}@{}}{%
2.8 ppm, 7\%, years} \\
\multicolumn{6}{@{}>{\raggedright\arraybackslash}p{(\linewidth - 10\tabcolsep) * \real{0.9648} + 10\tabcolsep}@{}}{%
The payback period is the number of years at which discounted benefits first equal costs. Capital is paid in year 0, and the cost per metric ton does not depend on ammonia slip.} \\
\end{longtable}
}

\section*{Supplementary Data}

\textbf{Data S1.} Provided as an ancillary file with this arXiv submission (\texttt{Data\_S1.xlsx}, listed under Ancillary files on the arXiv abstract page).

Multi-sheet Excel workbook containing the tabulated data behind Figs. 1 to 4 and figs. S1 to S4, the paired TEMPO scans, unit-level operating records for June 2026, daily scheduled gas deliveries, and a statistics sheet giving estimates with 95\% intervals, sample sizes, and \emph{P} values. The first worksheet gives the figure or table and the contents of each worksheet.

\end{document}